\pdfoutput=1
\documentclass[aps,prb,reprint,superscriptaddress,amsmath,amssymb]{revtex4-2}
\usepackage{graphicx}
\usepackage{bm}
\usepackage{array}
\usepackage[dvipsnames]{xcolor}
\usepackage[colorlinks=true,linkcolor=blue,citecolor=blue,urlcolor=blue]{hyperref}

\newcommand{\bk}{\mathbf{k}}
\newcommand{\be}{\hat{\mathbf e}}
\newcommand{\jpar}{j_\parallel}
\newcommand{\jperp}{j_\perp}
\newcommand{\Ncal}{\mathcal{N}}
\newcommand{\SC}{\mathrm{SC}}
\newcommand{\Q}{\mathrm{Q}}

\begin{document}

\title{Symmetric and Hall-type transverse harmonics in strained honeycomb lattices:\\
adiabatic and nonperturbative corrections to Berry-curvature semiclassics}

\author{G. F. Mkrtchian}
\email{mkrtchian@ysu.am}
\affiliation{Centre of Strong Fields Physics at Research Institute of Physics, Yerevan State University,
Yerevan 0025, Armenia}

\date{\today}

\begin{abstract}
Transverse high-harmonic currents in solids are commonly interpreted in terms of the Berry-curvature
anomalous velocity. However, a transverse nonlinear current need not be of Hall type, and a Hall-type
response need not be fully described by semiclassical dynamics. We examine both distinctions by
comparing coherent quantum evolution with collisionless Berry-curvature semiclassics in strained gapped
honeycomb lattices. Building on the decomposition of nonlinear response tensors into fully symmetric
and Hall-type sectors, we develop an angular projection for harmonic currents and extend it
nonperturbatively to finite driving amplitudes; within semiclassics the extended projection recovers
the Berry-curvature current exactly. The coherent response contains a fully symmetric component,
reactive and, at low frequency, linear in frequency, that semiclassics lacks by time-reversal symmetry. An adiabatic
expansion of the quantum dynamics identifies its origin in the nonlinear momentum dependence of the
Bloch Hamiltonian, since a Hamiltonian linear in momentum produces no second-order symmetric response,
and yields the leading coherent corrections to the Hall-type current. Analytical results for a
massive Dirac model and numerical calculations for lattice Hamiltonians quantify the departure of the
projected quantum response from the Berry-curvature multipole description. When the field drives
electrons through the Dirac point, Landau--Zener excitation leaves real interband population and
adiabatic semiclassics breaks down. Identifying the Hall-type component of a transverse harmonic is
therefore necessary but does not by itself establish a semiclassical Berry-curvature origin.
\end{abstract}

\maketitle

\section{Introduction}

In the semiclassical theory of Bloch electrons, a wave packet acquires an anomalous velocity
$-\dot{\bk}\times\bm\Omega$ proportional to the Berry curvature
$\bm\Omega$~\cite{Karplus1954,Chang1996,Sundaram1999,Xiao2010RMP}. In time-reversal-invariant crystals
with broken inversion, the anomalous velocity produces a second-order transverse current controlled
by the Berry-curvature dipole (BCD) of the Fermi sea~\cite{Moore2010,Sodemann2015}. This nonlinear Hall
effect has been observed in few-layer WTe$_2$~\cite{Ma2019,Kang2019} and reviewed
extensively~\cite{Du2021,Ortix2021}. Uniaxial strain is a natural way to generate a BCD in honeycomb
systems, since it removes the threefold rotation that forces the dipole to
vanish~\cite{Battilomo2019,Ortix2021}.

The present work starts from the microscopic semiclassical theory of Berry-curvature-induced
high-harmonic generation in time-reversal-invariant systems with broken inversion symmetry,
developed by Avetissian and Mkrtchian~\cite{AvetissianMkrtchian2020}. In that theory the anomalous
current of a tilted massive Dirac system was obtained to all orders in the field. Its expansion in the
field-induced momentum displacement relates the anomalous even harmonics to odd-order
Berry-curvature multipoles. At leading perturbative order the $2n$th harmonic is determined by the
$(2n-1)$th multipole: the dipole for the second harmonic, the octupole for the fourth, and so on. At
stronger fields a given harmonic also receives contributions from higher
multipoles~\cite{AvetissianMkrtchian2020}. Later work introduced Berry-curvature multipole moments in
a general nonlinear-response theory, with the dipole and the next moment accessible through the second
and fourth harmonics~\cite{Zhang2021}. Real-time simulations showed that the second-harmonic Hall
response of time-reversal-symmetric insulators follows an interband BCD, with interband multipoles
entering at strong fields and surviving when threefold symmetry removes the dipole~\cite{Okyay2022}.
This raises a basic question in nonlinear optical response: what can be inferred about the microscopic
origin of a transverse harmonic from its dependence on the orientation of the driving field, and when
is its identification with the Berry-curvature anomalous velocity justified? We address this question
by separating the fully symmetric and Hall-type contributions to the current and comparing them, as
obtained from the coherent quantum dynamics including interband polarization, with Berry-curvature
semiclassics and its multipole hierarchy, from the perturbative regime to finite driving amplitudes.

The anomalous-velocity picture is also used to interpret transverse and polarization-resolved
high-harmonic generation (HHG) in
solids~\cite{Ghimire2011,Schubert2014,Vampa2014,Liu2017MoS2,Yoshikawa2017,Ghimire2019,Kruchinin2018},
including attempts to read Berry curvature or band topology from harmonic
spectra~\cite{Luu2018,Silva2019,Chacon2020,Bauer2018}. Neufeld \emph{et al.} concluded that universal
topological signatures in HHG are probably absent~\cite{Neufeld2023}.

Two complications limit such interpretations. First, the anomalous velocity is the leading term of an
adiabatic expansion; at finite frequency and field, interband coherence contributes to the current,
as known from the theory of nonlinear optical
susceptibilities~\cite{Aversa1995,Sipe2000,Ventura2017,Parker2019} and from finite-frequency
extensions of the nonlinear Hall
response~\cite{Gao2014,Morimoto2016,Matsyshyn2019,deJuan2020,Kaplan2024}. Second, a measured transverse
current need not be of Hall type: a current that derives from a scalar potential in field space is
generally not parallel to the field. A comparison with the multipole hierarchy therefore requires a
symmetry-consistent decomposition of the response tensor at each field order. At second order such a
split is standard~\cite{Sodemann2015,Du2021,Ortix2021}; coherent strong-field currents mix orders.
For dc responses, Tsirkin and Souza showed that the current at any order in the field admits a unique,
gauge-invariant separation into a Hall part, transverse to the field for every orientation, and an
Ohmic part described by the fully symmetrized conductivity~\cite{TsirkinSouza2022}. We use the same
tensor decomposition, give it an explicit angular form for two-dimensional systems, and extend it to
the harmonic response of coherent pulsed dynamics and beyond perturbation theory.

The angular projection removes the fully symmetric part of a field-order-$p$ transverse
current exactly in two dimensions. We validate it at $p=2$ and $p=4$, and apply it to coherent dynamics in
strained gapped honeycomb lattices~\cite{Pereira2009,Ribeiro2009,Ortix2021}, compared with collisionless
semiclassics evaluated with identical Hamiltonians, initial states, and quadratures. Our principal
findings are:

(i) The coherent second-order response contains a fully symmetric coefficient that collisionless
semiclassics lacks by time reversal. It is linear in frequency, $s/\hbar\omega$ being constant to
$0.7\%$ over a fourfold frequency range. It is the Kleinman-symmetric part of the reactive
$O(\omega)$ term, which far below resonance in insulators derives from an energy~\cite{HeLaw2026}. In the doped lattice it is a small remainder of large, nearly cancelling insulator and
Pauli-blocked pocket contributions. It is the entire response of the unstrained lattice.

(ii) After projection, the Hall-type coefficient exceeds the BCD term in phase. For a tilted massive
Dirac model the relative in-phase correction of the Hall coefficient, $\delta=(K_\Q-K_\SC)/K_\SC$, follows in
closed form from resonance denominators at the local gap, $\delta=c_2\eta^2+O(\eta^4)$ with $c_2=1.294$, $\eta=\hbar\omega/E_{\rm ref}$. It is of the same
order in $\omega$ as the symmetric coefficient.

(iii) At fourth order the Hall-type quantum coefficient deviates from the Berry-curvature octupole
term, which we evaluate directly as a Brillouin-zone integral, by several percent; the symmetric
sector contributes below $1\%$.

(iv) At intermediate fields the semiclassical amplitude series is consistent with convergence, but its
low-order truncation fails; the nonperturbative projection shows that at $F_0=0.3$ the Hall-type quantum
second harmonic differs from the Berry current by $-7\%$ ($\theta_0=0$) and $+2\%$ ($\theta_0=\pi/4$),
consistent within the fit uncertainty with the fourth-order deviation. When the field sweeps the Dirac point the Landau-Zener parameter is of order
unity, real interband excitation is left after the pulse, and adiabatic semiclassics fails.

The Berry-curvature multipole relation is not a result of this paper; it is the semiclassical
reference established in Ref.~\onlinecite{AvetissianMkrtchian2020}. Nor are the symmetric/Hall
split at each order~\cite{TsirkinSouza2022} or its second-order connection to Kleinman symmetry new: He and Law showed that the
second-harmonic Hall response is the non-Kleinman part of the susceptibility, forbidden far below
resonance in insulators and generated by frequency dispersion, with a Fermi-surface term that reduces
to the BCD in metals~\cite{HeLaw2026}. Our contributions are an explicit angular form of the
symmetric/Hall decomposition~\cite{TsirkinSouza2022} for two-dimensional harmonic response and a
nonperturbative generalization of it; its application to a doped lattice, where the symmetric coefficient is
a Pauli-blocking remainder and is verified to scale as $\omega$; a compact closed form of the
in-phase correction for the linear tilted Dirac model, an explicit special case of the general formula
of Ref.~\onlinecite{HeLaw2026}; an adiabatic expansion of the Bloch equation that derives the semiclassical current and accounts quantitatively for both second-order lattice corrections; the demonstration that the nonperturbative projection reproduces the semiclassical Berry
current exactly at any field strength; and the order-resolved, intermediate-field and strong-field
comparison of coherent lattice dynamics with the multipole hierarchy. Neither the Hall-type coefficient nor its correction is a topological invariant; a
spin-conserving Kane--Mele model illustrates this (Sec.~\ref{sec:km}).

\section{Model Hamiltonians and symmetries}\label{sec:model}

\subsection{Continuum two-band model}
The low-energy reference is a two-valley massive Dirac Hamiltonian with a valley-odd tilt,
\begin{equation}
h_\tau(\mathbf q)=\alpha\tau q_x\,\sigma_0+v(\tau q_x\sigma_x+q_y\sigma_y)+m\sigma_z,
\label{eq:cont}
\end{equation}
with valley index $\tau=\pm1$. Energies are in units of a reference energy $E_{\rm ref}$, equal to
the chemical potential measured from the Dirac point, and momenta in units of $E_{\rm ref}/v$. The
baseline parameters are $\alpha/v=0.2$, $m=0.2E_{\rm ref}$, and $k_BT=0.02E_{\rm ref}$. The
conduction-band Berry curvature, defined as
$\Omega_c=-\hat{\mathbf d}\cdot(\partial_x\hat{\mathbf d}\times\partial_y\hat{\mathbf d})/2$ for
$h=d_0\sigma_0+\mathbf d\cdot\bm\sigma$, is $\Omega_c=-\tau mv^2/[2(v^2q^2+m^2)^{3/2}]$. The tilt makes
the Fermi surface asymmetric and produces a BCD along $x$.

\subsection{Strained honeycomb lattices}
The lattice models use an embedded-orbital Bloch basis. Sites $A$ sit at $\mathbf R$ and sites $B$
at $\mathbf R-(0,a_y)$, and
\begin{equation}
H(\bk)=d_0(\bk)\sigma_0+\mathrm{Re}f\,\sigma_x+\mathrm{Im}f\,\sigma_y+M\sigma_z,
\label{eq:Hlat}
\end{equation}
\begin{equation}
f=\sum_{l=1}^3t_le^{i\bk\cdot\bm\delta_l},\qquad d_0=2\sum_{l=1}^3t'_l\cos(\bk\cdot\mathbf R_l).
\end{equation}
The NN vectors are $\bm\delta_1=(0,a_y)$ and $\bm\delta_{2,3}=(\pm\sqrt3a_x/2,-a_y/2)$; the NNN
vectors are $\mathbf R_{1,2}=(\sqrt3a_x/2,\pm3a_y/2)$ and $\mathbf R_3=(\sqrt3a_x,0)$. The staggered
mass is $M=0.02$ (direct gap 0.04). Energies are in units of the unstrained NN hopping $t_{\rm ref}$,
lengths in units of the unstrained NN distance $a_0$, and $\hbar=e=1$. Armchair strain $\varepsilon$
gives $a_x=1-\nu\varepsilon$ and $a_y=1+\varepsilon$, with $\nu=0.165$~\cite{Pereira2009}.

Model I contains NN hopping only, with the linearized law of Ref.~\onlinecite{Ortix2021} (Sec.~4.1
there), $t_1=1-\beta\varepsilon_{yy}$ and $t_2=t_3=1-\beta(3\varepsilon_{xx}+\varepsilon_{yy})/4$, with
$\beta=3.37$~\cite{Pereira2009,Ribeiro2009}. Its local Dirac tilt is identically zero, and the BCD
arises from the trigonal warping of the Fermi contour~\cite{Battilomo2019}. Model II adds a scalar
NNN dispersion, $t'(d)=\exp[-\beta(d/a_0-1)]$ applied to the deformed NNN lengths, giving
$t'_p=t'_h=0.0848$ at $\varepsilon=0$ and a small tilt $\alpha/v_x=-0.0045$ at $\varepsilon=1.3\%$;
this NNN strain law is not calibrated by first-principles calculations. An engineered model III
($t_1=0.9$, $t_2=1$, $t'_p=0.118$, $t'_h=0.02$, undeformed) has $\alpha/v_x=0.2$ and serves only as a
large-tilt lattice reference matching the continuum tilt.

The chemical potential is fixed at $\mu=E_D+0.1$ above the local Dirac energy $E_D$. With
$E_{\rm ref}=0.1$, $k_BT=0.002$, and $\hbar\omega=0.01$, this reproduces the continuum values
$m/E_{\rm ref}=0.2$, $k_BT/E_{\rm ref}=0.02$, and $\eta\equiv\hbar\omega/E_{\rm ref}=0.1$.

\begin{figure*}[t]
\includegraphics[width=0.86\textwidth]{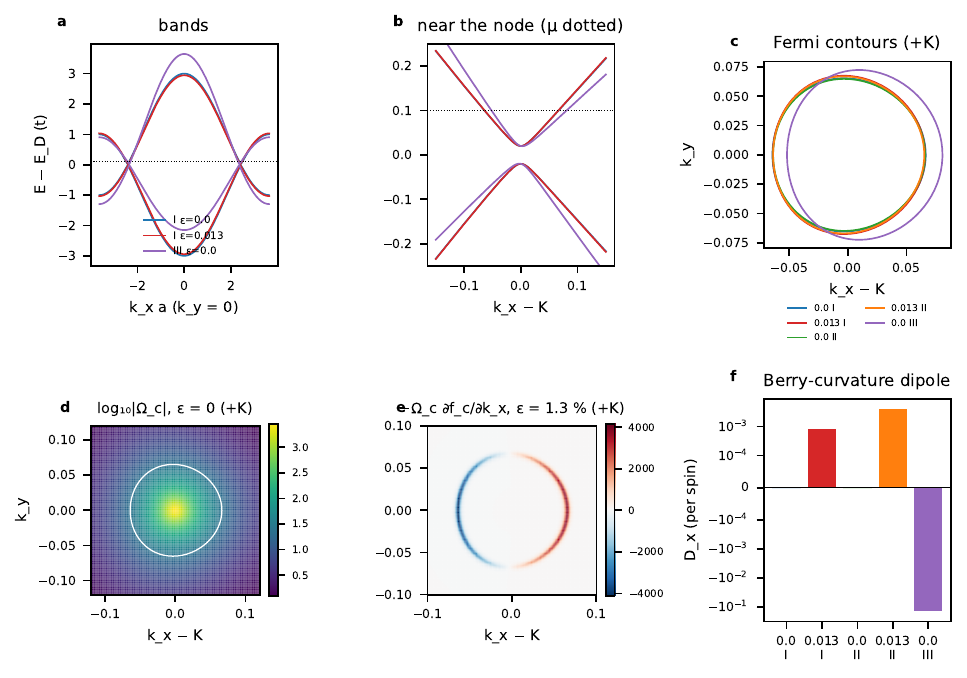}
\caption{Lattice models. (a)~Bands along $k_y=0$ for model I at $\varepsilon=0$ and $1.3\%$, and for
the engineered model III; the dotted line is $\mu=E_D+0.1$. (b)~Zoom near the Dirac point.
(c)~Fermi contours around $K_+$. (d)~$\log_{10}|\Omega_c|$ near $K_+$ at $\varepsilon=0$, with the
Fermi contour in white. (e)~Fermi-surface BCD integrand $-\Omega_c\partial_{k_x}f_c$ at
$\varepsilon=1.3\%$; strain makes the two lobes unequal. (f)~Per-spin BCD $D_x$ on a symmetric-log
scale; it vanishes at $\varepsilon=0$ by $C_3$ symmetry. Units: $t_{\rm ref}=a_0=\hbar=1$.}
\label{fig:model}
\end{figure*}

\emph{Physical scales.} For $t_{\rm ref}=2.7$~eV and $a_0=1.42$~\AA, the gap is $108$~meV, $E_{\rm ref}$
is $270$~meV, the temperature is $63$~K, and $\hbar\omega=27$~meV corresponds to $6.5$~THz. The field
unit is $t_{\rm ref}/(ea_0)=190$~MV/cm, so the reference field $E_0=1/1500$ is $0.13$~MV/cm. The
drive is therefore in the terahertz regime, and ``harmonics'' here means orders $n\le8$.
Figure~\ref{fig:model} shows bands, Fermi contours, Berry curvature, and BCD.

\subsection{Brillouin-zone cell and symmetries}
For the deformed lattice the rectangle $[-L,L]\times[-H,H]$, $L=2\pi/(\sqrt3a_x)$, $H=\pi/(3a_y)$,
is a fundamental domain of the reciprocal lattice under $(k_x,k_y)\sim(k_x+L,k_y+2H)$. All
gauge-invariant integrands are periodic, so integrals over it are exact Brillouin-zone integrals
including both valleys. We write $\int[d\bk]\equiv\int d^2k/(2\pi)^2$; densities and currents are per
unit area and per spin. Quadratures use valley-centered polar grids refined at the Fermi
radius~\cite{SM}.

The Hamiltonian is time-reversal invariant, $H(-\bk)=H(\bk)^*$, and mirror symmetric,
$H(-k_x,k_y)=H(k_x,k_y)$. At $\varepsilon=0$ the threefold rotation acts trivially in the embedded
gauge. Strain breaks $C_3$, and the staggered mass breaks inversion. These relations hold numerically
to $\lesssim10^{-15}$.

\section{Quantum and semiclassical dynamics}\label{sec:dyn}

\subsection{Pulse, minimal coupling, and observables}
A uniform field $\mathbf E(t)=E(t)\be$ with $\be=(\cos\theta,\sin\theta)$ enters through
$\bk(t)=\bk_0-A(t)\be$, $\dot A=E$; the electron charge is $-1$. The pulse is
\begin{equation}
E(u)=F_0E_0\sin^2\!\left(\frac{u}{2C}\right)\cos u,\qquad u=\omega t\in[0,2\pi C],
\label{eq:pulse}
\end{equation}
with $C=20$ cycles. The reference field $E_0=\hbar\omega E_{\rm ref}/(ev_0a_0)=1/1500$, $v_0=3/2$, is
the field whose excursion $E_0/\omega$ equals the Fermi-momentum scale $E_{\rm ref}/v_0$. At $F_0=1$
the oscillating Fermi pocket therefore reaches the Dirac point. The pulse is cyclic.

Harmonic coefficients are Hann-windowed Fourier amplitudes,
\begin{equation}
J_n=\frac{\int_0^{2\pi C}du\,w(u)\,j(u)\,e^{inu}}{\int_0^{2\pi C}du\,w(u)},\qquad w=\sin^2\!\frac{u}{2C},
\label{eq:harm}
\end{equation}
and we report $J_n\times\Ncal/(F_0E_0)^2$ with $\Ncal=(2\pi)^2\cdot1.5/0.1^2$. The factor $\Ncal$
converts a per-spin lattice current density into the natural current unit of the continuum model,
$eE_{\rm ref}^2/[(2\pi)^2\hbar v_0]$ in lattice units, so continuum and lattice coefficients are
directly comparable. Because the second-order signal carries the envelope $\sin^4$ weighted by
$w=\sin^2$, a steady-state monochromatic coefficient equals the Hann value divided by
$\langle\sin^6\rangle/\langle\sin^2\rangle=0.625$; ratios such as $\delta$ below do not depend on this
factor. A monochromatic current $j=\mathrm{Re}\,[Xe^{-inu}]$ has $J_n=X/2$.

Field reversal is performed in the laboratory frame: $\mathbf J^{(e,o)}=\frac12[\mathbf J(\theta)\pm\mathbf
J(\theta+\pi)]$, then $\jpar=\be\cdot\mathbf J$ and $\jperp=\hat{\mathbf e}_\perp\cdot\mathbf J$ with
$\hat{\mathbf e}_\perp=(-\sin\theta,\cos\theta)$. All currents are per spin, except in the Kane--Mele
model, where the charge current sums both spin blocks.

\subsection{Coherent density-matrix dynamics}
Each crystal momentum carries a two-level density matrix obeying $i\hbar\dot\rho_{\bk}=[H(\bk(t)),\rho_{\bk}]$.
With $H=d_0\sigma_0+\mathbf d\cdot\bm\sigma$ and $\rho=\frac12(n\sigma_0+g\,\mathbf S\cdot\bm\sigma)$,
where $n=f_c+f_v$ and $g=f_c-f_v$ are fixed by the initial Fermi functions,
\begin{equation}
\frac{d\mathbf S}{du}=\frac{2}{\hbar\omega}\,\mathbf d(\bk(u))\times\mathbf S,\qquad
\mathbf S(0)=\hat{\mathbf d}(\bk_0),
\label{eq:bloch}
\end{equation}
which contains all interband coherence. The current is
\begin{equation}
\mathbf j(t)=-\int[d\bk]\left[n\,\partial_{\bk}d_0+g\,(\partial_{\bk}\mathbf d)\cdot\mathbf S\right]_{\bk=\bk(t)}.
\label{eq:current}
\end{equation}
All states, including the entire valence band, are propagated with exact midpoint rotations, which
conserve $|\mathbf S|$ to $\lesssim5\times10^{-12}$.

\subsection{Collisionless semiclassics}
In adiabatic semiclassics the occupations $f_b(\bk_0)$ are carried along $\bk(t)$, and the anomalous
velocity is $-E\Omega_b\hat{\mathbf e}_\perp$. Hence
\begin{equation}
\begin{split}
\mathbf j_\SC={}&-\sum_b\int[d\bk]f_b(\bk_0)\,\partial_{\bk}\varepsilon_b\\
&+E(t)\,\hat{\mathbf e}_\perp\sum_b\int[d\bk]f_b(\bk_0)\Omega_b,
\end{split}
\label{eq:sc}
\end{equation}
with band quantities at $\bk(t)$. The second term is the Berry current $\mathbf j_B$. Filled-band
contributions vanish exactly. All lattice comparisons are collisionless.

Equation~\eqref{eq:sc} follows from Eqs.~\eqref{eq:bloch} and \eqref{eq:current} as the first two
orders of an adiabatic expansion~\cite{SM}. Taking $\hat{\mathbf d}\times$ of Eq.~\eqref{eq:bloch} gives
the exact relation $\mathbf S_\perp=-(\hbar/2|\mathbf d|)\,\hat{\mathbf d}\times\dot{\mathbf S}$. Its zeroth
iterate, $\mathbf S=\hat{\mathbf d}$, yields the band current. The first,
$\mathbf S_1=(\hbar E/2|\mathbf d|)\,\hat{\mathbf d}\times\partial_e\hat{\mathbf d}$, yields the Berry current
with its sign. The second iterate gives the interband polarization current, with polarizability
$\propto\mathcal G/|\mathbf d|$, where $\mathcal G$ is the quantum metric. It is field-odd at second order in
the field. The field-even quantum corrections first appear at third order (Sec.~\ref{sec:res-weak}). The
local expansion parameter is $\hbar v\,eE/|\mathbf d(\bk(t))|^2$: about $F_0\eta$ at the Fermi surface,
but $1/\delta_{\rm LZ}$ for states dragged to the Dirac point (Sec.~\ref{sec:strong}).

\subsection{Berry-curvature multipole hierarchy}\label{sec:multipole}
Following Ref.~\onlinecite{AvetissianMkrtchian2020}, we expand the curvature in the field-induced
displacement. For a drive along the mirror axis ($\theta=0$, $\hat{\mathbf e}_\perp=\hat y$),
time reversal ($f_b$ even, $\Omega_b$ odd in $\bk$) leaves only odd orders:
\begin{equation}
\begin{split}
j_{B,y}(t)&=-E(t)\sum_{n\ge1}\frac{A(t)^{2n-1}}{(2n-1)!}\,\mathcal D^{(2n-1)}_x,\\
\mathcal D^{(2n-1)}_x&=\sum_b\int[d\bk]\,f_b\,\partial_{k_x}^{2n-1}\Omega_b.
\end{split}
\label{eq:multipole}
\end{equation}
The $\mathcal D^{(2n-1)}_x$ are occupation-weighted momentum derivatives of the curvature,
generalized odd-order multipoles. $\mathcal D^{(1)}_x$ is the BCD, whose finite-pulse coefficient is
$K_\SC$ below; $\mathcal D^{(3)}_x$ is the octupole.

In the notation of Ref.~\onlinecite{AvetissianMkrtchian2020} (momentum $\mathbf p$, charge $-e$,
conduction occupation $N^{(0)}_{c1}$ and curvature $B_{1z}$ of one valley, degeneracy $g$), the
multipoles are [Eq.~(26) there]
\begin{equation}
D_{{\rm AM},x}^{(2n-1)}=g\int\frac{d^2p}{(2\pi\hbar)^2}\,N_{c1}^{(0)}(\mathbf p)\,\frac{\partial^{2n-1}B_{1z}(\mathbf p)}{\partial p_x^{2n-1}},
\end{equation}
integrated over one valley; the time-reversed valley enters there through an explicit factor 2. For
$E_x=E_0\cos\omega t$, $A=(E_0/\omega)\sin\omega t$, the order-$2n$ term of Eq.~\eqref{eq:multipole}
is $\propto EA^{2n-1}$. With $\sin^{2n-1}x\cos x=(2n)^{-1}d(\sin^{2n}x)/dx$ and the top-harmonic part
$(-1)^n\cos(2nx)/2^{2n-1}$ of $\sin^{2n}x$, its leading-order $2n$th harmonic is
\begin{equation}
j_{B,y}^{(2n),\,\rm top}(t)=\frac{(-1)^n\,\mathcal D^{(2n-1)}_x\,E_0^{2n}}{2^{2n-1}\,(2n-1)!\,\omega^{2n-1}}\,\sin(2n\omega t)
\label{eq:AM27}
\end{equation}
per spin, both valleys. Equation~\eqref{eq:AM27} is Eq.~(27) of
Ref.~\onlinecite{AvetissianMkrtchian2020} rewritten in the present normalization, in which
$\mathcal D$ is per spin and sums both valleys; the explicit valley factor of
Ref.~\onlinecite{AvetissianMkrtchian2020} is therefore absorbed into $\mathcal D$
[$\mathcal D=(2/g)D_{\rm AM}$]. The sign $(-1)^n$ follows from Eq.~\eqref{eq:multipole}; Eq.~(27) of
Ref.~\onlinecite{AvetissianMkrtchian2020} is printed without it~\cite{SM}.

A term of field order $p$ involves $\mathcal D^{(p-1)}$ and, through $EA^{p-1}$, contributes to all
even harmonics up to $p$. The $2n$th harmonic is controlled by $\mathcal D^{(2n-1)}$ only at its
leading order $p=2n$~\cite{AvetissianMkrtchian2020}. Because the Berry current is purely Hall-type at
every order (Sec.~\ref{sec:proj}), the order-$p$ Hall-type sector of semiclassics is exactly the
multipole term. The question is whether the coherent Hall-type current keeps this description.

\section{Symmetric and Hall-type nonlinear response}\label{sec:proj}

\subsection{Second order}
With $M_x$ symmetry the second-order current has three coefficients,
\begin{equation}
j_x=2a\,E_xE_y,\qquad j_y=b\,E_x^2+c\,E_y^2,
\label{eq:tensor}
\end{equation}
$a=\chi_{xxy}$, $b=\chi_{yxx}$, $c=\chi_{yyy}$. The tensor splits uniquely into a fully symmetric part,
$s_{xxy}=s\equiv(2a+b)/3$ and $s_{yyy}=c$, and a Hall-type part
\begin{equation}
\mathbf j_H=K\,(-E_xE_y,\,E_x^2)=K|\mathbf E|^2\cos\theta\,\hat{\mathbf e}_\perp,\ \ K=\tfrac23(b-a),
\label{eq:hall}
\end{equation}
perpendicular to $\mathbf E$ for every direction. The symmetric current is a field-space gradient,
$\mathbf j_{\rm sym}=\nabla_{\mathbf E}\Phi_3$, perpendicular to the (noncircular) level sets of the
cubic $\Phi_3$; it has a transverse component without being of Hall type. In polar coordinates
$\jpar^{\rm sym}=3\Phi_3/|\mathbf E|$ and $\jperp^{\rm sym}=\partial_\theta\Phi_3/|\mathbf E|$. For
$\mathbf E\parallel\hat x$ the transverse current is $(K+s)E_x^2$: one transverse measurement cannot
separate $K$ from $s$. The angular dependence does:
\begin{equation}
H^{(2)}(\theta)\equiv\jperp-\tfrac13\partial_\theta\jpar=K\cos\theta,
\label{eq:H2}
\end{equation}
for arbitrary $a,b,c$ and, since all angular dependence sits in the unit vectors, pointwise in time
for any second-order time kernel. At $\varepsilon=0$, $C_3$ imposes $a=b=-c$, so $K=0$ but $s=a\neq0$.

\subsection{Time reversal: no field-even symmetric current in collisionless semiclassics}
In collisionless semiclassics the occupations and band energies are even in $\bk_0$ and the curvature
is odd. The band current is therefore odd in $A$, and the Berry current is even in the field and
perpendicular to it. The semiclassical field-even current is entirely Hall-type at every order,
\begin{equation}
s_\SC\equiv0.
\end{equation}
The symmetric response of the coherent dynamics (Sec.~\ref{sec:res-weak}) arises from interband
coherence; Sec.~\ref{sec:lowfreq} identifies its form.

\subsection{Arbitrary field order}\label{sec:Hp}
A contribution of definite even field order $p$ to the field-even current is a rank-$(p+1)$ tensor,
symmetric in its $p$ field indices and contracted with $\be^{\otimes p}$ (each component may carry
its own time kernel). In two dimensions
$V\otimes\mathrm{Sym}^pV=\mathrm{Sym}^{p+1}V\oplus\mathrm{Sym}^{p-1}V$: a fully symmetric part
$\mathbf j_{\rm sym}=\nabla_{\mathbf E}\Phi_{p+1}$ and a Hall-type part
$\mathbf j_H=\hat z\times\mathbf E\,Q_{p-1}(\mathbf E)$. Since $\jpar^{\rm sym}=(p+1)\Phi_{p+1}/|\mathbf E|$,
$\jperp^{\rm sym}=\partial_\theta\Phi_{p+1}/|\mathbf E|$, and $\jpar^{\rm H}=0$,
\begin{equation}
H^{(p)}(\theta)=\jperp^{(p)}(\theta)-\frac{1}{p+1}\,\partial_\theta\jpar^{(p)}(\theta)=|\mathbf E|\,Q_{p-1}(\be)
\label{eq:Hp}
\end{equation}
removes the symmetric part exactly. At each order this coincides with the Hall/Ohmic separation of
Ref.~\onlinecite{TsirkinSouza2022}: $\mathrm{Sym}^{p+1}V$ is the fully symmetrized part of the
order-$(p+1)$ tensor. In the dc setting of Ref.~\onlinecite{TsirkinSouza2022} the symmetric part is
Ohmic, i.e., dissipative. In the collisionless coherent dynamics studied here, the symmetric component
of a harmonic is reactive rather than dissipative [Eq.~\eqref{eq:lowfreq}], so we call it ``symmetric'' rather than ``Ohmic'';
the Hall part remains the component that is transverse for every field direction. The derivation assumes a contribution of definite field order,
linear polarization at fixed waveform, two dimensions, and knowledge of $\jpar(\theta)$; point-group
symmetry is not required. With $M_x$, a $\cos5\theta$ component of $H^{(4)}$ signals a failure of
order separation. Applied to a finite-field response that mixes orders, a single factor $\frac13$
leaves the error $(\frac{1}{p+1}-\frac13)\partial_\theta\jpar^{(p)}$ at each order.

In semiclassics $H^{(p)}$ is the $\mathcal D^{(p-1)}$ term of Eq.~\eqref{eq:multipole}; in general it
also contains coherent contributions not of anomalous-velocity form.

\subsection{Nonperturbative projection}\label{sec:Hnp}
Equation~\eqref{eq:Hp} requires a contribution of definite field order. A finite-field current
mixes orders, and the truncated amplitude series loses accuracy already at $F_0\simeq0.3$
(Sec.~\ref{sec:inter}). The order-resolved projection can, however, be resummed without
using the series. Consider the family of drives $\mathbf E(t;F,\theta)=F f(t)\,\be(\theta)$ with
a fixed waveform $f(t)$, and regard the current at fixed $t$ as a vector field on the
two-dimensional field plane, $\mathbf E=F\be$. Define
\begin{equation}
\Phi(F,\theta,t)=\int_0^F\jpar(F',\theta,t)\,dF'=\int_0^1\mathbf E\cdot\mathbf j(s\mathbf E,t)\,ds
\label{eq:Phi}
\end{equation}
and
\begin{equation}
\begin{split}
H_{\rm NP}(F,\theta,t)&=\jperp(F,\theta,t)-\frac1F\,\partial_\theta\Phi(F,\theta,t)\\
&=\jperp-\partial_\theta\!\int_0^1\!\jpar(sF,\theta,t)\,ds .
\end{split}
\label{eq:Hnp}
\end{equation}
The second form of Eq.~\eqref{eq:Phi} is the homotopy (Poincar\'e-lemma) potential of
$\mathbf j$ along rays from $\mathbf E=0$.

\emph{Resummation.} Wherever $\mathbf j=\sum_pF^p\mathbf j^{(p)}$ exists, Eq.~\eqref{eq:Hnp} gives
$H_{\rm NP}=\sum_pF^pH^{(p)}$, with $H^{(p)}$ from Eq.~\eqref{eq:Hp}: each order receives its own
factor $1/(p+1)$. The construction itself does not require the series to converge; it requires
only that $\mathbf j$ be integrable in $F$ and regular at zero amplitude, $\mathbf j=O(F)$, so
that $H_{\rm NP}$ is regular.

\emph{Uniqueness.} In polar field coordinates
$\nabla_{\mathbf E}\Phi=(\partial_F\Phi,\,F^{-1}\partial_\theta\Phi)=(\jpar,\,F^{-1}\partial_\theta\Phi)$, so
\begin{equation}
\mathbf j=\nabla_{\mathbf E}\Phi+H_{\rm NP}\,\hat{\mathbf e}_\perp .
\label{eq:split}
\end{equation}
Equation~\eqref{eq:split} is the unique decomposition of $\mathbf j$ into a field-space gradient
regular at $\mathbf E=0$ and a part perpendicular to $\mathbf E$. If two such decompositions
existed, their difference would be a gradient with no component along $\be$. Its potential
would then depend on $\theta$ alone, and regularity at $\mathbf E=0$ forces it to be constant.
Order by order, Eq.~\eqref{eq:split} reduces to the decomposition
$\mathrm{Sym}^{p+1}V\oplus\mathrm{Sym}^{p-1}V$ of Sec.~\ref{sec:Hp}.

\emph{Collisionless semiclassics.} With $A(t)=Fa(t)$, $a=\int f\,dt$, the band term of
Eq.~\eqref{eq:sc} is an exact field-space gradient at every amplitude,
\begin{equation}
\begin{split}
&-\sum_b\int[d\bk]\,f_b\,\nabla_{\bk}\varepsilon_b(\bk_0-a\mathbf E)\\
&\qquad=\frac{1}{a(t)}\,\nabla_{\mathbf E}\sum_b\int[d\bk]\,f_b\,\varepsilon_b(\bk_0-a\mathbf E),
\end{split}
\end{equation}
while the Berry term is perpendicular to $\mathbf E$ and contributes nothing to $\jpar$. Hence
\begin{equation}
H_{\rm NP}^{\SC}(F,\theta,t)=j_B(F,\theta,t)\qquad\text{for all }F,\theta,t.
\label{eq:HnpSC}
\end{equation}
Equation~\eqref{eq:HnpSC} is exact: it needs neither weak fields nor the multipole expansion.
The comparison $H_{\rm NP}^{\Q}$ versus $j_B$ is therefore a nonperturbative test of the
Berry-curvature description, valid also where the order-resolved analysis is unavailable.

\emph{Comparison with a work-based construction.} An alternative separation uses the accumulated work
$W(\theta,t)=\int_0^t\jpar(\theta,t')\,dA(t')$. For adiabatic band currents $W$ is an instantaneous
function of the displacement $A$, and the subtraction $\jperp-A^{-1}\partial_\theta W|_A$ removes the
band-velocity contribution exactly, leaving the Berry-curvature current; there it carries the same
information as Eq.~\eqref{eq:Phi}, and both return $j_B$. In coherent dynamics, however, a field-even
current with a time-reversal-invariant initial state must involve time derivatives of the field, $W$
then depends on the history of the pulse, and the subtraction does not isolate the fully symmetric
sector~\cite{SM}. We therefore use it only as a negative control, while the angular and
nonperturbative projections provide the framework of the analysis.

\emph{Frequency domain.} Because Eq.~\eqref{eq:Hnp} is linear in the current and acts on the drive
parameters at fixed time, it commutes with Fourier transformation in $t$. It can therefore be applied
harmonic by harmonic to complex amplitudes,
$H_{\rm NP}(F,\theta;n\omega)=\jperp(F,\theta;n\omega)-\partial_\theta\int_0^1\jpar(sF,\theta;n\omega)\,ds$,
provided the waveform is fixed and harmonic phases are referenced to it; within semiclassics
$H_{\rm NP}^\SC(n\omega)=j_B(n\omega)$ for each harmonic. The harmonic ratios reported in
Sec.~\ref{sec:inter} are of this form.

\emph{What the projection determines.} Ordinary harmonic spectroscopy measures intensities at fixed
polarization, in which symmetric and Hall-type contributions to the same harmonic, and often to the
same angular harmonic (both are $\cos3\theta$ in the unstrained lattice), are inseparable. $H_{\rm NP}$
uses the complex harmonic amplitudes of both in-plane components as functions of drive amplitude and
orientation. From these it extracts, at finite amplitude and without an order expansion, the part of
the response that is not a regular field-space gradient. It thereby removes the band-velocity current
and all symmetric coherent contributions at once, and within semiclassics it returns the
Berry-curvature current exactly. A nonzero $H_{\rm NP}-j_B$ therefore measures Hall-type
contributions beyond the anomalous velocity, independently of any multipole truncation. The
projection does not identify the microscopic origin of the Hall-type part, which requires a model such
as $j_B$ for comparison; it does not separate anomalous-velocity and coherent Hall-type terms within
that sector; and beyond perturbation theory its potential $\Phi$ is not an energy, and the angular
selection rules of Sec.~\ref{sec:Hp} no longer apply. It also presupposes linear polarization at fixed
waveform, two dimensions, and regularity at zero amplitude.

\subsection{Low-frequency structure of the second-order response}\label{sec:lowfreq}
Consider the second-harmonic response $\sigma(\omega)\equiv\sigma^{(2)}(2\omega;\omega,\omega)$ of a
time-reversal-invariant system in the collisionless limit ($\tau\to\infty$ at fixed nonzero $\omega$,
before $\omega\to0$), with $\omega$ and $2\omega$ below the relevant interband absorption thresholds,
so that an adiabatically switched drive leaves no real excitation. The dynamics is then reversible
from a time-reversal-invariant equilibrium state, and time reversal relates the responses to
$\mathbf E(t)$ and $\mathbf E(-t)$: $\mathbf j[\mathbf E_r](t)=-\mathbf j[\mathbf E](-t)$ with
$\mathbf E_r(t)=\mathbf E(-t)$. For the second harmonic this gives $\sigma(-\omega)=-\sigma(\omega)$;
with reality, $\sigma(-\omega)=\sigma(\omega)^*$, $\sigma$ is purely imaginary. Assuming analyticity
apart from the Fermi-surface pole,
\begin{equation}
\sigma(\omega)=\frac{\sigma_{-1}}{\omega}+\sigma_1\,\omega+O(\omega^3),
\label{eq:lowfreq}
\end{equation}
with $\sigma_{-1}$ and $\sigma_1$ purely imaginary. The $\sigma_{-1}/\omega$ term is the Fermi-surface
BCD current, $j=EA\,\mathcal D^{(1)}$ with $A\propto E/\omega$, and is purely Hall-type. The even powers
that time-reversal-breaking systems can admit are excluded, such as the nonlinear Drude term
($\propto\tau^2$ in dc transport, $\propto\omega^{-2}$ in the collisionless limit) and the
$\tau$-independent intrinsic quantum-metric term ($\propto\omega^0$), both of which require broken time
reversal~\cite{Kaplan2024}. The frequency parity thus requires both time-reversal symmetry and the
absence of real excitation; neither condition alone suffices~\cite{SM}. The $O(\omega)$ term contains a reactive polarization
current $\mathbf j=\partial_t\mathbf P^{(2)}$ with $P^{(2)}_i=\chi_{ijk}E_jE_k$. Far from resonance the
Kleinman-symmetric part of $\chi_{ijk}$ derives from an energy, $\mathbf P=-\partial F/\partial\mathbf E$
with $F$ cubic in $\mathbf E$; its tensor is fully symmetric, i.e., it lies entirely in the
$\nabla_{\mathbf E}\Phi_3$ sector. This is the Kleinman-symmetry argument of He and
Law~\cite{HeLaw2026}. Their second-harmonic Hall conductivity
$\sigma^\perp_{qij}=[2\sigma_{qij}-\sigma_{iqj}-\sigma_{jiq}]/3$ is the non-Kleinman part of the
tensor and coincides with the Hall-type part used here (with $M_x$, $\sigma^\perp_{yxx}=\frac23(b-a)=K$).
It vanishes in insulators far below resonance and is generated by frequency dispersion; in metals
their Fermi-surface term reduces to the BCD as $\omega\to0$~\cite{HeLaw2026}. Okyay \emph{et al.}
related the two-band second-harmonic Hall term of insulators to an interband BCD~\cite{Okyay2022}. In
this language $s$ is the Kleinman-symmetric part of the reactive term. In an insulator far below
resonance this part derives from an energy; whether that identification carries over to the doped
system, where $s$ is a Pauli-blocking remainder, would require the static hyperpolarizability, which we
have not computed.
At $\varepsilon=0$ the pattern $a=b=-c$ is the standard $D_{3h}$ second-harmonic tensor of gapped
honeycomb layers. Both $s$ and $\delta K$ arise at the third order of the adiabatic expansion of
Eq.~\eqref{eq:bloch}, through terms $\propto E\dot E$ and $\ddot EA$, which reproduce the exact lattice values to
$O(\eta^4)$~\cite{SM}.

In our normalization by $E_0^2$, $s\propto\hbar\omega$ while $K_\SC\propto1/\omega$, so $s/K\propto\eta^2$.
The in-phase correction $\delta K=K_\Q-K_\SC$, of relative order $\eta^2$, is therefore of the
\emph{same} order of the low-frequency expansion as $s$: $s$ is its Kleinman-symmetric part,
and $\delta K$ is the dispersive, non-Kleinman part of the $O(\omega)$ term~\cite{HeLaw2026}. What
the present calculation adds is the doped-metal setting, where $s$ is a small Pauli-blocking
remainder, and an explicit check of $s\propto\omega$.

\section{Results}\label{sec:res}

\subsection{Second order: symmetric coefficient and quantum correction}\label{sec:res-weak}
\begin{figure*}[t]
\includegraphics[width=0.84\textwidth]{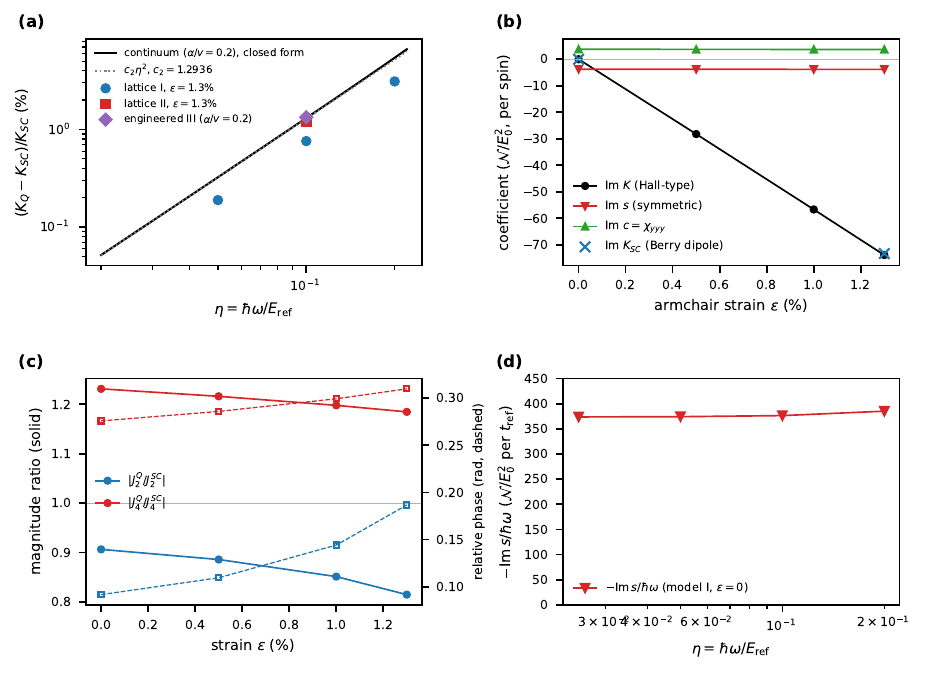}
\caption{(a)~In-phase quantum correction $(K_\Q-K_\SC)/K_\SC$ versus $\eta=\hbar\omega/E_{\rm ref}$.
Line: closed form, Eq.~\eqref{eq:closed}, for the continuum tilted Dirac model ($\alpha/v=0.2$);
dotted: its leading term $c_2\eta^2$. Symbols: exact lattice values for model I ($\varepsilon=1.3\%$, three
frequencies), model II ($\varepsilon=1.3\%$), and the engineered model III ($\varepsilon=0$). (b)~Second-order coefficients of model I versus strain, with
the BCD prediction $K_\SC$ (crosses). (c)~Finite-field ($F_0=1$) quantum/semiclassical magnitude ratios
(solid) and relative phases (dashed) of the second and fourth transverse harmonics versus strain.
(d)~$-\mathrm{Im}\,s/\hbar\omega$ versus $\eta$ for model I at $\varepsilon=0$. All quantities per
spin, Hann normalization.}
\label{fig:corr}
\end{figure*}

\begin{figure}[t]
\includegraphics[width=\columnwidth]{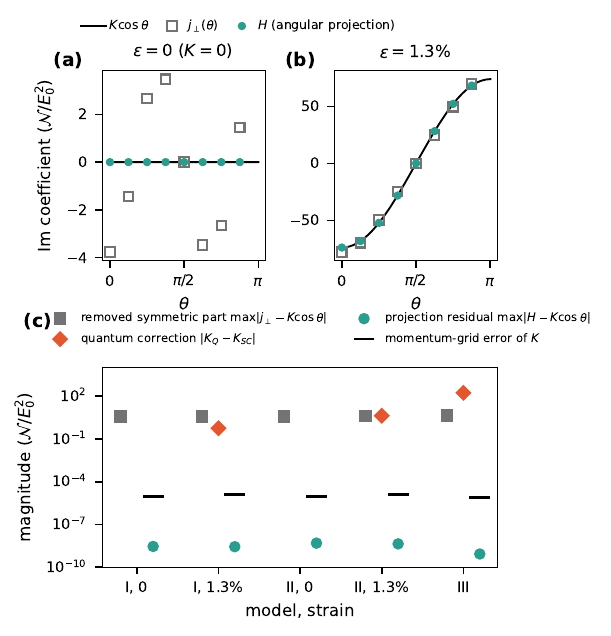}
\caption{Angular projection of the exact second-order tensor. (a,b)~$\jperp(\theta)$ (open squares)
and $H=\jperp-\frac13\partial_\theta\jpar$ (filled circles) compared with $K\cos\theta$ (line), at
$\varepsilon=0$ ($K=0$) and $\varepsilon=1.3\%$ (model I). (c)~For all models: largest removed
symmetric transverse contribution, quantum correction $|K_\Q-K_\SC|$, projection residual, and
momentum-grid error of $K$. Hann values $\times\Ncal/E_0^2$, per spin.}
\label{fig:weak}
\end{figure}

The exact second-order coefficients are zero-field Taylor coefficients of Eq.~\eqref{eq:bloch} with
analytic time integration over the finite pulse, evaluated at eight drive angles
(Table~\ref{tab:weak}, Fig.~\ref{fig:corr}). The strain-induced coefficient $K$ reaches $-73.88i$ (I)
and $-363.4i$ (II) at $\varepsilon=1.3\%$, while $s\simeq-3.8i$ and $c$ are nearly independent of
strain and model. Since $s=0$ for any affine $\mathbf d(\bk)$ with time-reversal-symmetric occupations, exactly at
second order in the field and at all frequencies below the interband resonances~\cite{SM},
the symmetric coefficient requires a nonlinear $\mathbf d(\bk)$, such as trigonal warping; the linear
Dirac part of the bands contributes nothing to it. All coefficients are imaginary, i.e., in phase with $-E(u)A(u)$. Mirror-forbidden
components vanish to $\le2\times10^{-9}$, $a=b=-c$ holds at $\varepsilon=0$ to $10^{-9}$, and
Eq.~\eqref{eq:H2} holds at all angles to $\le5\times10^{-9}$ while removing a symmetric transverse part
of 3.8--4.3 [Fig.~\ref{fig:weak}].

\begin{table}[t]
\caption{Exact second-order coefficients at $\hbar\omega=0.01$ and fixed $\mu$: imaginary parts of the
Hann second harmonic $\times\Ncal/E_0^2$, per spin (divide by $0.625$ for monochromatic values). Real
parts are below $10^{-8}$. $\delta=(K_\Q-K_\SC)/K_\SC$, with $K_\SC$ from the BCD. Momentum-grid errors
of $K$ are $\le1.2\times10^{-5}$.}
\label{tab:weak}
\begin{ruledtabular}
\begin{tabular}{lrrrrrr}
model, $\varepsilon$ & $a$ & $b$ & $c$ & $K$ & $s$ & $\delta$ (\%)\\
\hline
I, 0 & $-3.764$ & $-3.764$ & $3.764$ & $0$ & $-3.764$ & --\\
I, 0.5\% & $10.324$ & $-32.007$ & $3.730$ & $-28.220$ & $-3.786$ & --\\
I, 1.3\% & $33.114$ & $-77.709$ & $3.676$ & $-73.882$ & $-3.827$ & $0.760$\\
II, 1.3\% & $177.79$ & $-367.35$ & $3.766$ & $-363.43$ & $-3.920$ & $1.187$\\
III, 0 & $-6553.5$ & $13094.1$ & $3.565$ & $13098.4$ & $-4.311$ & $1.333$\\
\end{tabular}
\end{ruledtabular}
\end{table}

\emph{Symmetric coefficient.} For model I at $\varepsilon=0$, $s$ is $-0.934i$, $-1.871i$, $-3.764i$, and
$-7.704i$ at $\hbar\omega=0.0025$, $0.005$, $0.01$, and $0.02$ [Fig.~\ref{fig:corr}(d)]. Thus
$s/\hbar\omega=-373.7i$, $-374.2i$, $-376.4i$, constant to $0.7\%$ up to $\eta=0.1$, rising by $3\%$ at
$\eta=0.2$ as expected from the next, $O(\eta^2)$, correction. The linear frequency dependence
confirms Eq.~\eqref{eq:lowfreq}. At $\hbar\omega=0.01$ the completely filled valence band alone gives
$-566.9i$, and the Pauli-blocked pocket gives $+563.2i$: $s$ is a small remainder of two large
contributions. The insulator part alone is not linear in $\omega$ in this range, since its gap $2M$
places a two-photon resonance at $\hbar\omega=0.02$; Pauli blocking moves the relevant gap of the
doped system to $2|\mathbf d|\approx2E_{\rm ref}$ at the Fermi surface.

\emph{Quantum correction to $K$: closed form.} We write the relative in-phase correction as
$\delta\equiv\delta K/K_\SC=(K_\Q-K_\SC)/K_\SC$, where $K_\Q$ is the exact Hall-type coefficient after angular
projection and $K_\SC$ its BCD value. For the linear continuum model the correction can be
obtained analytically. On the subspace transverse to $\hat{\mathbf d}$, $L=2[\mathbf d]_\times$ acts as
$2|\mathbf d|\hat{\mathbf d}\times$ with $(\hat{\mathbf d}\times)^2=-1$, so
\begin{equation}
(-i\Omega-L)^{-1}\mathbf F=\frac{\mathbf F_\parallel}{-i\Omega}+\frac{-i\Omega\mathbf F_\perp+2|\mathbf d|\hat{\mathbf d}\times\mathbf F}{4|\mathbf d|^2-\Omega^2}.
\end{equation}
For $\mathbf d=(\tau vq_x,vq_y,m)$ all second derivatives of $\mathbf d$ vanish, so there are no vertex
corrections. Iterating Eq.~\eqref{eq:bloch} twice for the $e^{-2i\omega t}$ component of an $x$ drive
and discarding terms odd in $q_y$, the Hall current per $\mathbf q$ is (with $\hbar=1$)
\begin{equation}
j_y^H(\mathbf q)=\frac{j_y^\SC(\mathbf q)}{(1-\omega^2/4|\mathbf d|^2)(1-\omega^2/|\mathbf d|^2)},
\end{equation}
where $j_y^\SC=(i\omega/4)\partial_x\Omega_c$ (unit cosine drive) is the BCD integrand. The two factors
are the one- and two-photon resonance denominators at the local gap $2|\mathbf d|$. Hence
\begin{equation}
\begin{split}
\delta(\eta)&=\frac{\big\langle\partial_x\Omega_c\big[\big(1-\frac{\eta^2}{4|\mathbf d|^2}\big)^{-1}\big(1-\frac{\eta^2}{|\mathbf d|^2}\big)^{-1}-1\big]\big\rangle}{\langle\partial_x\Omega_c\rangle},\\
c_2&=\frac54\,\frac{\langle\partial_x\Omega_c\,|\mathbf d|^{-2}\rangle}{\langle\partial_x\Omega_c\rangle},
\end{split}
\label{eq:closed}
\end{equation}
with $\langle X\rangle=\sum_\tau\int f_cX$ and $|\mathbf d|$ in units of $E_{\rm ref}$. The filled valence
band of each valley contributes nothing, because the linear valley is rotation invariant and an
isotropic rank-three tensor vanishes in two dimensions; consequently only $f_c$ enters, and $s=0$
identically in this model. We verified the per-$\mathbf q$ formula in exact rational arithmetic and
against a brute-force resolvent calculation~\cite{SM}. It gives $c_2=1.29361$ and $\delta=1.30842\%$
at $\eta=0.1$, in agreement with the earlier direct second-order calculation
($c_2=1.2936\pm0.0001$). The relation $\delta=c_2\eta^2+O(\eta^4)$ is thus analytic, and the
collisionless correction is in phase automatically, by time reversal and the absence of real
excitation [Eq.~\eqref{eq:lowfreq}]; the absence of a resolved phase difference ($<10^{-9}$~rad) in the lattice is a consistency check.
Equation~\eqref{eq:closed} is the explicit linear-Dirac form of the general two-band second-harmonic
Hall formula of He and Law (Supplemental Material of Ref.~\onlinecite{HeLaw2026}). Evaluated for this
model, their Fermi-surface term alone gives $c_2=1.29361$. Their occupied-state terms enter only at
$O(\eta^4)$, and the sum reproduces Eq.~\eqref{eq:closed} to all printed digits for $\eta\le0.2$~\cite{SM}.
The agreement holds after integration, not pointwise: their Fermi-surface form and our Fermi-sea form
differ by an integration by parts. What Eq.~\eqref{eq:closed} adds is a compact closed form for this
model, in which the one- and two-photon resonance denominators at the local gap appear explicitly.

\emph{Lattice values.} For the lattice models $\delta$ is evaluated from the exact second-order
coefficients (Table~\ref{tab:weak}, Fig.~\ref{fig:corr}(a)), varying $\hbar\omega$ at fixed $\mu$. Model I
($\varepsilon=1.3\%$) gives $\delta=0.189\%$, $0.760\%$, and $3.105\%$ at $\eta=\hbar\omega/E_{\rm ref}=0.05$, $0.1$, and
$0.2$, i.e., $\delta/\eta^2=0.756$, $0.760$, and $0.776$, consistent with $\delta\propto\eta^2$ at small $\eta$;
models II ($\varepsilon=1.3\%$) and III ($\varepsilon=0$) give $1.187\%$ and $1.333\%$ at $\eta=0.1$, against
$1.308\%$ for the continuum model, Eq.~\eqref{eq:closed}. The third adiabatic order of
Eq.~\eqref{eq:bloch}, valid for any two-band $\mathbf d(\bk)$, accounts for both lattice corrections
quantitatively~\cite{SM}: after angular projection its $c_2$ agrees with fits of the exact values for
models I--III to $\le4.2\times10^{-5}$, and its $s/\hbar\omega$ equals the $\omega\to0$ limit of the exact
$s$ to $1.2\times10^{-8}$ (relative). The linear-Dirac resonance factor alone does not: weighted with the
lattice $f_c\partial_x\Omega_c$ and $|\mathbf d|$ it gives $\delta=56\%$ (I), $12.5\%$ (II), and
$0.51\%$ (III) at $\eta=0.1$, because the dipole is a small remainder of nearly cancelling lobes
($2.2\times10^{-4}$ of a lobe in model I). At $\eta=0.1$ the adiabatic order gives $s=-3.80i$ for
model I (exact $-3.83i$); the remaining $0.8\%$ is the next order. For the linear Dirac
model the same order reduces to a time-domain Hall polarization current,
$\mathbf j_H=\hat{\mathbf e}_\perp[-EA\,\mathcal D^{(1)}+\frac{\hbar^2}{4}(\ddot EA+2E\dot E)\tilde{\mathcal D}^{(1)}]$
with $\tilde{\mathcal D}^{(1)}=\sum_b\int f_b\,\partial_x(\Omega_b/|\mathbf d|^2)$, valid for any adiabatic pulse
shape and time-reversal-symmetric occupations; it reproduces $c_2$ of Eq.~\eqref{eq:closed} to $10^{-8}$.
For any affine $\mathbf d(\bk)$ the full second-order response is Hall-type pointwise in momentum, at
every frequency below the interband resonances, so $s=0$~\cite{SM}. The correction is a model-specific finite-frequency effect, not an additional
Berry-curvature or topological quantity.

\emph{Physical units.} For model I at $\varepsilon=1.3\%$ the monochromatic Hall coefficient
$|K|/0.625$ corresponds to a sheet second-order conductivity $\sigma^{(2)}=1.0\times10^{-15}$~A\,m/V$^2$
(both spins), i.e., a second-harmonic current density of $1.6\times10^{-3}$~A/m at $12.7$~kV/cm
($F_0=0.1$, within the perturbative range).

At finite field the transverse harmonics deviate more strongly [Fig.~\ref{fig:corr}(c)]: at $F_0=1$,
$|J_2^\Q/J_2^\SC|=0.81$--$0.91$ and $|J_4^\Q/J_4^\SC|=1.18$--$1.23$, with relative phases
$0.09$--$0.31$~rad, over $0\le\varepsilon\le1.3\%$. These harmonics were computed on a separately
validated full-zone momentum mesh; independent momentum and time refinements change them by at most
$0.003\%$~\cite{SM}.

\subsection{Weak field: what the projections isolate}
\begin{figure*}[t]
\includegraphics[width=0.88\textwidth]{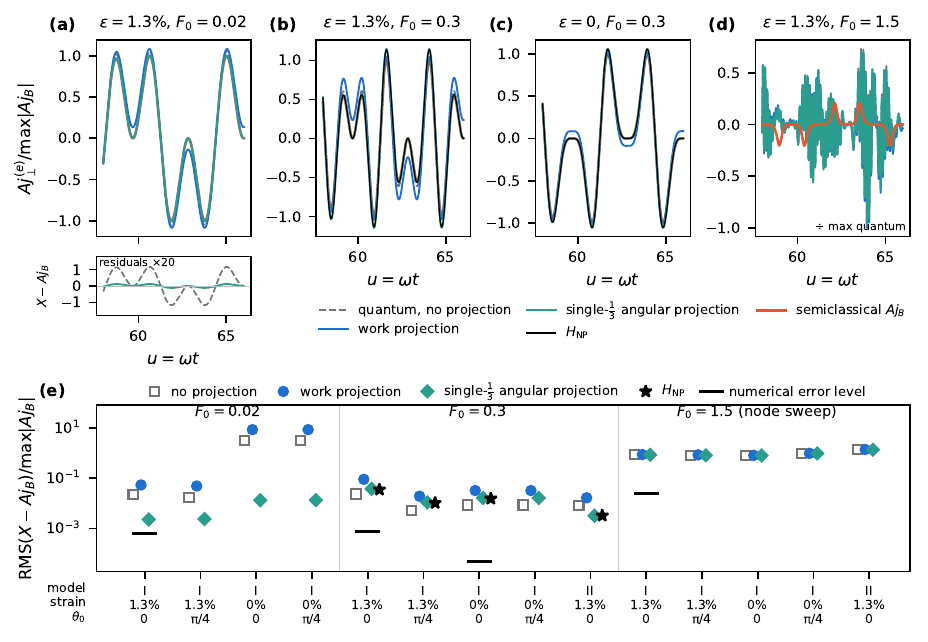}
\caption{Projections applied to coherent finite-pulse dynamics. (a)--(d)~Field-even transverse current
times $A$, the natural variable of the work-based construction~\cite{SM}, near the pulse center
($58<\omega t<66$), normalized to $\max|Aj_B|$ in that interval [in (d), to the maximum of the quantum
curves]. Dashed gray: quantum current without projection; blue: work projection; teal:
single-$\frac13$ angular projection, Eq.~\eqref{eq:H2}; black: nonperturbative projection
$H_{\rm NP}$, Eq.~\eqref{eq:Hnp}; red: semiclassical $Aj_B$. (a)~$\varepsilon=1.3\%$, $F_0=0.02$; lower
strip: residuals $X-Aj_B$ ($\times20$). (b)~$\varepsilon=1.3\%$, $F_0=0.3$. (c)~$\varepsilon=0$,
$F_0=0.3$. (d)~$\varepsilon=1.3\%$, $F_0=1.5$ (fine grid), where the field sweeps the gapped Dirac
region; the rapid oscillations are numerically converged (Sec.~\ref{sec:strong}); $H_{\rm NP}$ is not
shown, being not converged in the amplitude quadrature~\cite{SM}. (e)~RMS deviation from $Aj_B$
relative to $\max|Aj_B|$, over all times with $|A|>0.1A_{\max}$ ($X$: current times $A$). Black lines:
numerical error level of the refined cases, the largest summed change of the three projections under
time-step, momentum-grid, and angular-derivative or node-count refinement; in every case it is set by
the work projection.}
\label{fig:proj}
\end{figure*}

Figure~\ref{fig:proj} compares the coherent current and its two Hall-type projections with the Berry
current in the weak, intermediate, and strong regimes; the work-based construction, exact for adiabatic
band currents but not a valid separation in coherent dynamics, is shown as a negative control. At
$F_0=0.02$ [Fig.~\ref{fig:proj}(a)] with $\varepsilon=1.3\%$, $\theta_0=0$, the RMS deviation of the direct field-even transverse current from
the Berry current is $2.2\%$ of its peak, dominated by $s$. The single-$\frac13$ angular projection
reduces this to $0.23\%$; the work projection increases it to $5.4\%$~\cite{SM}. In the
residual strip of Fig.~\ref{fig:proj}(a) the direct current shows the symmetric contribution, which the
angular projection removes; the $0.23\%$ that remains is the in-phase Hall-type correction of
Sec.~\ref{sec:res-weak}, not projection error.

In the unstrained lattice [Fig.~\ref{fig:proj}(c)], $C_3$ removes the dipole but not the octupole, so
the Berry current starts at fourth order with a $\cos3\theta$ dependence, while the coherent current
also contains the symmetry-allowed second-order symmetric response. RMS values below are taken over
$|A|>0.1A_{\max}$ relative to $\max|Aj_B|$, as in Fig.~\ref{fig:proj}(e) and Table~\ref{tab:npsym}. At
$F_0=0.02$ the transverse current is almost entirely symmetric: its symmetric part has RMS $3.2$, and
its second harmonic is $-3.512i$, against $+0.236i$ for the Berry current; the single-$\frac13$ angular
projection gives $+0.244i$ and the work projection $-12.46i$. The symmetric term scales as $F_0^2$ and
the octupole Berry current as $F_0^4$, so their ratio changes by a factor of about $225$ between
$F_0=0.02$ and $0.3$, where the symmetric part has RMS $0.014$. At $F_0=0.3$ the direct current deviates
from $Aj_B$ by $0.85\%$, and the Hall-type residual of $H_{\rm NP}$, $1.5\%$, is far above the numerical
error ($5\times10^{-5}$). A finite transverse current without a Berry-curvature dipole is thus consistent
with the symmetry constraints, and its character changes from symmetric to Hall-type with increasing
amplitude.

\subsection{Fourth order: deviation from the octupole term}\label{sec:res-orders}
Order-resolved coefficients were obtained from sixteen drive directions $\theta_k=k\pi/8$ and ten
amplitudes $0.0025\le F_0\le0.3$ (models I and II, $\varepsilon=1.3\%$), fitting
$J/F_0^2=c_2+c_4F_0^2+c_6F_0^4$ and applying Eq.~\eqref{eq:Hp} with $\frac13$ to $c_2$ and $\frac15$ to
$c_4$; the semiclassical currents are processed identically~\cite{SM}. The semiclassical $j_B^{(4)}$
is the octupole term of Eq.~\eqref{eq:multipole}; a direct Brillouin-zone evaluation of
$\mathcal D^{(3)}_x$ with the finite-pulse functional reproduces the fitted $j_B^{(4)}$ to $0.009\%$ (I)
and $0.05\%$ (II). The octupole contributes to both the second and fourth harmonics.

\begin{figure*}[t]
\includegraphics[width=0.92\textwidth]{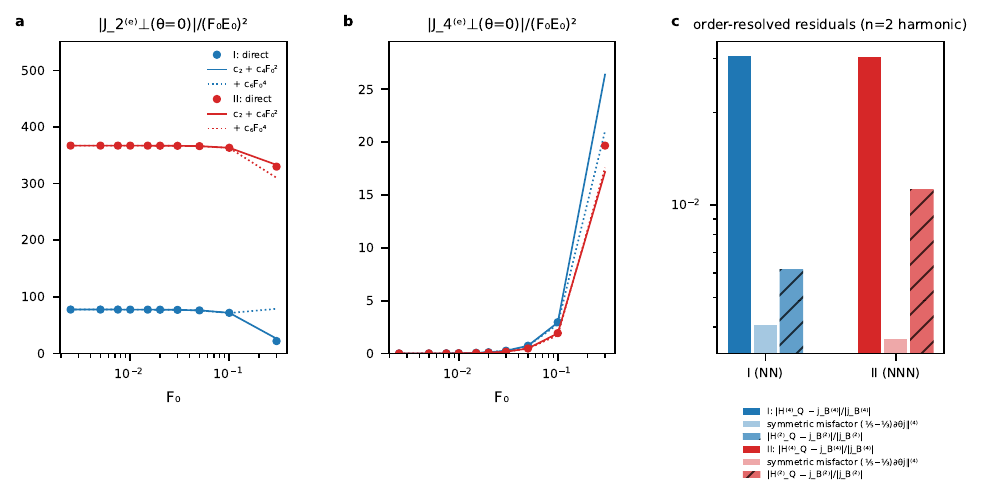}
\caption{Order separation of the field-even transverse response ($\varepsilon=1.3\%$, $\theta=0$).
(a,b)~Directly computed quantum harmonics $|J_n|/(F_0E_0)^2$ (symbols) and the two-term (solid) and
three-term (dotted) truncated quantum expansions, models I (blue) and II (red). (c)~Order-resolved
residuals in the second harmonic: fourth-order Hall-type deviation, symmetric misassignment, and
second-order deviation from finite-time dynamics (at $N=19200$).}
\label{fig:orders}
\end{figure*}

The quantum-semiclassical discrepancy at fourth order lies in the Hall-type sector
[Table~\ref{tab:orders}, Fig.~\ref{fig:orders}(c)]. With three time steps ($N=19200$, $38400$, $76800$;
model I, $\theta=0$), the Hall-type coefficient deviates from the octupole term by
$3.1\pm0.3\%$ in the second harmonic and $6.3\pm0.3\%$ in the fourth. The fourth-harmonic value
decreases monotonically, but with an observed order of about $0.75$ rather than the formal order two of
the integrator, so the sequence is not yet asymptotic; we quote the range spanned by the finest step and
the second-order and observed-order extrapolations. The second-harmonic value is not monotonic in $N$,
so we quote its spread~\cite{SM}. The symmetric misassignment that a single factor $\frac13$
would cause is $0.36$--$0.99\%$. Consistency checks (rank-five symmetrization, absence of
$\cos5\theta$, fit stability) are given in the Supplemental Material~\cite{SM}.

\begin{table*}[t]
\caption{Order-resolved Hall-type response at $\varepsilon=1.3\%$ (second harmonic, $\theta=0$,
imaginary parts; fourth-order coefficients per $(F_0/0.01)^2$). $H^{(2)}_\Q$ from finite-time
dynamics, with the exact Taylor value in parentheses. The octupole value is the direct
$\mathcal D^{(3)}_x$ evaluation. Deviations are computed from the imaginary (in-phase) parts. Model-I
deviations include time-step refinement to $N=76800$; model-II values are at $N=19200$ (preferred fit),
where the complex-modulus measure would give $2.8\%$ ($n=2$) because of a real fit residual.}
\label{tab:orders}
\begin{ruledtabular}
\begin{tabular}{lrr}
 & model I & model II\\
\hline
$H^{(2)}_\Q$ & $-73.78$ ($-73.88$) & $-363.19$ ($-363.43$)\\
$j_B^{(2)}$ (dipole) & $-73.32$ & $-359.16$\\
$H^{(4)}_\Q$ & $0.0572$ & $0.0380$\\
$j_B^{(4)}$ (octupole) & $0.0556$ & $0.0372$\\
deviation from octupole, $n=2$ / $n=4$ & $3.1\pm0.3\%$ / $6.3\pm0.3\%$ & $2.2\%$ / $5.2\%$\\
symmetric misassignment, $n=2$ / $n=4$ & $0.41\%$ / $0.99\%$ & $0.36\%$ / $0.91\%$\\
\end{tabular}
\end{ruledtabular}
\end{table*}

\subsection{Intermediate fields}\label{sec:inter}
At $F_0=0.3$ the drive does not reach the Dirac point and no energy remains after the pulse
(Sec.~\ref{sec:coh}), yet the two-term expansion misses the direct second harmonic by $19.5\%$ in model
I. The same test on the semiclassical harmonics, with identical fit windows and degrees, shows that
this is not a quantum effect: the semiclassical two-term error is $19.3\%$, and adding the third term
reduces it to $5\%$ ($1.5\%$ with the widest window). These results are consistent with
convergence of the semiclassical expansion at $F_0=0.3$, without determining its radius; the failure of
the two-term truncation reflects sixth- and higher-order multipoles. For the quantum harmonics the three-term truncation is
worse than the two-term one, but the sixth-order quantum coefficient is not resolved by the
weak-field fits: across fit windows it varies by more than a factor of ten (and changes sign for $n=4$), whereas the semiclassical one is stable to $2\%$. Whether the quantum
series has a smaller radius therefore cannot be decided from these data. The quantum/semiclassical
ratio of the transverse second harmonic changes from $1.061$ at $F_0=0.1$ to $1.128$ at $F_0=0.3$
(model I).

The nonperturbative projection, Eq.~\eqref{eq:Hnp}, does not use the series and applies directly at
$F_0=0.3$ (model I, $\varepsilon=1.3\%$; quadrature and tests in Ref.~\cite{SM}). Relative to $j_B$, the
second harmonic of $H_{\rm NP}^\Q$ has modulus ratio $0.927$ at $\theta_0=0$ and $1.019$ at
$\theta_0=\pi/4$, and the fourth harmonic $1.077$ and $1.081$; the relative phases vanish
($<10^{-8}$~rad). The numerical uncertainty is $0.4\%$ of the second harmonic (time step; node count and
grid change it by less than $10^{-6}$). At $\theta_0=0$ the difference is
$H^\Q_{{\rm NP},2}-j_{B,2}=+1.43i\pm0.07$ ($\times\Ncal/(F_0E_0)^2$, $j_{B,2}=-19.48i$). The
order-resolved coefficients predict $(H^{(2)}_\Q-H^{(2)}_\SC)+(F_0/0.01)^2(H^{(4)}_\Q-H^{(4)}_\SC)=+0.96i\pm0.68$
(Table~\ref{tab:orders}), where the uncertainty is the spread of the weak-field fits; the measured difference is consistent with
this within that uncertainty, which does not resolve the remaining orders. The fourth harmonic at
$\theta_0=0$ also agrees with the estimate ($-2.17i$ versus $-1.77i\pm0.64$). At $\theta_0=\pi/4$ the
second harmonic is likewise consistent with the estimate, whereas the fourth harmonic is
not~\cite{SM}. In the time domain
[Fig.~\ref{fig:proj}(b,e)] the quantum and semiclassical currents have similar overall temporal
structure, but their projected Hall-type components differ at the few-percent level:
RMS$[A(H_{\rm NP}^\Q-j_B)]/\max|Aj_B|=0.035$ ($\pm0.0006$), and $0.038$ for the single-$\frac13$
angular projection. The single-$\frac13$ projection removes the fully symmetric contribution only at
second field order, whereas the nonperturbative projection eliminates gradient-type contributions at
all orders. The unprojected current nevertheless lies closer to the Berry current ($0.023$) than either
projected result: over this window its symmetric part $\hat{\mathbf e}_\perp\cdot\nabla_{\mathbf E}\Phi$
(RMS $0.037$) is anticorrelated with the Hall-type deviation $H^\Q_{\rm NP}-j_B$ (overlap $-0.79$), so
the two partly cancel; at $F_0=0.01$ the same symmetric part dominates the unprojected residual
($0.020$ of $0.022$) and projection reduces it tenfold. The same anticorrelation occurs at
$\theta_0=\pi/4$ and in the unstrained lattice (Table~\ref{tab:npsym}), whereas in model II, where the
two are positively correlated, the unprojected current deviates more than either projected result
[Fig.~\ref{fig:proj}(e)]. The proximity of the unprojected current therefore reflects a partial
cancellation between the symmetric quantum contribution and the coherent correction within the
Hall-type sector, rather than improved validity of semiclassics. Only $H_{\rm NP}$ is comparable with
$j_B$. At $F_0=1$
($\delta_{\rm LZ}=0.40$; fine grid, numerical uncertainty of each harmonic $\le5\times10^{-5}$ of its
modulus~\cite{SM}) the second-harmonic ratio is $0.912$ with phase $0.10$~rad, and the ratios grow
to $3.6$ and $6.4$ for $n=6$ and $8$.

\subsection{Strong field: sweeping the Dirac point}\label{sec:strong}
At $F_0=1.5$ the excursion $A_{\max}=0.10$ is $1.56$ times the Fermi radius along the drive, so
singly occupied states are dragged through the Dirac point. The relevant adiabaticity parameter is
the Landau-Zener exponent $\delta_{\rm LZ}=M^2/(\hbar v_xeE)$ at the swept node~\cite{SM}: it is $0.40$
at $F_0=1$ and $0.27$ at $F_0=1.5$, giving single-passage transition probabilities
$e^{-\pi\delta_{\rm LZ}}=0.29$ and $0.43$. Collisionless adiabatic semiclassics, which assumes
$e^{-\pi\delta_{\rm LZ}}\to0$, must fail there. At $F_0=0.3$, by contrast, the node is not reached
(transition probability $\approx10^{-22}$~\cite{SM}).

\begin{figure*}[t]
\includegraphics[width=0.92\textwidth]{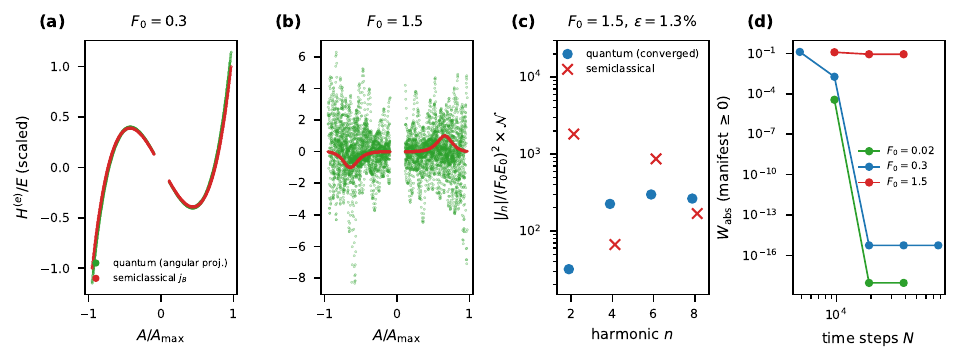}
\caption{Coherence diagnostics at $\varepsilon=1.3\%$ (model I, $\theta_0=0$). (a,b)~Angular-projected
field-even transverse current divided by $E$ versus $A$, scaled to the semiclassical maximum, for
$F_0=0.3$ and $1.5$; the semiclassical Berry current (red) is single-valued. Loops are a secondary
diagnostic (see text). (c)~Converged harmonic moduli at $F_0=1.5$, quantum versus semiclassical,
$\times\Ncal/(F_0E_0)^2$. (d)~Residual absorbed energy after the pulse, Eq.~\eqref{eq:passive}, versus
number of time steps ($\times\Ncal$, per spin); coarse steps produce spurious absorption.}
\label{fig:strong}
\end{figure*}

The coherent harmonics at $\varepsilon=1.3\%$, $\theta_0=0$ are
\begin{align}
J_2^\Q&=4.46+31.64i\ (\pm0.15),\quad J_2^\SC=1819.4i,\nonumber\\
J_4^\Q&=-218.5-54.9i\ (\pm0.07),\quad J_4^\SC=-66.7i,
\label{eq:strong}
\end{align}
with uncertainties from time refinement (momentum refinement changes $J_2^\Q$ by $\le0.01$). The
second harmonic is suppressed by a factor of 57 and the fourth enhanced by $3.4$ with a large real
part; over $n=4$--$8$ the quantum moduli lie between 225 and 300, whereas the semiclassical moduli,
which resum all multipoles~\cite{AvetissianMkrtchian2020}, alternate between 67 and 866
[Fig.~\ref{fig:strong}(c)]. In the time domain no projection brings the quantum current closer than
$79\%$ (RMS) to the Berry current. The situation thus changes qualitatively when the driving field
sweeps the gapped Dirac region [Fig.~\ref{fig:proj}(d)]. Nonadiabatic interband transitions
($\delta_{\rm LZ}<1$) invalidate the assumption of adiabatically transported occupations, and the
quantum response departs strongly from the semiclassical Berry current. All projections nearly
coincide: the symmetric part they remove is small compared with the current carried by real interband
population. The rapid oscillations are numerically converged (time-step and extra-fine-grid
refinements change the signal by at most $3\times10^{-4}$ of its RMS) and are consistent with interband
coherence at the local gap $2|\mathbf d|/\hbar$ of the states swept through the node, which ranges from
$4\omega$ to about $50\omega$~\cite{SM}. The nonperturbative projection is not quantitatively converged at this
amplitude.

\subsection{Residual excitation and hysteresis}\label{sec:coh}
Whether excitation remains is decided by the energy left in the electrons after the cyclic pulse. At
the end of the pulse, for each $\bk_0$,
\begin{equation}
\Delta E_{\bk}=-gR\,|\mathbf S-\hat{\mathbf d}|^2/2+gR\,(|\mathbf S|^2-1)/2,\qquad R=|\mathbf d|,
\label{eq:passive}
\end{equation}
with $g\le0$: the first term is manifestly non-negative, and the second is norm drift, zero for exact
unitary evolution. At $\varepsilon=1.3\%$ the residual absorbed energy is $9\times10^{-19}$,
$5.6\times10^{-16}$, and $8.7\times10^{-2}$ ($\times\Ncal$, per spin) at $F_0=0.02$, $0.3$, and $1.5$;
the last is $18\%$ of the maximal transient work, with excited population $5.9\times10^{-5}$ per spin.
At $F_0=0.3$ the transient work reaches $1.9\times10^{-2}$ but less than $3\times10^{-14}$ of it
remains. This is our primary diagnostic of real excitation, and it agrees with the Landau-Zener
estimates: reversible interband polarization at weak and intermediate fields, residual real
excitation under node sweep. The lattice dynamics are unitary, so residual excitation does not by
itself imply dissipation.

Hysteresis is a secondary diagnostic. A Berry current $E(t)\bar\Omega(A)$ makes $j_B/E$ single-valued
in $A$, but a reversible polarization current depends on $\dot E$ and is not of the form $E\,f(A)$, so
$H/E$ versus $A$ can show loops without any real excitation. At finite field the single
$\frac13$ projection also leaves fourth-order symmetric pieces. With these caveats, the angular-projected
current at $\varepsilon=1.3\%$ is single-valued within errors at $F_0=0.02$ and $0.3$ (median
significance $z=0.3$ and $0.8$) and strongly hysteretic at $F_0=1.5$ ($z\approx90$), consistent with
the absorbed energy. The nonperturbative projection $H_{\rm NP}$, which removes the symmetric part at all orders,
gives the same picture: $H^\Q_{\rm NP}/E$ is single-valued within errors at $\varepsilon=1.3\%$,
$F_0=0.3$ ($z=0.7$) and strongly hysteretic at $F_0=1$ (fine grid; $h=0.084$, $z\approx2\times10^3$). The unstrained lattice at $F_0=0.3$ shows weak
but significant loops, $h=5.6\times10^{-4}$, $z=14$ ($84\%$ of matched points with $z>3$), against
$h=6.2\times10^{-4}$, $z=16$ with the single-$\frac13$ angular projection. The loops therefore survive removal
of the symmetric sector at every order. No absorbed energy is resolved in this case
($|\int\jpar dA|\le4\times10^{-14}$, the quadrature error level). We therefore attribute these loops to
reversible currents in the Hall-type sector that are not of the form $E\,f(A)$, such as coherent
polarization, and not to real excitation; the loops alone do not identify that current~\cite{SM}.

\subsection{Valley additivity across a topological transition}\label{sec:km}
\begin{figure*}[t]
\includegraphics[width=0.88\textwidth]{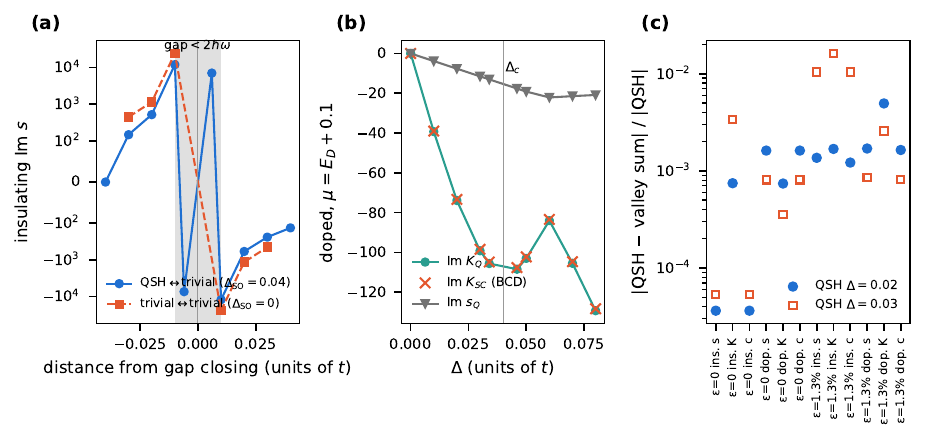}
\caption{Spin-conserving Kane--Mele model ($\Delta_{\rm SO}=0.04\,t$, $\varepsilon=1.3\%$, charge
response). (a)~Insulating symmetric coefficient versus distance from the gap closing, across the
topological transition (blue) and across a trivial inversion of all four cones (red); the gray band
marks the resonant window. (b)~Doped Hall-type coefficient $K_\Q$, BCD prediction $K_\SC$, and $s_\Q$
across the transition. (c)~Deviation of topological-phase coefficients from the valley-additive
prediction built from non-topological controls.}
\label{fig:km}
\end{figure*}

In a spin-conserving Kane--Mele model~\cite{Kane2005a,Kane2005b} built on the same lattice (illustrative
$\Delta_{\rm SO}=0.04\,t$, not a germanene parameter set~\cite{SM}), each spin block is a two-band
model with valley masses $\Delta\mp s\Delta_{\rm SO}$, and the topological transition corresponds to a
sign change of one valley mass per block. Topological and trivial partners at matched gap differ in
every nonlinear coefficient, and in the insulating state some coefficients reverse sign. Two
non-topological controls reproduce these differences. The controls lie in the same parameter space: a
trivial Kane--Mele layer with the roles of $\Delta$ and $\Delta_{\rm SO}$ interchanged, which has the
same valley-mass magnitudes as the topological partner but no mass inversion, and the reversal of the
staggered potential at $\Delta_{\rm SO}=0$, as realized by reversing the perpendicular field in a
buckled Xene or by the opposite alignment registry on hBN. Every coefficient obeys valley additivity
$\sum_v\mathrm{sgn}(m_v)F(|m_v|)$ to $0.004$--$1.6\%$ at second order [Fig.~\ref{fig:km}], and the
in-phase correction to $K$ is $0.748$--$0.760\%$ on both sides of the transition. The bulk Hall-type
harmonic coefficients thus distinguish the phases only through the local Dirac masses, consistent with
Ref.~\cite{Neufeld2023} and in contrast to topological signatures proposed for other
observables~\cite{Silva2019,Chacon2020}. Edge states, spin-mixing (Rashba) terms, and the
spin-resolved response, whose mirror-odd components cancel in the charge current, lie outside this
test. Details, including a germanene-motivated parameter set, are in the Supplemental
Material~\cite{SM}.

\section{Discussion}\label{sec:disc}

The results form a progression that begins with Ref.~\onlinecite{AvetissianMkrtchian2020}. In
semiclassics the anomalous even harmonics are set perturbatively by the Berry-curvature multipoles,
Eqs.~\eqref{eq:multipole} and \eqref{eq:AM27}. The coherent response adds a fully symmetric transverse
contribution, the Kleinman-symmetric reactive $O(\omega)$ term of Eq.~\eqref{eq:lowfreq}, which the
angular projection removes at each order; at weak fields this substantially improves agreement with
$Aj_B$, whereas the work projection, incorrect in the field-even channel~\cite{SM}, worsens
it [Fig.~\ref{fig:proj}(e)]. The remaining Hall-type response differs from the multipole result: at
second order by an in-phase $O(\eta^2)$ correction, given in closed form by resonance denominators for
the linear Dirac model; at fourth order by several percent relative to the directly evaluated
octupole; and qualitatively when $\delta_{\rm LZ}\lesssim1$, where no tested projection restores
agreement. Since $H_{\rm NP}$ is exact by construction, these residuals measure the limits of the
Berry-curvature approximation, not of the projection. Isolating the Hall-type sector is thus distinct
from recovering the semiclassical current.

These results are ordered by the two parameters of the adiabatic expansion~\cite{SM}: the frequency
ratio $\hbar\omega/|\mathbf d|$ and the field ratio $\hbar v\,eE/|\mathbf d|^2$. The first controls the
second-order quantum effects: $s$ and $\delta K$ appear at relative order $\eta^2$ and follow
quantitatively from the third adiabatic order. The second grows with amplitude and is of order
$F_0\eta$ at the Fermi surface. The fourth-order Hall-type deviation from the octupole term, and the
deviation of $H_{\rm NP}$ at $F_0=0.3$ that is consistent with it, are corrections of this kind,
although we have not derived them from the adiabatic expansion. In this regime the symmetric part and
the Hall-type correction are comparable and can partly cancel in the unprojected current, so a smaller
unprojected residual does not imply a more accurate Hall-type description, and only a projected
comparison is meaningful. For states dragged to the Dirac point the field ratio becomes
$1/\delta_{\rm LZ}\gtrsim1$, the expansion fails, and Landau--Zener excitation produces the
strong-field breakdown. The progression from weak to strong driving is thus a progression in
adiabaticity, and the projections determine which part of the deviation belongs to the Hall-type
sector.

Both second-order quantum effects, $s$ and $\delta K$, are of the same order in the low-frequency
expansion. A measured transverse harmonic at $\hbar\omega\ll E_{\rm ref}$ therefore equals the BCD
term up to corrections of relative order $\eta^2$, of which the symmetric one can be removed by the
angular dependence and the Hall-type one cannot. A measured Hall-type current is thus not by itself a
measurement of the BCD, and still less of band topology. The failure of the work projection in the
field-even channel is a symmetry statement and applies to any energy-based split of measured
strong-field currents into ``band'' and ``anomalous'' parts~\cite{SM}.

\emph{Model dependence.} The near independence of $s$ from strain and model and the size of $\delta$
relative to the continuum value are properties of the two-band models studied. The NNN strain laws
and the staggered mass are not calibrated by first-principles theory.

\emph{Open questions.} The third adiabatic order accounts for $s$ and $\delta K$ quantitatively; a static
hyperpolarizability calculation would still be needed to establish that $s$ derives from an energy in
the doped system. At $F_0=0.3$ the nonperturbative projection shows that the Hall-type
quantum second harmonic differs from $j_B$ by $-7\%$ ($\theta_0=0$) and $+2\%$ ($\theta_0=\pi/4$),
consistent within the fit uncertainty with the fourth-order deviation alone (the fourth harmonic at $\theta_0=\pi/4$ is not); whether higher quantum orders contribute, i.e., whether the quantum
amplitude series has a smaller radius than the semiclassical one, is not resolved. The reversible
Hall-type loops of the unstrained lattice also remain to be identified. Relaxation,
disorder, and interactions are absent from the lattice calculations.

\emph{Experimental relevance.} At model level: (i) a symmetric, $\cos3\theta$ transverse response
linear in frequency in unstrained gapped honeycomb layers, replaced by a $\cos\theta$ Hall-type term
under strain; (ii) an in-phase deviation from the BCD formula growing as $\omega^2$; (iii) residual
excitation once the field excursion drags carriers through the Dirac point, controlled by
$\delta_{\rm LZ}$. Since $H_{\rm NP}$ acts harmonic by harmonic, it can be applied directly to measured
complex harmonic fields, which for a thin film are proportional to the sheet current at $n\omega$
through an amplitude- and orientation-independent transfer factor (with the two field components
calibrated if the substrate or detection is anisotropic). This requires both in-plane
components with their phases at a fixed waveform, an amplitude scan starting near zero, and the
dependence on drive orientation near $\theta_0$, i.e., phase- and polarization-resolved harmonic
detection combined with intensity and polarization-rotation scans. Harmonic intensities alone are not
sufficient.

\section{Conclusions}\label{sec:concl}

Starting from the semiclassical Berry-curvature theory of anomalous harmonics~\cite{AvetissianMkrtchian2020},
we examined how the transverse response changes when coherent quantum dynamics is retained. Three
levels emerge.

\emph{(1) Perturbative semiclassical multipoles.} In collisionless semiclassics the field-even
current is purely Hall-type; its order $p$ is the $(p-1)$th multipole term, and at leading order the
$2n$th harmonic is set by the $(2n-1)$th multipole~\cite{AvetissianMkrtchian2020}.

\emph{(2) Order-resolved quantum response.} The coherent current of definite order separates exactly
into symmetric and Hall-type sectors, implemented at arbitrary order by Eq.~\eqref{eq:Hp}. The
symmetric sector is the reactive $O(\omega)$ term; the Hall-type sector exceeds the dipole term by an
in-phase $O(\eta^2)$ correction ($c_2=1.294$ in closed form for the tilted Dirac model, $0.76$--$1.33\%$
at $\eta=0.1$ on the lattices) and deviates from the octupole term by $3.1\%$ and $6.3\%$ in the second
and fourth harmonics. The symmetric coefficient and the in-phase correction, both at second order in
the field, follow quantitatively from the third adiabatic order of the Bloch equation; for any affine
$\mathbf d(\bk)$ the full second-order response is purely Hall-type. The nonperturbative resummation $H_{\rm NP}$ of the Hall-type sector, exact in semiclassics, gives at
$F_0=0.3$ a second-harmonic deviation of $-7\%$ ($\theta_0=0$) and $+2\%$ ($\theta_0=\pi/4$) from the
Berry current, consistent within the fit uncertainty with the fourth-order deviation.

\emph{(3) Nonperturbative coherent dynamics.} When the field drags carriers through the Dirac point
($\delta_{\rm LZ}\approx0.27$ at $F_0=1.5$), real excitation remains after the pulse and the harmonics
differ qualitatively from the resummed semiclassical result.

The three levels are successive regimes of adiabaticity, set by $\hbar\omega/|\mathbf d|$ and by the
field ratio $\hbar v\,eE/|\mathbf d|^2$, which reaches $1/\delta_{\rm LZ}$ for states swept through the
Dirac point.

In a spin-conserving Kane--Mele model, the bulk Hall-type harmonic coefficients are not a signature
of global topology: they follow the local Dirac masses.

\begin{acknowledgments}
The work was supported by the Science Committee of Republic of Armenia.
\end{acknowledgments}

\section*{Data availability}
The data that support the findings of this article are openly available~\cite{dataset}. The archive
contains the numerical results underlying all figures and tables, the Bloch-equation and zero-field
Taylor-kernel solvers, and the scripts that generate each figure, table and quoted number, with a README
mapping manuscript items to scripts. Intermediate time-evolution data and earlier development code are
available from the corresponding author upon reasonable request.

\bibliography{references,sm}

\clearpage
\onecolumngrid
\setcounter{section}{0}\setcounter{equation}{0}\setcounter{figure}{0}\setcounter{table}{0}
\renewcommand{\thesection}{S\arabic{section}}
\renewcommand{\theequation}{S\arabic{equation}}
\renewcommand{\thefigure}{S\arabic{figure}}
\renewcommand{\thetable}{S\arabic{table}}
\renewcommand{\theHsection}{S\arabic{section}}
\renewcommand{\theHequation}{S\arabic{equation}}
\renewcommand{\theHfigure}{S\arabic{figure}}
\renewcommand{\theHtable}{S\arabic{table}}
\begin{center}
{\large\bfseries Supplemental Material for ``Symmetric and Hall-type transverse harmonics in strained honeycomb lattices: adiabatic and nonperturbative corrections to Berry-curvature semiclassics''\par}
\vspace{1ex}
G. F. Mkrtchian\\[0.5ex]
{\small\itshape Centre of Strong Fields Physics at Research Institute of Physics, Yerevan State University, Yerevan 0025, Armenia}
\end{center}
\makeatletter
\def\SM@tocname{toc}\def\stocname{\tocname}\def\att@TOC{stoc}
\print@toc{stoc}
\let\SM@atc\addtocontents
\def\addtocontents#1{\def\SM@a{#1}\ifx\SM@a\SM@tocname\expandafter\@firstoftwo\else\expandafter\@secondoftwo\fi{\SM@atc{stoc}}{\SM@atc{#1}}}
\makeatother

\section{Units, parameters and normalization}
Lattice units are $t_{\rm ref}=a_0=\hbar=e=1$, and the electron charge is $-1$. The fixed physical
protocol, identical for all strains, is: $E_{\rm ref}=0.1$, staggered mass $M=0.02$ (direct gap
$0.04$), $k_BT=0.002$, $\hbar\omega=0.01$, $C=20$ cycles, and peak field $F_0E_0$ with
$E_0=1/1500$. The chemical potential is $\mu=E_D+0.1$ above the local Dirac energy $E_D$. The
excursion scale is $E_0/\omega=E_{\rm ref}/v_0$ with $v_0=3/2$, so $F_0=1$ means
$A_{\max}v_0\approx E_{\rm ref}=\mu-E_D$.

Integrals are $\int[d\bk]=\int d^2k/(2\pi)^2$ over the full primitive cell. Currents are per spin and
per unit area. Reported values are multiplied by $\mathcal N=(2\pi)^2\cdot1.5/0.1^2=5921.76$. With $v_0=3/2$ and
$E_{\rm ref}=0.1$, $\mathcal N=(2\pi)^2v_0/E_{\rm ref}^2$ converts a per-spin lattice current density
into the natural current unit $eE_{\rm ref}^2/[(2\pi)^2\hbar v_0]$ of the continuum model, in which
momenta are measured in $E_{\rm ref}/v_0$ and integrals are $\int d^2q$. Harmonic coefficients follow
Eq.~\eqref{eq:harm} of the main text, divided by $(F_0E_0)^2$. For second-order signals the Hann
average gives $\langle\sin^6\rangle/\langle\sin^2\rangle=0.625$ relative to a steady monochromatic
drive of the same peak amplitude, so monochromatic coefficients are the Hann values divided by
$0.625$; ratios such as $\delta$ are unaffected. A monochromatic current $\mathrm{Re}[Xe^{-inu}]$ has
$J_n=X/2$.

\paragraph{Physical units.} With $t_{\rm ref}=2.7$~eV and $a_0=1.42$~\AA: field unit
$t_{\rm ref}/(ea_0)=1.90\times10^{10}$~V/m, so $E_0=1/1500$ is $1.27\times10^7$~V/m ($0.127$~MV/cm);
current-density unit $et_{\rm ref}/(\hbar a_0)=4.63\times10^6$~A/m per spin; $\hbar\omega=27$~meV
($6.53$~THz); $k_BT=63$~K. The monochromatic Hall coefficient of model I at $\varepsilon=1.3\%$,
$|K|/0.625=118.2$, gives $\sigma^{(2)}=2\times118.2/\mathcal N$ per spin in lattice units, i.e.,
$1.02\times10^{-15}$~A\,m/V$^2$ for both spins, and a second-harmonic current amplitude
$1.6\times10^{-3}$~A/m at $12.7$~kV/cm ($F_0=0.1$). Spin degeneracy is not included for the spinless models. In the
Kane--Mele model the charge current is the explicit sum over the two spin blocks.

All numbers, tables and figures of the main text and of this Supplement can be regenerated from the
archive of Ref.~\cite{S-dataset}; its README lists, for each item, the generating script and data file.

\section{Continuum-to-lattice expansion}
Near $\bk=(-\tau K,0)$, with $\cos(\sqrt3a_xK/2)=-t_1/(2t_2)$, the NN form factor expands as
\begin{equation}
d_x=\tau v_xq_x+\tfrac38a_x^2t_1q_x^2-\tfrac38a_y^2t_1q_y^2+O(q^3),\qquad
d_y=v_yq_y-\tau\tfrac12a_yv_xq_xq_y+O(q^3),
\end{equation}
with $v_x=\sqrt3a_xt_2\sin(\sqrt3a_xK/2)$ and $v_y=\tfrac32a_yt_1$. Strain makes $v_x\neq v_y$ and
the trigonal warping anisotropic; the warping produces the BCD of model I. The scalar
NNN term gives the tilt
$\alpha=2\sqrt3a_x\sin(\sqrt3a_xK/2)[t'_p+2t'_h\cos(\sqrt3a_xK/2)]$, which vanishes for
$t'_p=(t_1/t_2)t'_h$ and equals $-0.0045v_x$ for model II at $\varepsilon=1.3\%$. The
continuum model of the main text, Eq.~\eqref{eq:cont}, keeps the linear Dirac part with $\alpha/v=0.2$; this tilt
is realized on the lattice only by the engineered model III. The quadratic Taylor residuals scale
cubically and reproduce the strained-graphene warping form of Ref.~\cite{S-Ortix2021} after the
coordinate and orbital conventions are mapped.

\section{Symmetry analysis}\label{sec:symSM}
\begin{enumerate}
\item \emph{Time reversal}, $H(-\bk)=H(\bk)^*$. In collisionless semiclassics it implies that the
band current is odd and the Berry current even in the field, at all orders. Hence the field-even
semiclassical current is purely Hall-type, $s_{\rm SC}\equiv0$, and the field-odd band energy
vanishes, $\Phi^{(o)}\equiv0$. In the coherent dynamics an instantaneous function of $A$ is odd in
$A$, so field-even currents carry at least one time derivative.
\item \emph{Mirror} $M_x$: $H(-k_x,k_y)=H(\bk)$. It eliminates the tensor components with an odd
number of $x$ indices. For $x$ drive ($\theta=0$), $J_y$ is exactly field-even and $J_x$ exactly
field-odd at all orders, so $W^{(o)}(\theta=0)=0$ but $\partial_\theta W^{(o)}(0)\neq0$. More
generally $W(\pi-\theta)=W(\theta)$.
\item \emph{$C_3$} ($\varepsilon=0$): $H(R\bk)=H(\bk)$ in the embedded gauge, since rotation permutes
the bonds. It imposes $a=b=-c$, i.e., $K=0$. The field-even transverse current at higher order
varies as $\cos3\theta$; numerically the $\theta=\pi/4$ values equal $-0.7071$ times the $\theta=0$
values.
\item \emph{Inversion} is broken by $M$. In the Kane--Mele model it is restored at $\Delta=0$, where
all second-order responses vanish.
\end{enumerate}
The residuals of all these relations are $\le10^{-15}$ for the Hamiltonian and $\le2\times10^{-9}$
for the response tensors (Table~\ref{tab:checks}).

\section{Quantum propagator, current and filled-band subtraction}
\paragraph{Propagator.} Over a step $\Delta u$ we rotate $\mathbf S$ by the angle
$\phi=2|\mathbf m|\Delta u/\omega$ about $\hat{\mathbf m}$, where
$\mathbf m=\mathbf d(\bk(u+\Delta u/2))$:
\begin{equation}
\mathbf S'=\cos\phi\,\mathbf S+\sin\phi\,\hat{\mathbf m}\times\mathbf S+(1-\cos\phi)(\hat{\mathbf m}\cdot\mathbf S)\hat{\mathbf m}.
\end{equation}
This map is norm-preserving to roundoff and second-order accurate. For states far from the Fermi
surface the precession angle per step is large ($2|\mathbf d|\Delta u/\omega\sim4$ at $N=19200$). The
scheme then follows the adiabatic state only approximately, and it requires $N\gtrsim19200$ for
endpoint observables (Sec.~\ref{sec:energy}).

\paragraph{Current.} The current is $\mathbf j=-\int[d\bk]\{n\partial_{\bk}d_0+g\,\partial_{\bk}\mathbf d\cdot\mathbf S\}$,
with all bond derivatives evaluated analytically at $\bk(t)$. The valence reference state ($n=1$,
$g=-1$, $\mathbf S=\hat{\mathbf d}(\bk(t))$) carries $-\partial_{\bk}\varepsilon_v$, whose full-cell
integral is exactly zero. We integrate $\mathbf j-\mathbf j_{\rm ref}$ and
$E_{\rm el}-E_{\rm ref}$; no state is removed from the propagation.

\paragraph{Validation.} The independent solver reproduces the zero-field Taylor coefficients $a$ and
$b$ to $0.13\%$ and $0.15\%$ after Richardson extrapolation in $F_0$. The remaining offset is
temporal (Sec.~\ref{sec:orders}). Equilibrium currents vanish to $\le2\times10^{-15}$, and the
norm is conserved to $\le5\times10^{-12}$.

\section{Semiclassical equations}
The semiclassical solver is Lagrangian: $f_b(\bk_0)$ is fixed, and velocities and Berry curvature
are evaluated at $\bk_0-A\be$. The conduction-band curvature is
$\Omega_c=-\mathbf d\cdot(\partial_x\mathbf d\times\partial_y\mathbf d)/(2|\mathbf d|^3)$, with
$\Omega_v=-\Omega_c$; this sign was fixed against an independent implementation. Occupations
relative to the filled valence band ($f_c$, $f_v-1$) remove the vanishing filled-band integrals
exactly. The semiclassical band energy $\Phi=\sum_b\int f_b[\varepsilon_b(\bk_0-A\be)-\varepsilon_b(\bk_0)]$
obeys $A\jperp-\partial_\theta\Phi|_A=Aj_{B\perp}$ identically. Numerically this holds to
$\le1.2\times10^{-7}$ relative with exact $\Phi$ in all fourteen cases. The time-quadrature version
fails in the field-odd channel at $\theta=\pi/4$, at the $10^{-4}$--$10^{-3}$ level.

For the continuum model we also used generalized semiclassics with dephasing (kernel
$K_\Gamma=2m/[|\mathbf d|(4|\mathbf d|^2+\Gamma^2)]$) and relaxation-time occupations. The RTA factor
$1/(1+i\gamma_1/\omega)$ is exact in generalized semiclassics, and the coherent quantum calculation
violates it at order $0.5\eta^2\gamma_1/\omega$. Dephasing adds an out-of-phase correction
$\mathrm{Im}\,\delta\approx1.5\eta^2\gamma_2/\omega$. The limits $m\to0$ and $\Gamma\to0$ do not
commute.

\section{Berry-curvature multipole hierarchy}\label{sec:multipoleSM}
This section supplements Sec.~\ref{sec:multipole} of the main text. The multipole expansion of
the anomalous current and its perturbative harmonic content were derived by Avetissian and
Mkrtchian~\cite{S-AvetissianMkrtchian2020}. We restate the steps and verify the prefactor.

\paragraph{Conventions of Ref.~\cite{S-AvetissianMkrtchian2020}.} The electron charge is $-e$ with
$e>0$, the interaction is $\hat H_{\rm int}=e\mathbf E\cdot\hat{\mathbf r}$, and the current is
$\mathbf j=-e\langle d\hat{\mathbf r}/dt\rangle$. The field-induced momentum is
$\mathbf p_E(t,t')=e\int_{t'}^t\mathbf E(t'')\,dt''$. Relaxation to equilibrium at rate $\gamma$ gives
the anomalous current of Eq.~(20) there. For $\hat{\mathbf e}=\hat x$, a Taylor expansion of
$B_{1z}(p_x+p_E,p_y)$ together with time reversal gives [Eq.~(25) there]
\begin{equation}
j_{yT}(t)=-2\frac{e^2}{\hbar}\gamma\sum_{n\ge1}\frac{D_x^{(2n-1)}}{(2n-1)!}\int_{-\infty}^t dt'\,
e^{-\gamma(t-t')}E_x(t)\,p_E^{2n-1}(t,t').
\end{equation}
Here $D_x^{(2n-1)}$ [Eq.~(26) there] refers to one valley ($\tau=1$) and contains the degeneracy $g$.
The factor 2 accounts for the time-reversed valley.

\paragraph{Perturbative harmonic.} For $E_x=E_0\cos\omega t$ and $\gamma\ll\omega$,
\begin{equation}
p_E(t,t')=\frac{eE_0}{\omega}\,(\sin\omega t-\sin\omega t'),
\end{equation}
and the weighted integral $\gamma\int dt'\,e^{-\gamma(t-t')}(\cdots)$ becomes an average over the
phase $\omega t'$. The highest power of $\sin\omega t$ in
$p_E^{2n-1}$ is $(eE_0/\omega)^{2n-1}\sin^{2n-1}\omega t$, and only this term reaches the harmonic
$2n$ at order $E_0^{2n}$. Using
\begin{equation}
\sin^{2n-1}x=2^{2-2n}\sum_{k=0}^{n-1}(-1)^{k}\binom{2n-1}{n-1-k}\sin[(2k+1)x],
\end{equation}
whose top term ($k=n-1$) carries $(-1)^{n-1}2^{2-2n}$, together with
$\cos x\,\sin[(2n-1)x]=\frac12\sin(2nx)+\frac12\sin[(2n-2)x]$, the coefficient of $\sin(2n\omega t)$ is
\begin{equation}
(-1)^n\,\frac{e^{2n+1}D_x^{(2n-1)}E_0^{2n}}{\hbar\,\omega^{2n-1}\,2^{2n-2}\,(2n-1)!}.
\end{equation}
Its magnitude equals Eq.~(27) of Ref.~\cite{S-AvetissianMkrtchian2020}; Eq.~\eqref{eq:AM27} of the
main text is the same result in the present normalization. The alternating sign $(-1)^n$ follows from Eq.~(25) as printed; Eq.~(27) is printed
without it. This affects only the overall phase convention of each harmonic, not the multipole
correspondence or any magnitude. Lower powers of $\sin\omega t$, which multiply even powers of
$\sin\omega t'$ with nonzero phase average, contribute only to lower harmonics at this field order.

\paragraph{Relation to the present notation.} Our crystal momentum is $\bk=\bk_0-A\be$ with
$\dot A=E$ (charge $-1$, $\hbar=1$), so $-A$ plays the role of $p_E/e$. Our multipoles
$\mathcal D_x^{(2n-1)}=\sum_b\int[d\bk]f_b\partial^{2n-1}_{k_x}\Omega_b$ sum over both valleys and all
bands, per spin. In Ref.~\cite{S-AvetissianMkrtchian2020}, $D_x^{(2n-1)}$ is a one-valley integral
multiplied by the degeneracy $g$, which multiplies the explicit sum over valleys in Eq.~(11) there and
therefore counts spin, and the time-reversed valley enters through the factor 2 of Eq.~(25). Hence
$\mathcal D_x^{(2n-1)}=(2/g)\,D_x^{(2n-1)}$, which equals $D_x^{(2n-1)}$ for $g=2$. In this normalization
the explicit factor 2 is absorbed into $\mathcal D$, so the denominator $2^{2n-2}$ of Eq.~(27) there
becomes $2^{2n-1}$ in Eq.~\eqref{eq:AM27}; Eq.~(27) there is the current of both spins.
For $n=1$, Eq.~\eqref{eq:AM27} gives $-\mathcal D^{(1)}_xE_0^2\sin(2\omega t)/(2\omega)$, which is the
second-order term of Eq.~\eqref{eq:multipole}, $-EA\,\mathcal D^{(1)}_x$ with
$EA=(E_0^2/2\omega)\sin2\omega t$. The filled valence band contributes nothing because the derivatives of a periodic
function integrate to zero. The collisionless finite-pulse expression, Eq.~\eqref{eq:multipole},
contains no relaxation average. For a monochromatic field its leading harmonic content agrees with
the relaxation-time result at $\gamma\ll\omega$, because only the top power of $\sin\omega t$
contributes.

\paragraph{Consistency check.} The finite-pulse dipole coefficient $K_\SC$ computed from
$\mathcal D^{(1)}_x$ and the Hann functional of $-E(u)A(u)$ agrees with Richardson-extrapolated
finite-field semiclassical dynamics to $0.015\%$ (model I, $\varepsilon=1.3\%$), and with an
independent implementation of the same functional to $<10^{-9}$ relative.

\paragraph{Direct octupole.} We evaluated $\mathcal D^{(3)}_x$ as a Brillouin-zone integral in two
ways: directly, with a five-point finite difference for $\partial_x^3\Omega$, and by parts,
$-\int[d\bk]\,\partial_x(g+1)\,\partial_x^2\Omega_c$, a Fermi-surface form. On base, fine and extra-fine
grids and for steps $h=2\times10^{-4}$ and $10^{-4}$ all values agree to $5\times10^{-5}$ relative:
$\mathcal D^{(3)}_x=-20.535$ (model I) and $-13.709$ (model II), per spin, at $\varepsilon=1.3\%$. The
finite-pulse fourth-order coefficient is
$c_{4,n}=10^{-4}\mathcal N\,J_n[-e(u)a(u)^3/6]\,\mathcal D^{(3)}_x/E_0^2$ per $(F_0/0.01)^2$, with $e,a$
the field and vector potential at $F_0=1$. It equals $0.055649i$ ($n=2$) and $-0.027778i$ ($n=4$) for
model I, and $0.037151i$ and $-0.018544i$ for model II. The semiclassical $j_B^{(4)}$ extracted from
the amplitude fits agrees to $0.009\%$ and $0.005\%$ (I) and $0.02\%$ and $0.05\%$ (II).

\section{Frequency dependence of the symmetric coefficient}\label{sec:sw}
The exact zero-field Taylor kernel behind Table~\ref{tab:weak} of the main text was evaluated for
model I at $\varepsilon=0$ at four frequencies, on the base and fine momentum grids (identical to all
digits shown). The occupation weight $g=f_c-f_v$ was split into the completely filled valence band
($g=-1$ everywhere) and the Pauli-blocked remainder $g+1$, which is nonzero only in the doped pocket.
Table~\ref{tab:sw} lists $s=(2a+b)/3$ (Hann values; monochromatic values are larger by $1/0.625$).

\begin{table}[!htbp]
\caption{Symmetric coefficient $s$ of model I at $\varepsilon=0$ (imaginary parts, Hann,
$\times\mathcal N/E_0^2$, per spin). Real parts are below $10^{-8}$, except that at $\hbar\omega=0.02$ the
insulator and pocket parts carry equal and opposite real parts $\pm2821.4$ (two-photon resonance of
the filled band), which cancel in $s$. Last column: relative in-phase correction
$\delta=(K_\Q-K_\SC)/K_\SC$ of model I at $\varepsilon=1.3\%$ for the same frequencies (fixed $\mu$, $E_{\rm ref}$ and $E_0$).}
\label{tab:sw}
\begin{ruledtabular}
\begin{tabular}{rrrrrr}
$\hbar\omega$ & $s$ & $s/\hbar\omega$ & insulator part & pocket part & $\delta$, I, $1.3\%$\\
\hline
0.0025 & $-0.934$ & $-373.7$ & $-116.4$ & $+115.5$ & $0.0472\%$\\
0.005 & $-1.871$ & $-374.2$ & $-241.5$ & $+239.6$ & $0.189\%$\\
0.01 & $-3.764$ & $-376.4$ & $-566.9$ & $+563.2$ & $0.760\%$\\
0.02 & $-7.704$ & $-385.2$ & $-5978.8$ & $+5971.1$ & $3.105\%$\\
\end{tabular}
\end{ruledtabular}
\end{table}

$s/\hbar\omega$ is constant to $0.7\%$ for $\hbar\omega\le0.01$ ($\eta\le0.1$) and rises by $3\%$ at
$\eta=0.2$. In monochromatic normalization $s/\hbar\omega=-598i$, $-599i$, $-602i$ for
$\hbar\omega=0.0025$--$0.01$, and at $\hbar\omega=0.01$ the insulator and pocket parts are $-907i$ and
$+901i$; these agree with an independent monochromatic calculation to better than $0.2\%$. The
insulator part grows faster than linearly because its gap $2M=0.04$ puts a two-photon resonance at
$\hbar\omega=0.02$; in the doped system Pauli blocking removes the states near the band edge, and the
total remains linear. We did not compute the static hyperpolarizability $\chi^{(2)}(0;0,0)$ of the
filled band plus Fermi sea. The Kleinman-symmetric $s$ would derive from an energy in an insulator far
below resonance~\cite{S-HeLaw2026}; whether it does in the doped lattice therefore remains open.

\paragraph{Frequency parity.} Equation~\eqref{eq:lowfreq} of the main text rests on the following
argument. The electrons evolve unitarily from a time-reversal-invariant equilibrium state $\rho_0$
under a drive switched on and off adiabatically, with all frequencies below the Pauli-allowed interband
thresholds, so that the final state $\rho(+\infty)$ equals $\rho_0$. Time reversal $\Theta$ maps the
evolution under $\mathbf E(t)$ onto the evolution under $\mathbf E_r(t)=\mathbf E(-t)$ [with
$A_r(t)=-A(-t)$ for a cyclic pulse and $\bk_0\to-\bk_0$], started from $\Theta\rho(+\infty)\Theta^{-1}$,
and the current is odd, $\mathbf j_r(t)=-\mathbf j(-t)$. With $\mathbf E(t)=\int d\omega\,\mathbf
E(\omega)e^{-i\omega t}$ one has $\mathbf E_r(\omega)=\mathbf E(-\omega)$, and the $2\omega$ component of
$\mathbf j_r(t)=-\mathbf j(-t)$ gives $\sigma(\omega)\mathbf E(-\omega)^2=-\sigma(-\omega)\mathbf
E(-\omega)^2$ (tensor indices suppressed), hence $\sigma(-\omega)=-\sigma(\omega)$; reality,
$\sigma(-\omega)=\sigma(\omega)^*$, then makes $\sigma$ purely imaginary. The two conditions enter at
different steps. The reversed evolution is the equilibrium response only if
$\Theta\rho(+\infty)\Theta^{-1}=\rho_0$, i.e., if no excitation remains, which is the quantity
$\Delta E_{\bk}$ of Eq.~\eqref{eq:passive}. The mapping onto the same Hamiltonian needs
$\Theta H\Theta^{-1}=H$; otherwise the reversed evolution is governed by the time-reversed Hamiltonian
(e.g., with reversed magnetization), and $\sigma$ itself is not constrained.

Collisionless semiclassics satisfies the relation explicitly. With $f_b$ and $\varepsilon_b$ even and
$\Omega_b$ odd in $\bk$, the band current $-\sum_b\int[d\bk]\,f_b(\bk_0)\,\partial_{\bk}\varepsilon_b(\bk_0-A\be)$
is odd in $A$, and the Berry current of Eq.~\eqref{eq:sc}, $\mathbf j_B=E\,\hat{\mathbf e}_\perp\bar\Omega(A)$
with $\bar\Omega(A)=\sum_b\int[d\bk]\,f_b(\bk_0)\,\Omega_b(\bk_0-A\be)=-\bar\Omega(-A)$, obeys
$\mathbf j_{B,r}(t)=E(-t)\,\hat{\mathbf e}_\perp\bar\Omega(-A(-t))=-\mathbf j_B(-t)$; both therefore satisfy
$\mathbf j_r(t)=-\mathbf j(-t)$, and the Fermi-surface pole $\sigma_{-1}/\omega$ is consistent with the
parity.

Neither condition alone suffices. With time reversal but resonant driving, real excitation remains
and real parts appear: in the insulating Kane--Mele state the gap lies below $2\hbar\omega$ and the
coefficients acquire large real parts (Sec.~\ref{sec:kmSM}), as does the filled band alone at its
two-photon resonance in Table~\ref{tab:sw}. Without time reversal, even powers are allowed in the
collisionless limit, such as the nonlinear Drude term $\propto\omega^{-2}$ and the $\tau$-independent
intrinsic quantum-metric term $\propto\omega^0$~\cite{S-Kaplan2024}. In the lattice calculations the
second-order coefficients are imaginary with no resolved phase difference ($<10^{-9}$~rad,
Sec.~\ref{sec:res-weak}), and $s/\hbar\omega$ is constant at low frequency (Table~\ref{tab:sw}).

\section{Closed form of the in-phase correction for the linear Dirac model}\label{sec:closed}
\paragraph{Resolvent.} With $\hbar=1$, Eq.~\eqref{eq:bloch} reads $\dot{\mathbf S}=L\mathbf S$ with
$L=2[\mathbf d]_\times$. For a component $\mathbf F$ write $\mathbf F=\mathbf F_\parallel+\mathbf F_\perp$
with respect to $\hat{\mathbf d}$. Since $L\mathbf F_\parallel=0$ and $L\mathbf F_\perp=2|\mathbf d|\hat{\mathbf d}\times\mathbf F_\perp$
with $(\hat{\mathbf d}\times)^2=-1$ on the transverse plane,
$(-i\Omega-L)^{-1}\mathbf F=\mathbf F_\parallel/(-i\Omega)+[-i\Omega\mathbf F_\perp+2|\mathbf d|\hat{\mathbf d}\times\mathbf F]/(4|\mathbf d|^2-\Omega^2)$.

\paragraph{Iteration.} For $\mathbf d=(\tau vq_x,vq_y,m)$ and $\bk=\bk_0-A\hat x$,
$A=\mathrm{Re}\,A_0e^{-i\omega t}$, $\mathbf d(\bk)=\mathbf d(\bk_0)-A\mathbf u$ with
$\mathbf u=(\tau v,0,0)$ exactly, because $\mathbf d$ is linear. At first order
$(-i\omega-L)\mathbf s_1=-2A_0\mathbf u\times\mathbf S_0$; the second-harmonic part of the second order
obeys $(-2i\omega-L)\mathbf s_2=-A_0\mathbf u\times\mathbf s_1$. The current vertex
$\partial_{\bk}\mathbf d$ is constant, so $j_y=-g\,v\,\hat y\cdot\mathbf s_2$ without vertex
corrections. Symmetrizing in $q_y\to-q_y$ (odd terms integrate to zero) gives
\begin{equation}
j^H_y(\mathbf q)=\frac{j_y^\SC(\mathbf q)}{(1-\omega^2/4|\mathbf d|^2)(1-\omega^2/|\mathbf d|^2)},\qquad
j_y^\SC=g\,\frac{i\omega A_0^2}{2}\,\partial_{q_x}\Omega_c,
\end{equation}
with $\Omega_c=-\tau mv^2/(2|\mathbf d|^3)$; $j^\SC$ is the BCD integrand ($EA\to-i\omega A_0^2/2$ for
the second harmonic). We verified this identity in exact rational arithmetic at 60 points with
rational $|\mathbf d|$, both valleys and generic $\omega<|\mathbf d|$ (all 60 exact), and against a
brute-force numerical resolvent calculation of $K$, $s$, $c$ for the full continuum model, which
reproduces the closed form to all printed digits and gives $s=c=0$ to roundoff ($10^{-16}$).

\paragraph{Valence band.} With $f_v=1$ everywhere, each valley's valence contribution is a rank-three
tensor invariant under the continuous rotation symmetry of the linear valley (combined with a
pseudospin rotation about $z$), and such a tensor vanishes in two dimensions. Only $f_c$ enters, so the
symmetric part vanishes identically; the tilt enters only through $f_c$.

\paragraph{Values.} For $\alpha/v=0.2$, $m=0.2$, $\mu=1$, $k_BT=0.02$, both valleys (two independent
polar quadratures, identical to $10^{-15}$):
\begin{center}
\begin{tabular}{rrr}
\hline
$\eta$ & $\delta$ (closed form) & $\delta/\eta^2$\\
\hline
0.025 & $0.08091\%$ & $1.2945$\\
0.05 & $0.32432\%$ & $1.2973$\\
0.1 & $1.30842\%$ & $1.3084$\\
0.2 & $5.41993\%$ & $1.3550$\\
\hline
\end{tabular}
\end{center}
and $c_2=1.29361$, in agreement with the earlier direct second-order calculation
($c_2=1.2936\pm0.0001$, $\delta(0.1)=1.308\%$).

\paragraph{Relation to the general formula of He and Law.} The Supplemental Material of
Ref.~\cite{S-HeLaw2026} gives the second-harmonic Hall conductivity of a time-reversal-invariant system as
$\sigma^\perp_{qij}=\epsilon_{ljq}[\gamma^{(1)}_{il}+\gamma^{(2)}_{il}+\gamma^{(3)}_{il}+\gamma^{(4)}_{il}]+(i\leftrightarrow j)$,
their Eqs.~(23)--(31) in the arXiv version (arXiv:2411.07456). $\gamma^{(1)}$ is a Fermi-surface term with form factor
$g^{(1)}=\frac{e^3}{2\hbar}[\tau/\hbar(1-i\omega\tau)+i\hbar\bar\omega/(\hbar^2\bar\omega^2-\Delta E^2)-4i\bar\omega/(4\hbar^2\bar\omega^2-\Delta E^2)]$.
$\gamma^{(2)}$ and $\gamma^{(3)}$ are occupied-state terms with form factors of order $\bar\omega^3$, and
$\gamma^{(4)}$ requires three bands. We evaluated $\gamma^{(1)}+\gamma^{(2)}+\gamma^{(3)}$ for the
continuum model of Eq.~\eqref{eq:cont} in the clean limit ($\tau\to\infty$). The non-Abelian connections
were computed from analytic spinors by finite differences; the $\gamma^{(2)}$ integrand is gauge invariant
to $3\times10^{-8}$ under random smooth phase changes. The in-phase correction was formed relative to the
$i/\omega$ (BCD) part of $\gamma^{(1)}$, which makes it free of normalization conventions. On two
quadrature grids the result is
\begin{center}
\begin{tabular}{rrrr}
\hline
$\eta$ & $\gamma^{(1)}$ only & $\gamma^{(1)}+\gamma^{(2)}+\gamma^{(3)}$ & Eq.~\eqref{eq:closed}\\
\hline
0.025 & $0.08095\%$ & $0.08091\%$ & $0.08091\%$\\
0.05 & $0.32493\%$ & $0.32432\%$ & $0.32432\%$\\
0.1 & $1.31839\%$ & $1.30842\%$ & $1.30842\%$\\
0.2 & $5.59010\%$ & $5.41993\%$ & $5.41993\%$\\
\hline
\end{tabular}
\end{center}
The leading coefficient from $\gamma^{(1)}$ alone, extracted by Richardson extrapolation from
$\eta=0.0125$ and $0.025$, is $c_2=1.293605$, against $1.2936056$ from Eq.~\eqref{eq:closed}. The
occupied-state terms scale as $\eta^4$ and bring the full sum into exact agreement with
Eq.~\eqref{eq:closed}. Equation~\eqref{eq:closed} is therefore the explicit linear-Dirac form of
their general result. The correspondence holds after integration, not pointwise. For two bands their
$g^{(1)}$ factor relative to the BCD term is $(1-2x+4x^2)/[(1-x)(1-4x)]$, with $x=\omega^2/\Delta E^2$,
whereas Eq.~\eqref{eq:closed} carries $1/[(1-x)(1-4x)]$. The difference is compensated by the
integration by parts that relates their Fermi-surface form to our Fermi-sea form, together with the
occupied-state terms.

\paragraph{Lattice analogue of the resonance factor.} Weighting the same factor alone with the lattice
$(g+1)\,\partial_x\Omega_c$ and $|\mathbf d|$ (fine grid) gives, at $\eta=0.1$, $\delta=56.0\%$ (model I),
$12.5\%$ (model II), and $0.51\%$ (model III), i.e., ``$c_2$'' $=43.6$, $9.93$, $0.676$, against the exact
lattice values $0.760\%$, $1.187\%$, $1.333\%$. Table~\ref{tab:lobes} lists the net dipole
$\langle\partial_x\Omega_c\rangle=\sum_b\int[d\bk](g+1)\partial_x\Omega_c$ (per spin) and its positive and
negative Fermi-surface lobes. In every model the dipole is a small remainder of two nearly cancelling
lobes: $2.2\times10^{-4}$ of a lobe in model I (warping), $1.1\times10^{-3}$ in model II, and
$3.9\times10^{-2}$ in model III (tilt). A $\bk$-dependent weight such as $|\mathbf d|^{-2}$ therefore
changes the ratio greatly, most strongly in model I. The exact correction remains small because the
vertex terms generated by nonlinear $\mathbf d(\bk)$ follow the same cancellation. The resonance factor
alone fails for all three lattice models, most strongly for model I.

\begin{table}[!htbp]
\caption{Net Berry-curvature dipole and its Fermi-surface lobes ($\varepsilon=1.3\%$ for I and II;
per spin, $\int[d\bk]$; fine grid).}
\label{tab:lobes}
\begin{ruledtabular}
\begin{tabular}{lrrrr}
model & $\langle\partial_x\Omega_c\rangle$ & positive lobe & negative lobe & net/positive\\
\hline
I & $7.917\times10^{-4}$ & $3.5909$ & $-3.5901$ & $2.2\times10^{-4}$\\
II & $3.878\times10^{-3}$ & $3.5888$ & $-3.5849$ & $1.1\times10^{-3}$\\
III & $-0.1396$ & $3.6061$ & $-3.7457$ & $-3.9\times10^{-2}$\\
\end{tabular}
\end{ruledtabular}
\end{table}

\section{Adiabatic expansion of the Bloch equation}\label{sec:adiabatic}
This section derives the semiclassical current, Eq.~\eqref{eq:sc}, from Eqs.~\eqref{eq:bloch} and
\eqref{eq:current}, and continues the expansion to the order at which the quantum corrections of the
main text appear. We write $\partial_e\equiv\be\cdot\nabla_{\bk}$, $\hat{\mathbf e}_\perp=\hat z\times\be$, and
keep $\hbar$ explicit; the drive enters only through $\bk(t)=\bk_0-A(t)\be$, so a time derivative of a
band quantity is $-E\,\partial_e$ of it.
Powers of $\hbar$ are those of the dimensional Bloch equation with $e=1$; since each adiabatic order
carries one power of $\hbar/|\mathbf d|$ per time derivative, they also serve as order markers. All
numerical evaluations use $\hbar=1$, as in the main text.

\paragraph{Exact recursion.} In time $t$, Eq.~\eqref{eq:bloch} reads
$\dot{\mathbf S}=(2/\hbar)\,\mathbf d\times\mathbf S$. Taking $\hat{\mathbf d}\times$ of both sides gives the
exact relations
\begin{equation}
\mathbf S_\perp=-\frac{\hbar}{2|\mathbf d|}\,\hat{\mathbf d}\times\dot{\mathbf S},\qquad
S_\parallel=\big(1-\mathbf S_\perp^2\big)^{1/2},
\label{eq:recursion}
\end{equation}
with $\perp$ and $\parallel$ taken relative to $\hat{\mathbf d}(\bk(t))$. Iterating from
$\mathbf S_0=\hat{\mathbf d}$ produces $\mathbf S=\sum_k\mathbf S_k$, where $\mathbf S_k$ carries $k$ factors of
$\hbar/2|\mathbf d|$ and $k$ time derivatives. The expansion parameters are $\hbar\omega/|\mathbf d|$ and
$\hbar|\partial_t\hat{\mathbf d}|/|\mathbf d|\sim\hbar v\,eE/|\mathbf d|^2$. For the pulse of Eq.~\eqref{eq:pulse},
$E(0)=\dot E(0)=0$ but $\ddot E(0)=F_0E_0\omega^2/(2C^2)\neq0$. The exact initial condition
$\mathbf S(0)=\hat{\mathbf d}(\bk_0)$ therefore coincides with the adiabatic series through second order and
differs from it at third order by a term $\propto\ddot E(0)$. This mismatch is perturbative. It is
removed by a homogeneous solution of Eq.~\eqref{eq:bloch}, a free precession at the local interband
frequency $2|\mathbf d|/\hbar$, which after one further interaction with the drive contributes at
$2|\mathbf d|/\hbar\pm\omega$. For every state with $g\neq0$ these frequencies lie far above the harmonics
analysed ($|\mathbf d|\gtrsim E_{\rm ref}\gg\hbar\omega$ in the doped system). Their contribution to the
Hann-windowed harmonics is further suppressed by $1/C^2$ and by the window. The exact Taylor kernel,
which imposes $\mathbf S(0)=\hat{\mathbf d}$, agrees with the adiabatic coefficients to the precision stated
in Table~\ref{tab:adiabatic}, which bounds this effect for the quantities reported.

\paragraph{Orders zero and one: Eq.~\eqref{eq:sc}.} For $\mathbf S_0$,
$(\partial_{\bk}\mathbf d)\cdot\hat{\mathbf d}=\partial_{\bk}|\mathbf d|$, and with
$\varepsilon_{c,v}=d_0\pm|\mathbf d|$,
$n\,\partial d_0+g\,\partial|\mathbf d|=\sum_bf_b\,\partial_{\bk}\varepsilon_b$: the band current. The first
iterate is
\begin{equation}
\mathbf S_1=\frac{\hbar E}{2|\mathbf d|}\,\hat{\mathbf d}\times\partial_e\hat{\mathbf d}.
\end{equation}
With $\partial_i\mathbf d=(\partial_i|\mathbf d|)\hat{\mathbf d}+|\mathbf d|\partial_i\hat{\mathbf d}$ and
$\hat{\mathbf d}\cdot(\partial_e\hat{\mathbf d}\times\partial_i\hat{\mathbf d})=-2\Omega_c(\hat{\mathbf e}_\perp)_i$,
\begin{equation}
-g\,\partial_i\mathbf d\cdot\mathbf S_1=\hbar E\,g\,\Omega_c(\hat{\mathbf e}_\perp)_i
=\hbar E\,(\hat{\mathbf e}_\perp)_i\sum_bf_b\Omega_b ,
\end{equation}
the Berry current. Equation~\eqref{eq:sc} is therefore exactly the $O(\hbar^0)+O(\hbar^1)$ truncation of
Eqs.~\eqref{eq:bloch}--\eqref{eq:current}.

\paragraph{Order two: interband polarization.} With $\mathbf V\equiv\hat{\mathbf d}\times\partial_e\hat{\mathbf d}/(2|\mathbf d|)$
and $\mathbf W\equiv\partial_e\hat{\mathbf d}/(4|\mathbf d|^2)$,
\begin{equation}
\mathbf S_2=\hbar^2\dot E\,\mathbf W+\frac{\hbar^2E^2}{2|\mathbf d|}\,\hat{\mathbf d}\times\partial_e\mathbf V
-\frac{\hbar^2E^2}{2}\,|\mathbf V|^2\,\hat{\mathbf d}.
\end{equation}
The $\dot E$ term gives
$j_i=-\hbar^2\dot E\int[d\bk]\,g\,\mathcal G_{ie}/|\mathbf d|$, with the two-band quantum metric
$\mathcal G_{ij}=\frac14\partial_i\hat{\mathbf d}\cdot\partial_j\hat{\mathbf d}$. This is the interband
polarization current $\partial_t\mathbf P$, a symmetric rank-two (field-space gradient) response; it is
the anisotropic linear polarization of Sec.~\ref{sec:addproj}. The $E^2$ terms are the second-order Stark shift and
the field-induced curvature. At second order in the field all $\mathbf S_2$ contributions are field-odd
or vanish, because $f_b$ and $\mathcal G$ are even and $\Omega_b$ is odd under time reversal. For the
continuum model they vanish numerically to $2\times10^{-11}$ of the BCD term.

\paragraph{Order three: the quantum corrections.} Field-even corrections at second order first appear
in $\mathbf S_3$. Keeping terms of second order in the field (where $S_{3\parallel}=-\mathbf S_1\cdot\mathbf S_{2\perp}$
vanishes, since $\mathbf S_1\perp\mathbf W$), Eq.~\eqref{eq:current} gives, for any two-band $\mathbf d(\bk)$,
\begin{equation}
\begin{split}
\mathbf j^{(3)}={}&-\frac{\hbar^3}{4}\,\ddot E\,\hat{\mathbf e}_\perp\int[d\bk]\,g\,\frac{\Omega_c}{|\mathbf d|^2}\Big|_{\bk(t)}\\
&-\frac{\hbar^3}{2}\,E\dot E\int[d\bk]\,g\,\Big[\partial_i\hat{\mathbf d}\cdot(\hat{\mathbf d}\times\partial_e\mathbf W)
+\frac{\partial_i\hat{\mathbf d}\cdot\partial_e\mathbf V}{|\mathbf d|}\Big]_{\bk_0}\hat{\mathbf x}_i .
\end{split}
\label{eq:S3current}
\end{equation}
The $\ddot E$ term is purely Hall-type at every order in $A$. At second order it contributes
$+\frac{\hbar^3}{4}\ddot E A\,\hat{\mathbf e}_\perp\int g\,\partial_e(\Omega_c/|\mathbf d|^2)$. The $E\dot E$ term contains
both the symmetric and the Hall-type sectors. For a monochromatic drive, $\ddot EA$ and $E\dot E$ have the
same second harmonic, $-\omega^2$ times that of $EA$. So $\mathbf j^{(3)}$ is of relative order $\eta^2$
with respect to the BCD current $-EA\,\hat{\mathbf e}_\perp\mathcal D^{(1)}_e$, for both $s$ and $\delta K$,
as anticipated by Eq.~\eqref{eq:lowfreq}.

\paragraph{Affine models: a pointwise identity.} Let $\mathbf d(\bk)=\mathbf d_*+k_x\mathbf w_x+k_y\mathbf w_y$
be affine, with $R=|\mathbf d|$, constant $\mathbf u=\partial_e\mathbf d$ and $\mathbf w_i=\partial_i\mathbf d$.
Then $\mathbf V=\mathbf d\times\mathbf u/(2R^3)$. With $a=\mathbf d\cdot\mathbf u$ (so $\partial_eR=a/R$),
\begin{equation}
\partial_e\mathbf W=\frac{-4a\,\mathbf u-u^2\mathbf d+5a^2\mathbf d/R^2}{4R^5},\qquad
\partial_e\mathbf V=-\frac{3a\,\mathbf d\times\mathbf u}{2R^5},\qquad
\partial_i\hat{\mathbf d}=\frac{\mathbf w_i}{R}-\frac{(\mathbf d\cdot\mathbf w_i)\mathbf d}{R^3}.
\end{equation}
Terms along $\mathbf d$ drop out of the triple and dot products. With
$T_i\equiv\mathbf d\cdot(\mathbf u\times\mathbf w_i)$,
\begin{equation}
\partial_i\hat{\mathbf d}\cdot(\hat{\mathbf d}\times\partial_e\mathbf W)+\frac1R\,\partial_i\hat{\mathbf d}\cdot\partial_e\mathbf V
=-\frac{a\,T_i}{R^7}-\frac{3a\,T_i}{2R^7}=-\frac{5a\,T_i}{2R^7}.
\end{equation}
Define $\Omega_{ei}=-T_i/(2R^3)$; in two dimensions $\Omega_{ei}=(\hat{\mathbf e}_\perp)_i\Omega_c$. Because
$\partial_eT_i=\mathbf u\cdot(\mathbf u\times\mathbf w_i)=0$ for affine $\mathbf d$,
$\partial_e(\Omega_{ei}/R^2)=5aT_i/(2R^7)$. Hence, pointwise in $\bk$,
\begin{equation}
\partial_i\hat{\mathbf d}\cdot(\hat{\mathbf d}\times\partial_e\mathbf W)+\frac1R\,\partial_i\hat{\mathbf d}\cdot\partial_e\mathbf V
=-(\hat{\mathbf e}_\perp)_i\,\partial_e\!\Big(\frac{\Omega_c}{R^2}\Big).
\label{eq:pointwise}
\end{equation}
A closed-form numerical check at $10^5$ random points (generic $\mathbf d_*$, non-orthogonal
$\mathbf w_x,\mathbf w_y$, random $\be$, both components) gives a worst relative deviation of
$1.1\times10^{-10}$ (roundoff). By Eq.~\eqref{eq:pointwise} the $E\dot E$ term of Eq.~\eqref{eq:S3current}
is purely transverse before momentum integration, and it equals twice the second-order part of the
$\ddot E$ term.

\paragraph{Affine models: exact statement.} The identity above is the low-frequency limit of an exact
result at second order in the field. For affine $\mathbf d(\bk)=\mathbf d_*+k_x\mathbf w_x+k_y\mathbf w_y$
the second-order equations are solved exactly by the resolvent of Sec.~\ref{sec:closed}. Let $\hbar=1$,
$A=A_0\cos\omega t$ with $A_0=1$, and let $\mathbf S_2$ be the $e^{-2i\omega t}$ Fourier component of
$\mathbf S$. With $\mathbf u=\partial_e\mathbf d$, $\hat{\mathbf n}=\hat{\mathbf d}$, $R=|\mathbf d|$,
$u_n=\hat{\mathbf n}\cdot\mathbf u$, $\mathbf c=\mathbf u\times\hat{\mathbf n}$,
$\mathbf c_2=\mathbf u-\hat{\mathbf n}u_n$, $\Delta_1=4R^2-\omega^2$ and $\Delta_2=4R^2-4\omega^2$,
\begin{equation}
\mathbf S_2=\hat{\mathbf n}\,\frac{u_n^2-u^2}{2\Delta_1}+u_n\,\frac{-(4R^2+2\omega^2)\,\mathbf c_2+6i\omega R\,\mathbf c}{\Delta_1\Delta_2},
\label{eq:S2affine}
\end{equation}
with $j_i=-g\,\mathbf w_i\cdot\mathbf S_2$. Time reversal, $H(-\bk)=H(\bk)^*$, i.e.,
$\mathbf d(-\bk)=\mathcal R\,\mathbf d(\bk)$ with $\mathcal R=\mathrm{diag}(1,-1,1)$ and $\det\mathcal R=-1$,
maps the state at $\mathbf q$ to its partner at $-\mathbf q$ (in the same patch, or in the time-reversed
valley) with $\hat{\mathbf n}\to\mathcal R\hat{\mathbf n}$, $\mathbf u\to-\mathcal R\mathbf u$ and
$\mathbf w_i\to-\mathcal R\mathbf w_i$. Hence $u_n\to-u_n$, $\mathbf c\to\mathcal R\mathbf c$ and
$\mathbf c_2\to-\mathcal R\mathbf c_2$. For time-reversal-symmetric occupations the pair carries equal
weight $g$, the $\hat{\mathbf n}$ and $\mathbf c_2$ terms cancel, and the paired current is
\begin{equation}
\mathbf j_i^{\rm pair}=-\frac{12\,i\omega R\,g\,u_n\,(\mathbf w_i\cdot\mathbf c)}{\Delta_1\Delta_2}.
\label{eq:jpair}
\end{equation}
For every drive direction $\sum_ie_i\,\mathbf j_i^{\rm pair}\propto\mathbf u\cdot(\mathbf u\times\hat{\mathbf n})=0$.
The paired current thus has no component along the field for any $\be$, so the fully symmetric part
of its rank-three tensor vanishes. The second-order response of any affine two-band model with
time-reversal-symmetric occupations is therefore Hall-type pointwise in momentum, at every frequency
below the resonances, $\hbar\omega<|\mathbf d|$, so $s=0$ exactly at second order in the field. The
pairing is pointwise, so the result does not depend on how the filled band is regularized. With
$\Omega_{ei}=-\mathbf d\cdot(\mathbf u\times\mathbf w_i)/(2R^3)$, Eq.~\eqref{eq:jpair} equals the paired BCD
integrand, $-2[EA]_2\,g\,\partial_e\Omega_{ei}$ with $[EA]_2=-i\omega/4$, times the resonance factor
$[(1-\omega^2/4R^2)(1-\omega^2/R^2)]^{-1}$ of Sec.~\ref{sec:closed}, now for any affine $\mathbf d$; its
$O(\omega^2)$ relative term is the Hall current of Eq.~\eqref{eq:Dirac3}, consistent with
Eq.~\eqref{eq:pointwise}. No statement is made beyond second order in the field. At $5000$ random
(affine $\mathbf d$, $\mathbf q$, $\be$, $\omega\le0.95R$) Eq.~\eqref{eq:S2affine} agrees with the numerical
resolvent to $4\times10^{-14}$, and Eq.~\eqref{eq:jpair}, the vanishing longitudinal component and the
resonance-factor form hold to $2\times10^{-9}$. The worst $|\mathrm{Sym}|/|\chi|$ of the paired per-$\mathbf q$
tensor at $2000$ further random points is below $10^{-9}$. Integrated full-tensor tests (all six
components, no mirror symmetry assumed), with isotropic and random affine $\mathbf d$ and thermal and
non-thermal time-reversal-paired occupations, at $\omega=0.02$ and $0.1$, give
$|\mathrm{Sym}|/|\mathrm{Hall}|$ below $2\times10^{-10}$. On the lattice, the
symmetric coefficient therefore arises entirely from the nonlinear terms of $\mathbf d(\bk)$, such as
trigonal warping. This is consistent with its near
independence of strain and model (Table~\ref{tab:weak}).

\paragraph{Linear Dirac model.} For affine $\mathbf d$ and time-reversal-symmetric occupations, with an
arbitrary tilt $d_0$ (which enters only through $f_b$) and arbitrary temperature, the second-order Hall
current takes the time-domain form
\begin{equation}
\mathbf j_H=\hat{\mathbf e}_\perp\Big[-EA\,\mathcal D^{(1)}_e+\frac{\hbar^2}{4}\,(\ddot EA+2E\dot E)\,
\tilde{\mathcal D}^{(1)}_e\Big],\qquad
\tilde{\mathcal D}^{(1)}_e=\sum_b\int[d\bk]\,f_b\,\partial_e\!\Big(\frac{\Omega_b}{|\mathbf d|^2}\Big),
\label{eq:Dirac3}
\end{equation}
valid for any adiabatically varying pulse. For a constant-amplitude drive $A=A_0\cos\omega t$,
$\ddot EA=E\dot E=-\omega^2EA$ pointwise in time; with a pulse envelope this holds for the second
harmonic up to envelope corrections. Since $EA=\partial_t(A^2/2)$ and
$\ddot EA+2E\dot E=\partial_t(\dot EA+E^2/2)$, both terms are time derivatives of a transverse
polarization. The correction is a reactive Hall polarization current, not a modification of the
anomalous velocity, and it transfers no net transverse charge over a cyclic pulse. For a monochromatic
drive, $\delta=\frac34\,\hbar^2\omega^2\tilde{\mathcal D}^{(1)}_x/\mathcal D^{(1)}_x$. Because
$\Omega_c\propto|\mathbf d|^{-3}$, $\partial_x(\Omega_c/|\mathbf d|^2)=\frac53|\mathbf d|^{-2}\partial_x\Omega_c$,
and this is $c_2$ of Eq.~\eqref{eq:closed}. Numerically the third-order term gives $c_2=1.29360557$ with
nested finite differences and $1.2936055591$ with analytic derivatives, against $1.2936055591$ from
Eq.~\eqref{eq:closed}.

\paragraph{Lattice.} Equation~\eqref{eq:S3current} does not assume a linear $\mathbf d(\bk)$, so it
applies to the lattice models. We evaluated it on the fine full-zone polar grid of the Taylor-kernel
calculation (Table~\ref{tab:weak}), with analytic lattice derivatives (bond sums for $\mathbf d$ and its first
and second derivatives, and the chain rule for $\hat{\mathbf d}$), for models I and II at $\varepsilon=1.3\%$,
model III, and model I at $\varepsilon=0$. The analytic integrands agree with Richardson-extrapolated
central differences to $4\times10^{-8}$, and the base grid changes the results by at most
$2\times10^{-7}$ relative. On the lattice the $x$-drive transverse coefficient contains $s$, so we apply
the angular projection, Eq.~\eqref{eq:H2}, to the third-order coefficients ($b$ from an $x$ drive, $a$ from
a $\pi/4$ drive). The Hann coefficients use the second harmonics of $\ddot EA$ and $E\dot E$ for the actual
pulse, which differ from $-\omega^2$ times that of $EA$ by $+0.2\%$ and $-0.1\%$. The exact values come from the
Taylor kernel at $\hbar\omega=0.0025$, $0.005$, $0.01$ and $0.02$ on the same grid. Table~\ref{tab:adiabatic}
lists the results. The unprojected $x$-drive value would give
``$c_2$''$=5.93$ for model I, dominated by $s$ relative to the nearly cancelling dipole.

\begin{table}[!htbp]
\caption{Third-order adiabatic coefficients [Eq.~\eqref{eq:S3current}, angular projection] against the exact second-order Taylor kernel of Table~\ref{tab:weak} (same fine momentum grid). Exact $c_2$: fit $\delta=c_2\eta^2+c_4\eta^4$ to $\hbar\omega=0.0025$, $0.005$, $0.01$; the number in parentheses is the uncertainty in the last digits (spread over fit forms and momentum grids). Exact $\lim s/\hbar\omega$: extrapolation $s/\hbar\omega=a_0+a_2\eta^2+a_4\eta^4$ through the same three frequencies. Adiabatic Hann coefficients use the second harmonics of $\ddot EA$ and $E\dot E$ for the actual pulse; the same procedure reproduces $K_\SC$ of Table~\ref{tab:weak} to $10^{-8}$ relative. Coefficients $\times\Ncal/E_0^2$ per spin.}\label{tab:adiabatic}
\begin{ruledtabular}
\begin{tabular}{lll}
quantity & third-order adiabatic & exact\\
\hline
$c_2$, continuum & $1.29360557$ & $1.29360556$ [Eq.~\eqref{eq:closed}]\\
$c_2$, model I, $\varepsilon=1.3\%$ & $0.7548$ & $0.7548\,(1)$\\
$c_2$, model II, $\varepsilon=1.3\%$ & $1.17524$ & $1.17521\,(3)$\\
$c_2$, model III & $1.31773$ & $1.31769\,(4)$\\
$s$, model I, $\varepsilon=1.3\%$, $\eta=0.1$ & $-3.798i$ & $-3.827i$\\
$s$, model I, $\varepsilon=0$, $\eta=0.1$ & $-3.735i$ & $-3.764i$\\
$s$, model II, $\varepsilon=1.3\%$, $\eta=0.1$ & $-3.890i$ & $-3.920i$\\
$s$, model III, $\eta=0.1$ & $-4.274i$ & $-4.311i$\\
$\lim s/\hbar\omega$, model I, $\varepsilon=1.3\%$ & $-379.76i$ & $-379.76i$ ($-379.94i$ at $\hbar\omega=0.0025$)\\
$\lim s/\hbar\omega$, model I, $\varepsilon=0$ & $-373.51i$ & $-373.51i$ ($-373.68i$ at $\hbar\omega=0.0025$)\\
$\lim s/\hbar\omega$, model II, $\varepsilon=1.3\%$ & $-389.00i$ & $-389.00i$ ($-389.19i$ at $\hbar\omega=0.0025$)\\
$\lim s/\hbar\omega$, model III & $-427.40i$ & $-427.40i$ ($-427.64i$ at $\hbar\omega=0.0025$)\\
\end{tabular}
\end{ruledtabular}
\end{table}

For all three lattice models the adiabatic $c_2$ agrees with the exact value within the fit uncertainty
(differences $\le4.2\times10^{-5}$). The exact $s/\hbar\omega$, extrapolated to $\omega\to0$, equals the adiabatic value to
$1.2\times10^{-8}$ relative. At finite frequency the exact $s$ exceeds it by $0.76\%$ at $\eta=0.1$ and
$3.1\%$ at $\eta=0.2$ (model I, $\varepsilon=0$; Table~\ref{tab:sw}), the $O(\eta^2)$ relative correction of
the next adiabatic order. Both quantum corrections of the lattice second-order response are therefore
given quantitatively by the third adiabatic order of Eq.~\eqref{eq:bloch}. The failure of the resonance
factor of Eq.~\eqref{eq:closed} on the lattice (Sec.~\ref{sec:closed}) reflects the terms of
Eq.~\eqref{eq:S3current} beyond the linear-Dirac reduction, not a breakdown of the expansion.

\paragraph{Breakdown.} The local expansion parameter is $\hbar v\,eE/|\mathbf d(\bk(t))|^2$. With
$E_0v_0=\hbar\omega E_{\rm ref}$, at the Fermi surface ($|\mathbf d|\approx E_{\rm ref}$) it is about
$F_0\eta$, i.e., $0.15$ at $F_0=1.5$. For states dragged to the node ($|\mathbf d|=M$) it becomes
$\hbar v\,eE/M^2=1/\delta_{\rm LZ}$ (Sec.~\ref{sec:strong}), which is $3.8$ at $F_0=1.5$. The
adiabatic series, and with it Eq.~\eqref{eq:sc}, fails when occupied-boundary states reach the node,
not when the field is merely large.

\section{Tensor projection at arbitrary order and rank-five symmetrization}\label{sec:orders}
\paragraph{Decomposition.} Let $\chi^{(p)}_{i\,j_1\dots j_p}$ be symmetric in its $p$ field indices. In
2D, $V\otimes\mathrm{Sym}^p=\mathrm{Sym}^{p+1}\oplus\mathrm{Sym}^{p-1}$. The remainder
$\chi-\mathrm{Sym}(\chi)$ is of Hall type, $\epsilon_{ik}E_k\,Q_{p-1}(\mathbf E)$. The component
counts check this: with $M_x$ at $p=4$ there are five components, three fully symmetric and two
Hall-type. For the symmetric part, $\jpar=(p+1)\Phi_{p+1}/|\mathbf E|$ and
$\jperp=\partial_\theta\Phi_{p+1}/|\mathbf E|$, which gives Eq.~\eqref{eq:Hp} of the main text.

\paragraph{Amplitude fits.} Sixteen directions $\theta_k=k\pi/8$, including the reversed fields, are
used, with ten amplitudes $F_0\in\{0.0025,0.005,0.0075,0.01,0.015,0.02,0.03,0.05,0.1,0.3\}$ and
$N=19200$. Since $\jpar^{(e)}(\theta+\pi)=-\jpar^{(e)}(\theta)$, sixteen samples resolve angular
harmonics up to seven. For each harmonic $n\in\{2,4\}$ and each angle we fit
$J/F_0^2=\sum_{q=0}^{d}c_{2q+2}(F_0/0.01)^{2q}$ with $d\in\{1,2,3\}$, using the smallest $4$--$7$
amplitudes (eleven combinations of window and degree). Angular derivatives are spectral on the
antiperiodic extension, applied to the complex coefficients. These finite-field coefficients are
numerical estimates, distinct from the exact Taylor coefficients of Table~\ref{tab:weak}.

\paragraph{Evidence for the fourth-order coefficients.} (i) Equation~\eqref{eq:Hp} is an algebraic
identity once the order-four coefficient is known. (ii) An independent fit of the five mirror-allowed
components of the rank-five tensor (residual $2$--$4\times10^{-4}$ of the scale), with its weight
distributed over the $\binom4a$ index orderings and averaged over all $5!$ permutations, reproduces
$H^{(4)}$ to $1.0$--$1.2\%$ of its scale; the remainder has no longitudinal part
($\le2\times10^{-17}$), and $H^{(4)}$ contains $\le0.18\%$ of $\cos5\theta$. This agreement is at the
level of the fit residual and checks consistency rather than the identity. (iii) The spread of
$H^{(4)}$ over fit windows and degrees is below $3\%$ of its value in six of eight cases and
$3.4$--$3.5\%$ in the other two. (iv) Momentum-grid refinement changes the projected second-order
coefficient at $F_0=0.01$ by $4\times10^{-5}$ (absolute, $5\times10^{-7}$ relative); it was not repeated
for the full fourth-order fit. (v) Time steps: Table~\ref{tab:time4}.

\paragraph{Fourth-order time-step convergence.} The deviation $|H^{(4)}_\Q-j_B^{(4)}|/|j_B^{(4)}|$ at
$\theta=0$ (model I) was obtained from a two-point fit ($F_0=0.005,0.01$) at $N=19200$, $38400$ and
$76800$, with $j_B^{(4)}$ the direct octupole value (Sec.~\ref{sec:multipoleSM}).
\begin{table}[!htbp]
\caption{Fourth-order time-step check (two-point fit at $F_0=0.005,0.01$, $\theta=0$, model I; deviations
$|H^{(4)}_\Q-j_B^{(4)}|/|j_B^{(4)}|$ relative to the direct octupole value; the imaginary parts alone give the same values to
$0.01$ percentage points).}
\label{tab:time4}
\begin{ruledtabular}
\begin{tabular}{lcccc}
$N$ & $H^{(2)}_{\rm Q}$ ($n=2$) & $H^{(4)}_{\rm Q}$ ($n=2$) & dev.\ $n=2$ & dev.\ $n=4$\\
\hline
19200 & $-73.777i$ & $0.057165i$ & $2.73\%$ & $7.36\%$\\
38400 & $-73.856i$ & $0.057516i$ & $3.36\%$ & $6.82\%$\\
76800 & $-73.876i$ & $0.057394i$ & $3.14\%$ & $6.51\%$\\
Richardson ($\Delta t^2$, last two) & & $0.057354i$ & $3.06\%$ & $6.40\%$\\
Richardson (observed order $0.75$) & & & -- & $6.04\%$\\
\end{tabular}
\end{ruledtabular}
\end{table}
The fourth-harmonic deviation decreases monotonically (successive changes $-0.53$ and $-0.32$
percentage points). The observed order computed from the deviations themselves is $0.75$ ($0.74$ from the
imaginary parts of $H^{(4)}_\Q$), well below the formal order two of the midpoint-rotation integrator, so the sequence
is not in the asymptotic regime. (Differences of the complex coefficients would suggest an order of $1.3$, but their real parts
are fit residuals of order $10^{-4}$ that change sign between runs.) Richardson extrapolation gives $6.40\%$
with order two and $6.04\%$ with the observed order; with the finest-step value $6.51\%$ these span $6.0$--$6.5\%$, and
we quote $6.3\pm0.3\%$. The second-harmonic deviation is not monotonic in $N$
($2.73$, $3.36$, $3.14\%$), so Richardson extrapolation is not justified; we quote $3.1\pm0.3\%$,
covering the two finer values and the formally extrapolated $3.06\%$. The semiclassical fit reproduces
the direct octupole to $\le2\times10^{-4}$ at all $N$. The projected second-order coefficient approaches
the Taylor value $-73.882i$ ($-73.876i$ at $N=76800$).

\paragraph{Second-order time offset.} At $F_0=0.01$ the projected second-order coefficient is
$-73.7198i$ ($N=19200$), $-73.7984i$ ($N=38400$), and $-73.7199i$ (fine grid). Richardson
extrapolation in the time step gives $-73.825i$; subtracting the fourth-order term at $F_0=0.01$
($+0.057i$) gives $-73.882i$, the exact Taylor value. Finite-time dynamics at $N=19200$ therefore
underestimate $K_\Q-K_\SC$ by about $0.10$ (about $20\%$ of the correction), and the Taylor values are
the reference. The Lagrangian semiclassics have no time-step error. As an example of the order
budget, the residual of the single-$\frac13$ angular projection in the second harmonic at $F_0=0.1$, $\theta=0$
(model I), splits into the second-order quantum correction to $K$ ($0.45$ at $N=19200$), the
fourth-order Hall-type deviation ($0.17$) and the fourth-order symmetric misassignment ($0.02$).

\paragraph{Range of validity and the semiclassical series.} Table~\ref{tab:series} compares the
truncated quantum and semiclassical series at $\theta=0$, using identical fit windows and degrees
(preferred fit: six smallest amplitudes, degree 2). The two-term quantum series reproduces the direct
second harmonic to $0.02\%$ ($F_0=0.05$) and $0.11\%$ ($F_0=0.1$) in model I. At $F_0=0.3$ the two-term
truncation fails equally for both. The three-term semiclassical series reduces the error, consistent
with convergence of the semiclassical expansion without determining its radius; the three-term quantum
series does not. This is consistent with the Lagrangian form of Eq.~\eqref{eq:multipole}: the
occupations stay at $\bk_0$ and only the smooth curvature is expanded, so the radius in $A$ is expected
to be set by the distance from the occupied region to the Dirac point rather than by $k_BT$. The fitted sixth-order quantum coefficient ($n=2$, model I) is
$1.6\times10^{-4}i$, $4.2\times10^{-5}i$, $1.0\times10^{-5}i$ and $7.9\times10^{-5}i$ for the windows
5/6/7 points (degree 2) and 7 points (degree 3); the semiclassical one is $4.26$--$4.36\times10^{-6}i$
for the same windows. The quantum sixth-order coefficient is therefore not resolved by these
fits, and we draw no conclusion about the convergence radius of the quantum series. For the small fourth harmonic at
$F_0\le0.005$ the series test is at the noise level. At $F_0=0.3$ the RMS deviations of the field-even
transverse current from the Berry current (model I) are $0.5$--$2.3\%$ without projection,
$0.3$--$3.8\%$ with the single-$\frac13$ angular projection, and $1.7$--$9\%$ with the work projection; model II,
where $|K|\gg|s|$, is brought to $0.32\%$.

\begin{table}[!htbp]
\caption{Relative error of the truncated amplitude series at $F_0=0.3$ ($\theta=0$, $j_\perp$, preferred
fit; in parentheses the widest window, 7 points, degree 2).}
\label{tab:series}
\begin{ruledtabular}
\begin{tabular}{llrr}
model, $n$ & & two-term & three-term\\
\hline
I, 2 & quantum & $19.5\%$ & $382\%$ ($57\%$)\\
I, 2 & semiclassical & $19.3\%$ & $5.0\%$ ($1.5\%$)\\
I, 4 & quantum & $11.8\%$ & $84\%$ ($15\%$)\\
I, 4 & semiclassical & $10.9\%$ & $1.8\%$ ($0.9\%$)\\
II, 2 & quantum & $0.96\%$ & $26\%$ ($3.9\%$)\\
II, 2 & semiclassical & $0.85\%$ & $0.56\%$ ($0.11\%$)\\
II, 4 & quantum & $12.7\%$ & $127\%$ ($22\%$)\\
II, 4 & semiclassical & $11.8\%$ & $11\%$ ($1.8\%$)\\
\end{tabular}
\end{ruledtabular}
\end{table}

\section{Angular derivatives}
\paragraph{Complex step.} The drive angle enters only through $\be(\theta)$ in the bond phases. Every
operation of the propagator is holomorphic in $\theta$: trigonometric functions, sums, products,
$\sqrt{\mathbf d\cdot\mathbf d}$ with $\mathrm{Re}(\mathbf d\cdot\mathbf d)\ge M^2>0$, and the rotation
map. No absolute value, complex conjugate, or comparison is applied to a complex quantity. With
$h=10^{-20}$, $\mathrm{Im}\,F(\theta+ih)/h=F'(\theta)+O(h^2)$ without cancellation~\cite{S-Squire1998,S-Martins2003},
where $F$ is the \emph{discretized} evolution. The derivative is taken on the real time-domain current
before the (linear) Fourier transform, so complex harmonic coefficients are unaffected.

\begin{figure}[!htbp]
\includegraphics[width=0.85\textwidth]{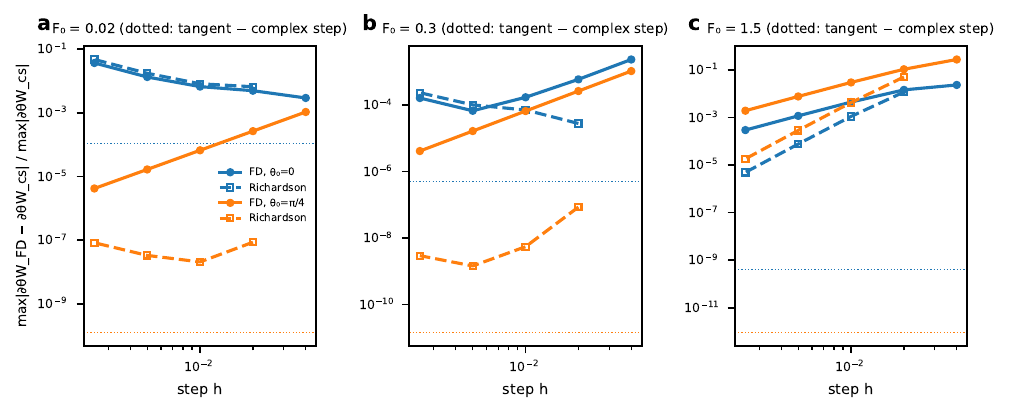}
\caption{Finite-difference (circles) and Richardson (squares) errors of $\partial_\theta W$ relative
to the complex step, versus the step $h$, for $F_0=0.02$, $0.3$, $1.5$ and $\theta_0=0$, $\pi/4$.
Dotted lines show the difference between the tangent propagation and the complex step.}
\label{fig:ang}
\end{figure}

\paragraph{Tangent propagation.} Independently, $\mathbf T=\partial_\theta\mathbf S$ is propagated
through the analytic derivative of the rotation map,
\begin{equation}
\mathbf T'=R\mathbf T+\delta\phi[-\sin\phi\,\mathbf S+\cos\phi\,\hat{\mathbf m}\times\mathbf S+\sin\phi(\hat{\mathbf m}\cdot\mathbf S)\hat{\mathbf m}]
+\sin\phi\,\delta\hat{\mathbf m}\times\mathbf S+(1-\cos\phi)[(\delta\hat{\mathbf m}\cdot\mathbf S)\hat{\mathbf m}+(\hat{\mathbf m}\cdot\mathbf S)\delta\hat{\mathbf m}],
\end{equation}
with $\delta\mathbf m=-A_{\rm mid}(\hat{\mathbf e}_\perp\cdot\nabla)\mathbf d$,
$\delta\hat{\mathbf m}=[\delta\mathbf m-\hat{\mathbf m}(\hat{\mathbf m}\cdot\delta\mathbf m)]/|\mathbf m|$, and
$\delta\phi=2(\hat{\mathbf m}\cdot\delta\mathbf m)\Delta u/\omega$. The two methods agree to
$10^{-13}$--$10^{-10}$. At the mirror angle $\theta_0=0$ the agreement is $5\times10^{-7}$
($F_0=0.3$) and $1.1\times10^{-4}$ ($F_0=0.02$), where $\partial_\theta W$ is a field-odd remnant
$10^3$--$10^5$ times smaller than its constituents. Central finite differences converge as $h^2$,
with Richardson extrapolation reaching $10^{-8}$--$2\times10^{-5}$. At $\theta_0=0$ and weak field
they are roundoff-limited, with an error growing as $1/h$ (Fig.~\ref{fig:ang}).

\section{Quadrature and convergence}
\paragraph{Grids.} Valley-centered polar grids run to the exact rectangle boundary. The base grid has
$n_\theta=128$ and 10-point Gauss panels of width $0.03$ in $Q=kv/E_{\rm ref}$ ($3.4\times10^5$
nodes). The fine grid has $n_\theta=256$, 14-point panels, and width $0.015$ ($1.7\times10^6$ nodes).
The radial-only and angular-only refinements change one factor at a time. The extra-fine grid has
$n_\theta=384$, 18-point panels, and width $0.01$. Panel edges are placed at the Fermi radius
$\pm(1,3,6,12)k_BT/v$. The weights sum to $1/A_{\rm cell}$ to $\le1.3\times10^{-6}$.

\paragraph{Time steps.} $N=19200$ is the baseline, refined to 38400 and 76800; $N\le9600$ is
unusable for endpoint observables. Pointwise error budgets sum the time and grid changes and the
angular-derivative difference. The weak-field Taylor coefficients use analytic time integration and
a momentum envelope summed vertex by vertex.

\paragraph{Strain scan at $F_0=1$.} The harmonics of Fig.~\ref{fig:corr}(c) were computed with a
separate full-zone momentum mesh (rectangular reciprocal cell) at fixed field, frequency and
temperature. Independent simultaneous radial, angular and outer-region refinements together with
time refinement at $\varepsilon=0.5\%$, momentum refinement at the $1.3\%$ endpoint, and separate
momentum and time checks at $\varepsilon=0$ change the complex second and fourth harmonics by at most
$0.003\%$ (absolute $\le6.5\times10^{-9}$ in common units); the $1.0\%$ point uses the same mesh with
a bracketing error envelope.

\section{Work, energy and passivity}\label{sec:energy}
For unitary dynamics, $dE_{\rm el}/dt=\mathrm{Tr}(\rho\,\partial_tH)=\mathbf j\cdot\mathbf E$, so
$W(t)=\int_0^t\mathbf j\cdot\mathbf E\,dt'=E_{\rm el}(t)-E_{\rm el}(0)$. The pulse is cyclic: $A(U)=0$
analytically and $3.6\times10^{-18}$ numerically. For each $\bk_0$, $g\le0$, and
$\Delta E_{\bk}=gR(\mathbf S\cdot\hat{\mathbf d}-1)$ splits into a manifestly non-negative part and a
norm-drift part [Eq.~\eqref{eq:passive} of the main text]. Table~\ref{tab:energy} lists the endpoint diagnostics.
The summed $E_{\rm el}(U)-E_{\rm el}(0)$ equals the manifest absorption plus the norm term to
$<10^{-12}$ in all rows. Its sign at $F_0\le0.3$ follows the roundoff-dominated norm drift. The
quadrature value $\int\jpar dA$ has an error of order $10^{-14}$.

\begin{table}[!htbp]
\caption{Endpoint absorption at $\varepsilon=1.3\%$, $\theta_0=0$ (values $\times\mathcal N$; populations per spin).}
\label{tab:energy}
\begin{ruledtabular}
\begin{tabular}{rrlrrrr}
$F_0$ & $N$ & grid & $W_{\rm abs}$ (manifest) & norm term & $\int\jpar dA$ & excited pop.\\
\hline
0.3 & 4800 & base & $1.289\times10^{-1}$ & $-1.7\times10^{-11}$ & $1.289\times10^{-1}$ & $9.0\times10^{-6}$\\
0.3 & 9600 & base & $1.845\times10^{-3}$ & $-6.3\times10^{-10}$ & $1.845\times10^{-3}$ & $6.5\times10^{-8}$\\
0.3 & 19200 & base & $5.614\times10^{-16}$ & $-5.30\times10^{-10}$ & $-3.98\times10^{-14}$ & $6.21\times10^{-19}$\\
0.3 & 38400 & base & $5.618\times10^{-16}$ & $+1.99\times10^{-10}$ & $-2.58\times10^{-14}$ & $6.21\times10^{-19}$\\
0.3 & 76800 & base & $5.619\times10^{-16}$ & $+1.78\times10^{-10}$ & $-1.58\times10^{-14}$ & $6.21\times10^{-19}$\\
0.3 & 19200 & fine & $5.637\times10^{-16}$ & $-5.80\times10^{-10}$ & $-3.80\times10^{-14}$ & $6.23\times10^{-19}$\\
0.02 & 19200 & base & $9.115\times10^{-19}$ & $-3.83\times10^{-10}$ & $-4.3\times10^{-16}$ & $6.5\times10^{-22}$\\
0.02 & 38400 & base & $9.144\times10^{-19}$ & $+3.54\times10^{-10}$ & $-1.8\times10^{-16}$ & $6.5\times10^{-22}$\\
1.5 & 19200 & base & $8.6807\times10^{-2}$ & $-5.5\times10^{-10}$ & $8.6807\times10^{-2}$ & $5.87\times10^{-5}$\\
1.5 & 38400 & base & $8.6808\times10^{-2}$ & $+1.7\times10^{-10}$ & $8.6808\times10^{-2}$ & $5.87\times10^{-5}$\\
\end{tabular}
\end{ruledtabular}
\end{table}

\begin{figure}[!htbp]
\includegraphics[width=0.75\textwidth]{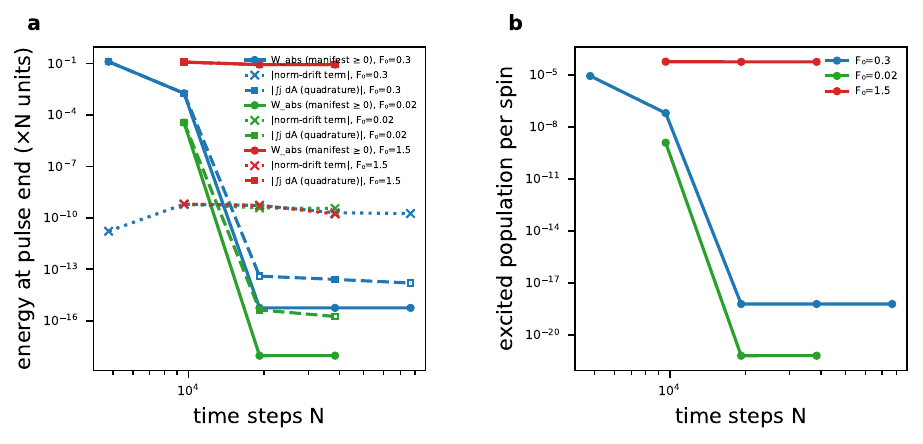}
\caption{(a) Endpoint energy diagnostics versus time steps: manifest absorption (circles), norm-drift
term (crosses), and quadrature work (squares). (b) Excited population.}
\end{figure}

\section{Strong-field angular and time convergence}
Table~\ref{tab:strong} lists the field-even transverse harmonics at $F_0=1.5$. Convergence requires
angular resolution of the momentum grid near the swept Dirac points: on the base grid $J_2^\Q$ is in
error by $10.8$ in absolute value; refining only the radial direction does not remove the error,
whereas refining only the angular direction does. The converged values are those of the main text.

\paragraph{Adiabaticity parameters.} Table~\ref{tab:lz} lists, for model I at $\varepsilon=1.3\%$ and
$x$ drive, the peak excursion $A_{\max}=F_0E_0/\omega$ relative to the Fermi radius
$k_F=0.0641$ (along $-x$; $0.0661$ along $+x$, from the exact band), the Landau-Zener exponent at the node
$\delta_{\rm LZ}=M^2/(v_xeE)$ with $v_x=1.507$ and $E=F_0E_0$ the nominal peak field, and the
closest approach of a singly occupied state to the node, $\max(0,k_F-A_{\max})$, with the local
half-gap $|\mathbf d|$ there (exact lattice $\mathbf d$) and the corresponding exponent
$\delta_{\rm eff}=|\mathbf d|^2/(v_xeE)$. Only states with exactly one occupied band contribute: a
$\bk_0$ with both bands occupied is inert. The Keldysh-type parameter is
$\gamma_K=\omega E_{\rm ref}/(v_xeE)$, with $E_{\rm ref}=|\mathbf d|$ at the Fermi surface.

\begin{table}[!htbp]
\caption{Adiabaticity parameters (model I, $\varepsilon=1.3\%$, $\theta_0=0$).}
\label{tab:lz}
\begin{ruledtabular}
\begin{tabular}{rrrrrrr}
$F_0$ & $A_{\max}/k_F$ & $\delta_{\rm LZ}$ (node) & $e^{-\pi\delta_{\rm LZ}}$ & closest $|\mathbf d|$ & $e^{-\pi\delta_{\rm eff}}$ & $\gamma_K$\\
\hline
0.3 & 0.31 & 1.33 & 0.015 & 0.070 & $6\times10^{-23}$ & 3.32\\
1.0 & 1.04 & 0.40 & 0.29 & 0.020 (node) & 0.29 & 1.00\\
1.5 & 1.56 & 0.27 & 0.43 & 0.020 (node) & 0.43 & 0.66\\
\end{tabular}
\end{ruledtabular}
\end{table}

The Keldysh-type parameter at the Fermi surface is $3.3$, $1.0$, and $0.66$ at $F_0=0.3$, $1$, and $1.5$.
At $F_0=0.3$ no singly occupied state reaches the node: the closest approach gives a local half-gap
$0.070$ and an exponent $\delta_{\rm eff}=16$, and the transition probability along the closest
trajectory is negligible, consistent with the residual absorbed energy of $5.6\times10^{-16}$.
At $F_0\ge1$ the node is swept with order-unity Landau-Zener probability, consistent with the
residual excitation at $F_0=1.5$. We did not compare with generalized semiclassics with tunnelling on
the lattice.

\paragraph{Time-domain convergence and oscillation frequency at $F_0=1.5$.} For the fine-grid signal of
Fig.~\ref{fig:proj}(d), $A\jperp^{(e)}$ over $|A|>0.1A_{\max}$, the time step ($N=38400$) and the
extra-fine grid change it by $1.0\times10^{-4}$ and $2.7\times10^{-4}$ of its RMS (pointwise at most
$6\times10^{-4}$ and $2.1\times10^{-3}$ of the RMS), so the rapid oscillations are reproduced point by point.
These refinement runs carry no angular derivative; for the projected signals only base versus fine
exists, and the projections differ from $A\jperp^{(e)}$ by a removed part of RMS $0.017$ of
$\max|Aj_B|$, small against the signal. The base grid itself differs from the fine grid by $2.9\%$
of the RMS, consistent with the base-grid harmonic error of Table~\ref{tab:strong}. The spectrum of
$A\jperp^{(e)}$ in the interval shown (after removing a cubic trend; Hann window) extends from about
$7\omega$ to $54\omega$ at half maximum; in one-cycle windows centered at $50\lesssim\omega t\lesssim75$,
which include the interval shown, its peak lies at $44$--$52\omega$, while toward the pulse edges a
component near $7\omega$ dominates. For the
Fermi-surface states that cross the node, the local gap $2|\mathbf d(\bk_0+\mathbf A(t))|/\hbar$ along
the drive ranges from $2M/\hbar=4\omega$ at the node to about $50\omega$ at the turning points of
$A(t)$, where it is stationary in time. The oscillations are therefore consistent with interband
coherence at the local gap frequency after Landau--Zener passage; this is a one-dimensional estimate,
not a mode assignment.

\begin{table}[!htbp]
\caption{Second and fourth harmonics at $F_0=1.5$, $\varepsilon=1.3\%$, $\theta_0=0$ ($\times\mathcal N/(F_0E_0)^2$).}
\label{tab:strong}
\begin{ruledtabular}
\begin{tabular}{lrrrr}
grid / $N$ & $J_2$ & $|\Delta J_2|$ & $J_4$ & $|\Delta J_4|$\\
\hline
base, 19200 & $4.02+20.87i$ & 10.78 & $-220.08-45.28i$ & 9.78\\
base, 38400 & $4.02+20.72i$ & 10.94 & $-220.06-45.31i$ & 9.74\\
radial-only, 19200 & $4.02+20.87i$ & 10.78 & $-220.08-45.28i$ & 9.78\\
angular-only, 19200 & $4.46+31.65i$ & 0.01 & $-218.52-54.87i$ & 0.07\\
fine, 19200 & $4.46+31.65i$ & 0.01 & $-218.52-54.87i$ & 0.07\\
fine, 38400 & $4.47+31.49i$ & 0.15 & $-218.50-54.90i$ & 0.03\\
extra-fine, 19200 & $4.46+31.64i$ & 0 & $-218.47-54.92i$ & 0\\
semiclassical & $1819.4i$ & -- & $-66.7i$ & --\\
\end{tabular}
\end{ruledtabular}
\end{table}

\begin{figure}[!htbp]
\includegraphics[width=0.7\textwidth]{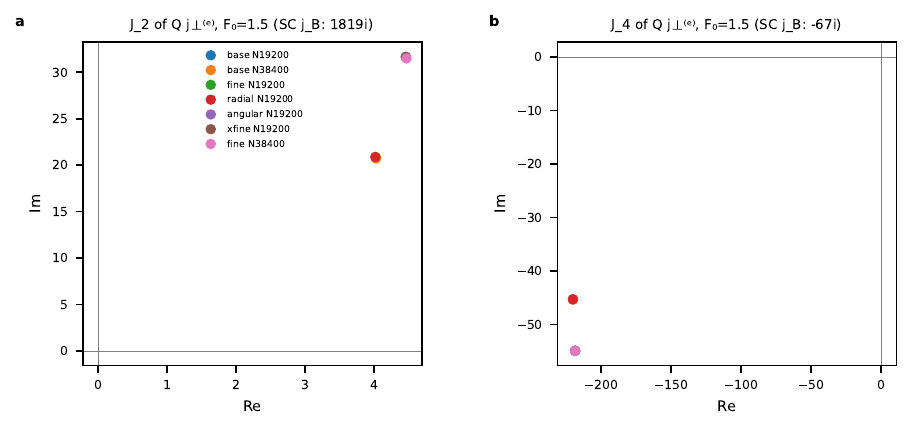}
\caption{Complex-plane convergence of $J_2$ and $J_4$ at $F_0=1.5$ over grids and time steps.}
\end{figure}

\section{Hysteresis of the unstrained lattice at $F_0=0.3$}\label{sec:hyst}
The rising/falling-branch estimator compares $H/E$ at matched $A$ over the pulse core
($|A|>0.1A_{\max}$, $|E|>0.3E_{\max}$) and quotes the RMS branch difference relative to the range ($h$)
and the median significance $z$ against pointwise errors (time refinement $N=38400$ plus grid
refinement, applied to every run entering the estimator). For model I at $\varepsilon=0$,
$\theta_0=0$, $F_0=0.3$:
\begin{itemize}
\item single-$\frac13$ angular projection: $h=6.21\times10^{-4}$, $z=16.4$ ($85\%$ of points with $z>3$);
\item mixed projection $H=\jperp-\frac13S_2-\frac15(\partial_\theta\jpar-S_2)$, with
$S_2=\partial_\theta\jpar(F_0=0.02)\times(0.3/0.02)^2$: $h=5.60\times10^{-4}$, $z=14.3$ ($84\%$);
\item nonperturbative projection $H_{\rm NP}$, Eq.~(\ref{eq:Hnp}), symmetric part removed at all
orders (Sec.~\ref{sec:npSM}; errors from $N=38400$, fine grid and node count at $F_0=0.3$ itself):
$h=5.57\times10^{-4}$, $z=14.3$ ($84\%$).
\end{itemize}
$S_2$ carries an order-four admixture of relative weight $(0.02/0.3)^2=0.4\%$, and symmetric terms of
order $\ge6$ still receive the wrong factor. The order-four symmetric correction
$(\frac15-\frac13)(\partial_\theta\jpar-S_2)$ is $0.31\%$ of $\max|H|$, and the two projections differ
by $0.10\%$ (RMS). The loops therefore survive removal of the order-four symmetric part, and,
with $H_{\rm NP}$, of the symmetric part at every order. The
refinement runs at $F_0=0.02$ ($N=38400$; fine grid) were computed for this test. No residual
absorption is resolved ($|\int\jpar dA|\le4\times10^{-14}$, at the quadrature error; the manifest form
of Eq.~\eqref{eq:passive} was not evaluated for this case). A reversible polarization current
depends on $\dot E$ and is not of the form $E\,f(A)$, so such loops do not require real excitation.

\section{Nonperturbative projection: quadrature and tests}\label{sec:npSM}
\emph{Implementation.} Equation~(\ref{eq:Hnp}) is evaluated in the field-even channel (which
it may be applied to without changing the normalization, since it is linear in $\mathbf j$). The
radial integral is a Gauss--Legendre sum over $N_s$ amplitudes $s_kF_0$ (Table~\ref{tab:npnodes}); at
every node and at the target $F_0$ one quantum run is made with the validated propagator at the
same $N$ and momentum grid as the existing runs. The angular derivative $\partial_\theta\jpar$ is taken
by the complex-step method on the drive pair $\{\theta_0,\theta_0+\pi\}$ (field-even and field-odd
channels from the sum and difference), and the semiclassical currents of the same runs, which are
holomorphic in $\theta$, give $H_{\rm NP}^\SC$ at the same nodes. The complex-step runs reproduce the
real-mode runs: the time-domain RMS values of the direct and single-$\frac13$ projected currents in
Fig.~\ref{fig:proj}(e) are recovered to all printed digits. Node sets: $N_s=8,12,16$ at
$F_0=0.3$ and $1$ ($\theta_0=0$); $N_s=8,12$ at $\theta_0=\pi/4$, $\varepsilon=0$ and model II; $N_s=4,8$
at $F_0\le0.1$ and at $F_0=1.5$ (fine grid). Since $N_s=8$ and $16$ agree to $7\times10^{-8}$ of
$\max|j_B|$ at $F_0=0.3$, the refinement runs ($N=38400$; fine grid) use $N_s=8$. One run costs the same
as a standard run, so $H_{\rm NP}$ at one $(F_0,\theta_0)$ costs $N_s+1$ runs.

\begin{table}[h]\caption{Gauss--Legendre nodes $s_k$ and weights $w_k$ on $[0,1]$ for $N_s=8$. The radial integral in Eq.~(\ref{eq:Hnp}) is $\sum_k w_k\,\partial_\theta j_\parallel(s_kF_0,\theta,t)$; $N_s=12$ and $16$ use the standard Gauss--Legendre rules mapped in the same way.}\label{tab:npnodes}
\begin{ruledtabular}\begin{tabular}{ccc|ccc}
$k$ & $s_k$ & $w_k$ & $k$ & $s_k$ & $w_k$\\\hline
1 & 0.0198550718 & 0.0506142681 & 5 & 0.5917173212 & 0.1813418917\\
2 & 0.1016667613 & 0.1111905172 & 6 & 0.7627662050 & 0.1568533229\\
3 & 0.2372337950 & 0.1568533229 & 7 & 0.8983332387 & 0.1111905172\\
4 & 0.4082826788 & 0.1813418917 & 8 & 0.9801449282 & 0.0506142681\\
\end{tabular}\end{ruledtabular}\end{table}
\begin{table}[h]\caption{Test A: semiclassical $H_{\rm NP}^{\rm SC}$ against $j_B$ (model I, $\varepsilon=1.3\%$), from the vectorized semiclassical evaluator at the same nodes, $N_s=8,12,16$ (largest value over $N_s$ shown). Even: $\max_t|H^{(e)}_{\rm NP}-j_B|/\max|j_B|$. Odd: field-odd channel, where the band current has a large symmetric part $\nabla_E\Phi$; ``removed'' is $\max|\partial_\theta\int_0^1 j_\parallel^{(o)}ds|/\max|j_B|$, and the last two columns give the residual transverse current after projection relative to the removed part, for $H_{\rm NP}$ and for the single-$\frac13$ angular projection.}\label{tab:nptestA}
\begin{ruledtabular}\begin{tabular}{ccccccc}
$F_0$ & $\theta_0$ & even & odd: removed & odd: $H_{\rm NP}$/removed & odd: single-$1/3$/removed\\\hline
0.02 & 0 & $5.6\times10^{-7}$ & $3.0\times10^{-7}$ & (trivial) & (trivial)\\
0.02 & $\pi/4$ & $3.8\times10^{-7}$ & $1.2\times10^{5}$ & $4.8\times10^{-12}$ & 0.333\\
0.3 & 0 & $2.4\times10^{-9}$ & $1.4\times10^{-9}$ & (trivial) & (trivial)\\
0.3 & $\pi/4$ & $2.1\times10^{-9}$ & $7.7\times10^{3}$ & $3.1\times10^{-13}$ & 0.333\\
1.0 & 0 & $2.3\times10^{-10}$ & $1.2\times10^{-10}$ & (trivial) & (trivial)\\
1.0 & $\pi/4$ & $1.8\times10^{-10}$ & $2.3\times10^{3}$ & $2.4\times10^{-13}$ & 0.333\\
1.5 & 0 & $1.1\times10^{-10}$ & $8.2\times10^{-11}$ & (trivial) & (trivial)\\
1.5 & $\pi/4$ & $9.2\times10^{-11}$ & $1.5\times10^{3}$ & $5.4\times10^{-14}$ & 0.333\\
\end{tabular}\end{ruledtabular}\end{table}
\begin{table}[h]\caption{Tests B and C: quantum $H_{\rm NP}$, $n=2$ (Hann, $\times\mathcal N/(F_0E_0)^2$, imaginary part; real parts $<10^{-5}$), model I, $\varepsilon=1.3\%$, $\theta_0=0$, against the order-resolved fit $H^{(2)}+(F_0/0.01)^2H^{(4)}$ (preferred fit); ``fit unc.'' is the spread of $H^{(2)}$ plus $(F_0/0.01)^2$ times the spread of $H^{(4)}$ over the fit family. $N_s=4$ and $8$ agree to the digits shown.}\label{tab:nptestBC}
\begin{ruledtabular}\begin{tabular}{cccccc}
$F_0$ & $H_{\rm NP}$ ($N_s=8$) & $N_s=4$ & fit & difference & fit unc.\\\hline
0.01 & $-73.7199i$ & $-73.7199i$ & $-73.7200i$ & $+0.0001i$ & 0.0009\\
0.05 & $-72.3408i$ & $-72.3408i$ & $-72.3466i$ & $+0.0059i$ & 0.0181\\
0.1 & $-67.9955i$ & $-67.9955i$ & $-68.0550i$ & $+0.0595i$ & 0.0718\\
\end{tabular}\end{ruledtabular}\end{table}
\begin{table*}[t]\caption{Quantum $H_{\rm NP}$: node convergence, numerical errors and time-domain residuals. All error columns use the metric of the error bar, RMS$[A\,\Delta H_{\rm NP}]/\max|Aj_B|$ over $|A|>0.1A_{\max}$ (the window of Fig.~4(e)): ``nodes'', the two largest node counts; ``time'', $N=38400$ vs $19200$; ``grid'', fine vs base (base-grid rows) or, for the fine-grid rows at $F_0=1$, angular-only grid vs fine for $H_{\rm NP}$ plus extra-fine vs fine for $j_\perp$ at the target amplitude ($\theta_0=0$; no grid or time-step check exists for the fine-grid row at $\theta_0=\pi/4$, and its error bar contains the node difference only); node and time entries of the fine rows ($^\dagger$) are the base-grid differences. RMS columns: residual of $X=H_{\rm NP}$, the single-$\frac13$ angular projection and the unprojected $j_\perp^{(e)}$; the error bar is the RMS of the pointwise sum of all listed differences, so it contains each of them. Pointwise maxima over the full pulse can be much larger: at $F_0=1$ base vs fine differs by $2.7\times10^{-1}$ of $\max|j_B|$ at $u=124$, in the post-pulse tail ($|A|\approx0$, outside the window, negligible Hann weight). $^\ddagger$\,$F_0=1.5$: not converged in $N_s$, not used.}\label{tab:npconv}
\begin{ruledtabular}\begin{tabular}{ccccccccccc}
model & $\varepsilon$ & $\theta_0$ & $F_0$ & $N_s$ & nodes & time & grid & RMS $H_{\rm NP}$ & RMS single-$1/3$ & RMS direct\\\hline
I & 1.3\% & 0 & 0.01 & 4,8 & $1.3\times10^{-5}$ & -- & -- & 0.0023 & 0.0023 & 0.0220\\
I & 1.3\% & 0 & 0.05 & 4,8 & $5.8\times10^{-7}$ & -- & -- & 0.0020 & 0.0020 & 0.0224\\
I & 1.3\% & 0 & 0.1 & 4,8 & $2.6\times10^{-7}$ & -- & -- & 0.0014 & 0.0015 & 0.0237\\
I & 1.3\% & 0 & 0.3 & 8,12,16 & $1.0\times10^{-8}$ & $6.4\times10^{-4}$ & $1.2\times10^{-5}$ & 0.0349 $\pm$ $6.4\times10^{-4}$ & 0.0375 & 0.0233\\
I & 1.3\% & $\pi/4$ & 0.3 & 8,12 & $1.9\times10^{-9}$ & -- & -- & 0.0103 & 0.0109 & 0.0052\\
I & 0\% & 0 & 0.3 & 8,12 & $1.1\times10^{-8}$ & $3.3\times10^{-5}$ & $3.8\times10^{-6}$ & 0.0152 $\pm$ $3.5\times10^{-5}$ & 0.0162 & 0.0085\\
II & 1.3\% & 0 & 0.3 & 8,12 & $1.3\times10^{-9}$ & -- & -- & 0.0032 & 0.0032 & 0.0082\\
I & 1.3\% & 0 & 1.5 (fine)$^\ddagger$ & 4,8 & $1.0\times10^{-2}$ & -- & -- & 0.8465 & 0.8448 & 0.8476\\
I & 1.3\% & 0 & 1.0 (fine) & 8 & $3.6\times10^{-5}$$^\dagger$ & $8.4\times10^{-6}$$^\dagger$ & $2.7\times10^{-4}$ & 0.1909 $\pm$ $2.8\times10^{-4}$ & 0.1960 & 0.1898\\
I & 1.3\% & $\pi/4$ & 1.0 (fine) & 8 & $3.8\times10^{-5}$$^\dagger$ & --$^\dagger$ & -- & 0.1854 $\pm$ $3.8\times10^{-5}$ & 0.1911 & 0.1842\\
\end{tabular}\end{ruledtabular}\end{table*}
\begin{table}[h]\caption{Harmonics of the quantum $H_{\rm NP}$ relative to $j_B$: modulus ratio $|H_{{\rm NP},n}/j_{B,n}|$ and phase $\arg(H_{{\rm NP},n}/j_{B,n})$ (rad), largest $N_s$; $F_0=1$ on the fine grid; $^\ddagger$\,$F_0=1.5$: not converged, not used. ``err'': summed absolute numerical change of $H_{{\rm NP},n}$ relative to $|H_{{\rm NP},n}|$ (where refinements exist).}\label{tab:npharm}
\begin{ruledtabular}\begin{tabular}{cccccccc}
model & $\varepsilon$ & $\theta_0$ & $F_0$ & $n$ & ratio & phase & err\\\hline
I & 1.3\% & 0 & 0.01 & 2 & 1.0062 & +0.000 & --\\
I & 1.3\% & 0 & 0.01 & 4 & 1.0624 & +0.001 & --\\
I & 1.3\% & 0 & 0.05 & 2 & 1.0057 & -0.000 & --\\
I & 1.3\% & 0 & 0.05 & 4 & 1.0628 & -0.000 & --\\
I & 1.3\% & 0 & 0.1 & 2 & 1.0041 & -0.000 & --\\
I & 1.3\% & 0 & 0.1 & 4 & 1.0639 & -0.000 & --\\
I & 1.3\% & 0 & 0.3 & 2 & 0.9268 & -0.000 & $4.0\times10^{-3}$\\
I & 1.3\% & 0 & 0.3 & 4 & 1.0774 & +0.000 & $1.1\times10^{-4}$\\
I & 1.3\% & $\pi/4$ & 0.3 & 2 & 1.0190 & +0.000 & --\\
I & 1.3\% & $\pi/4$ & 0.3 & 4 & 1.0805 & +0.000 & --\\
I & 0\% & 0 & 0.3 & 2 & 1.0351 & -0.000 & $1.1\times10^{-4}$\\
I & 0\% & 0 & 0.3 & 4 & 1.0792 & +0.000 & $1.0\times10^{-4}$\\
II & 1.3\% & 0 & 0.3 & 2 & 1.0091 & +0.000 & --\\
II & 1.3\% & 0 & 0.3 & 4 & 1.0558 & +0.000 & --\\
I & 1.3\% & 0 & 1.5 (fine)$^\ddagger$ & 2 & 0.0281 & -0.511 & --\\
I & 1.3\% & 0 & 1.5 (fine)$^\ddagger$ & 4 & 3.2735 & -1.142 & --\\
I & 1.3\% & 0 & 1.5 (fine)$^\ddagger$ & 6 & 0.2919 & +1.798 & --\\
I & 1.3\% & 0 & 1.5 (fine)$^\ddagger$ & 8 & 1.6901 & +1.481 & --\\
I & 1.3\% & 0 & 1.0 (fine) & 2 & 0.9123 & +0.096 & $4.3\times10^{-5}$\\
I & 1.3\% & 0 & 1.0 (fine) & 4 & 1.2458 & +0.270 & $2.5\times10^{-5}$\\
I & 1.3\% & 0 & 1.0 (fine) & 6 & 3.5932 & +1.079 & $1.8\times10^{-5}$\\
I & 1.3\% & 0 & 1.0 (fine) & 8 & 6.4157 & -0.616 & $4.7\times10^{-5}$\\
I & 1.3\% & $\pi/4$ & 1.0 (fine) & 2 & 0.9530 & +0.081 & $2.0\times10^{-7}$\\
I & 1.3\% & $\pi/4$ & 1.0 (fine) & 4 & 1.2914 & +0.257 & $1.4\times10^{-6}$\\
I & 1.3\% & $\pi/4$ & 1.0 (fine) & 6 & 3.4425 & +1.027 & $6.2\times10^{-6}$\\
I & 1.3\% & $\pi/4$ & 1.0 (fine) & 8 & 9.4873 & -0.668 & $1.9\times10^{-5}$\\
\end{tabular}\end{ruledtabular}\end{table}
\begin{table}[h]\caption{Hysteresis of $X/E$ versus $A$ (estimator of Sec.~\ref{sec:hyst}): $h$ and median $z$ (fraction of matched pairs with $z>3$). Errors are the pointwise sum of the node, time-step and grid differences (Table~\ref{tab:npconv}).}\label{tab:nphyst}
\begin{ruledtabular}\begin{tabular}{cccccc}
model & $\varepsilon$ & $\theta_0$ & $F_0$ & $X=H_{\rm NP}$: $h$, $z$ & single-$1/3$: $h$, $z$\\\hline
I & 1.3\% & 0 & 0.3 & $1.1\times10^{-3}$, 0.7 (12\%) & $1.3\times10^{-3}$, 0.8 (15\%)\\
I & 0\% & 0 & 0.3 & $5.6\times10^{-4}$, 14.3 (84\%) & $6.2\times10^{-4}$, 16.4 (86\%)\\
I & 1.3\% & 0 & 1.0 (fine) & $8.4\times10^{-2}$, 1916 (97\%) & $8.6\times10^{-2}$, 2065 (97\%)\\
\end{tabular}\end{ruledtabular}\end{table}
\begin{table}[h]\caption{Decomposition of the unprojected residual, $j_\perp-j_B=(H_{\rm NP}-j_B)+R$, with $R=\hat{\mathbf e}_\perp\cdot\nabla_{\mathbf E}\Phi$ the removed symmetric part: RMS of $AR$ and of $A(H_{\rm NP}-j_B)$ over the window (relative to $\max|Aj_B|$) and their normalized overlap $\langle AR,A(H_{\rm NP}-j_B)\rangle/(\|AR\|\|A(H_{\rm NP}-j_B)\|)$. A negative overlap makes the unprojected current look closer to $j_B$.}\label{tab:npsym}
\begin{ruledtabular}\begin{tabular}{ccccccc}
model & $\varepsilon$ & $\theta_0$ & $F_0$ & RMS $R$ & RMS $(H_{\rm NP}-j_B)$ & overlap\\\hline
I & 1.3\% & 0 & 0.01 & 0.0197 & 0.0023 & +1.00\\
I & 1.3\% & 0 & 0.05 & 0.0204 & 0.0020 & +0.99\\
I & 1.3\% & 0 & 0.1 & 0.0227 & 0.0014 & +0.65\\
I & 1.3\% & 0 & 0.3 & 0.0374 & 0.0349 & -0.79\\
I & 1.3\% & $\pi/4$ & 0.3 & 0.0084 & 0.0103 & -0.86\\
I & 0\% & 0 & 0.3 & 0.0137 & 0.0152 & -0.83\\
II & 1.3\% & 0 & 0.3 & 0.0053 & 0.0032 & +0.83\\
I & 1.3\% & 0 & 1.0 (fine) & 0.0042 & 0.1909 & -0.27\\
I & 1.3\% & $\pi/4$ & 1.0 (fine) & 0.0037 & 0.1854 & -0.34\\
\end{tabular}\end{ruledtabular}\end{table}
\begin{table}[h]\caption{$H_{\rm NP}-j_B$ at $F_0=0.3$ (model I, $\varepsilon=1.3\%$; imaginary parts, $\times\mathcal N/(F_0E_0)^2$) against the perturbative estimate $(H^{(2)}_\Q-H^{(2)}_\SC)+(F_0/0.01)^2(H^{(4)}_\Q-H^{(4)}_\SC)$ from the preferred weak-field fits; ``unc.'' is the spread over the fit family. The estimate omits $O(F_0^4)$ terms.}\label{tab:nppert}
\begin{ruledtabular}\begin{tabular}{cccccc}
$\theta_0$ & $n$ & $H_{\rm NP}-j_B$ & estimate & unc. & within unc.\\\hline
0 & 2 & $+1.426i$ & $+0.964i$ & 0.680 & yes\\
0 & 4 & $-2.172i$ & $-1.769i$ & 0.637 & yes\\
$\pi/4$ & 2 & $-1.759i$ & $-1.403i$ & 1.371 & yes\\
$\pi/4$ & 4 & $+1.710i$ & $+1.187i$ & 0.426 & no\\
\end{tabular}\end{ruledtabular}\end{table}

\emph{Test A (exactness in semiclassics).} Table~\ref{tab:nptestA}. In the field-even channel the
test is passed trivially, because the semiclassical field-even current is $j_B$ itself and its
longitudinal part vanishes (time reversal); at $\theta_0=0$ mirror symmetry also makes the field-odd
longitudinal part vanish. The decisive case is the field-odd channel at $\theta_0=\pi/4$, where the
band current carries a symmetric part $10^3$--$10^5$ times $\max|j_B|$: $H_{\rm NP}$ removes it to
$\le5\times10^{-12}$ of its size for every $N_s$, while the single-$\frac13$ angular projection leaves exactly
the expected fraction $(\frac12-\frac13)/\frac12=\frac13$. All cases meet the criterion
($10^{-6}$).

\emph{Tests B and C (perturbative limit).} Table~\ref{tab:nptestBC}. At $F_0=0.01$, $H_{\rm NP}=-73.7199i$ agrees with
$H^{(2)}+(F_0/0.01)^2H^{(4)}=-73.7200i$ to $1\times10^{-6}$ (relative); its $0.08\%$ difference from the
finite-time $H^{(2)}_\Q=-73.78i$ of Table~\ref{tab:orders} is the $F_0^2$ term. At $F_0=0.05$ and $0.1$
it agrees with $H^{(2)}+(F_0/0.01)^2H^{(4)}$ within the fit uncertainty. The remaining
difference at $F_0=0.1$ has the size and sign of the omitted $F_0^4$ term (the semiclassical
$c_6$ contributes $+0.043i$ there).

\emph{Test D (node convergence).} Table~\ref{tab:npconv}. The change between the two largest node
counts is below the time-step error at $F_0=0.3$ by four orders of magnitude. At $F_0=1$ the reported values
are on the fine grid ($N_s=8$); node and time-step differences there are taken from the base grid
($^\dagger$ in Table~\ref{tab:npconv}). On the base grid the $N_s=12\to16$ change ($7.8\times10^{-5}$ of
$\max|j_B|$, pointwise maximum) is comparable to the time-step difference ($6.5\times10^{-5}$) and smaller
in the RMS over the window ($5.5\times10^{-6}$ vs $8.4\times10^{-6}$); the criterion is therefore met
marginally in the window and not met in the pointwise maximum. The $N_s=8\to16$ change
($3.6\times10^{-5}$, window RMS) and the time-step difference are both below the grid uncertainty of the
fine-grid result, $2.7\times10^{-4}$, which is the difference to a grid with the base radial rule and the
fine angular resolution; the extra-fine grid changes $j_\perp$ at the target amplitude by only
$8\times10^{-7}$, so this bound is conservative. On the fine grid itself the amplitude quadrature at $F_0=1$,
$\theta_0=0$ was checked with a composite rule (Gauss--Legendre 8 nodes on $s\in[0,0.95]$ plus 4 on
$[0.95,1]$): it changes $H_{\rm NP}$ by $4.1\times10^{-5}$ (window RMS; pointwise window maximum
$4.0\times10^{-4}$ of $\max|j_B|$) and the harmonics $n=2$--$8$ by at most $1.6\times10^{-5}$ of their
moduli, within the quoted uncertainties, so the $N_s=8$ values are kept.

\emph{Test E (longitudinal remainder).} The component of $\mathbf j-\nabla_{\mathbf E}\Phi$ along $\be$
is $\jpar-\partial_F\Phi$, which vanishes identically because
$\Phi=F\int_0^1\jpar(sF)\,ds$; the code never forms $\partial_F\Phi$, and $H_{\rm NP}$ is the
$\hat{\mathbf e}_\perp$ component by construction. On the arrays used the remainder is exactly zero.
As the nontrivial part of this check we report the convergence of the removed radial integral
$\int_0^1\partial_\theta\jpar(sF)\,ds$ with $N_s$: its relative change between the two largest node
counts is $6\times10^{-7}$ at $F_0=0.3$ and $3\times10^{-3}$ at $F_0=1$, where $\partial_\theta\jpar$ varies
more rapidly with amplitude; the effect on $H_{\rm NP}$ is the node change of Table~\ref{tab:npconv}.

\emph{Results.} Tables~\ref{tab:npconv}--\ref{tab:nphyst}. For model I the time-domain residual of
$H_{\rm NP}$ lies slightly below that of the single-$\frac13$ angular projection at $0.1\le F_0\le1$ (at
$F_0=0.01$ and $0.05$ the two agree to $1\%$) and above that of the unprojected current at $F_0=0.3$ and
$1$; for model II at $F_0=0.3$ it lies $2\%$ above the single-$\frac13$ value and below the unprojected
one. Harmonic phases relative
to $j_B$ are below $10^{-7}$~rad at $F_0=0.1$ and $0.3$ and below $10^{-6}$~rad at $F_0=0.05$; at
$F_0=0.01$ the small fourth harmonic has phase $9\times10^{-4}$~rad. For the hysteresis of the unstrained lattice (Sec.~\ref{sec:hyst}), $H_{\rm NP}$ gives
$h$ and $z$ within $1\%$ of the mixed $\frac13/\frac15$ projection, so symmetric terms of order $\ge6$
do not produce the loops. At $F_0=1.5$ (fine grid; $\delta_{\rm LZ}=0.27$) the results are reported for
completeness only, and the amplitude quadrature is not converged there: $N_s=4\to8$ changes
$H_{\rm NP}$ by $6\%$ of $\max|j_B|$ and the removed radial integral by $72\%$, so the entries for
$F_0=1.5$ in Tables~\ref{tab:npconv} and \ref{tab:npharm} are not converged values and are not used.

\paragraph{Frequency domain.} Equation~\eqref{eq:Hnp} is linear in $\mathbf j$ and acts on the drive
parameters $(F,\theta)$ at fixed $t$, so a Fourier transform in $t$ with any fixed window $w(t)$ commutes
with it: $H_{\rm NP}(F,\theta;n\omega)=\jperp(F,\theta;n\omega)-\partial_\theta\int_0^1\jpar(sF,\theta;n\omega)\,ds$
for complex harmonic amplitudes, and $H_{\rm NP}^\SC(n\omega)=j_B(n\omega)$ for each harmonic separately.
This requires (1) the same waveform $f(t)$ for all amplitudes and orientations, i.e., a fixed envelope
and carrier phase relative to the time origin, so that harmonic phases at different $F$ are referenced
consistently; (2) complex amplitudes of both in-plane components, since the projection is linear in the
amplitudes, not in $|\mathbf j|^2$, so intensities or the modulus of one component do not suffice
(phase-resolved detection, for example electro-optic sampling in the terahertz range); and
(3) the orientation dependence around $\theta_0$ for $\partial_\theta$, from neighboring drive angles or a
full rotation of the linear polarization relative to the crystal axes. The harmonics in
Table~\ref{tab:npharm} and in the main text are Hann-windowed Fourier coefficients of $H_{\rm NP}(t)$.
Applying $H_{\rm NP}$ instead to the Hann-windowed harmonics of the individual runs gives the same values
to roundoff (largest relative difference $2\times10^{-11}$ over $n=2$--$8$ and the six cases at
$F_0=0.1$--$1$). For a thin film the emitted far field at $n\omega$ is proportional to the sheet current at
$n\omega$ through a transfer factor ($\propto in\omega$ times Fresnel and substrate factors) that does not
depend on $F$ or $\theta$; if it is also the same for both in-plane components, $H_{\rm NP}$ can be
applied directly to the measured complex harmonic fields, and otherwise the two components must first
be calibrated. The radial integral needs the harmonics at the quadrature amplitudes $s_kF_0$; $N_s=8$
Gauss--Legendre nodes suffice at $F_0=0.3$ (Table~\ref{tab:npconv}). An experiment can instead sample the
amplitude scan densely and integrate numerically, which requires data down to amplitudes where the
weak-field power law $j\propto F^p$ is resolved, or an extrapolation of that power law to zero
amplitude.

\section{Additional projections}\label{sec:addproj}
\paragraph{Work projection.} For unitary dynamics the work $W(\theta,t)=\int_0^t\jpar\,dA$ is the
change of electronic energy (Sec.~\ref{sec:energy}). The work projection is
$AH=A\jperp-\partial_\theta W|_A$. In adiabatic semiclassics $W$ is a function of $A$, and
$\jperp-A^{-1}\partial_\theta W|_A$ reproduces the Berry current exactly; the projection is exact for
any current that is a function of $A$, i.e., for adiabatic band currents. With a
time-reversal-invariant initial state, any current that is an instantaneous function of $A$ is odd in
$A$ (Sec.~\ref{sec:symSM}). A field-even current must therefore involve at least one time derivative,
$\mathbf j^{(e)}=\mathsf M(A)\mathbf E+\dots$, and the field-odd work $W^{(o)}=\int\jpar^{(e)}dA$ depends
on the history of the pulse. For a symmetric current $\propto EA$,
$A^{-1}\partial_\theta W^{(o)}\neq\frac13\partial_\theta\jpar^{(e)}$. In time-reversal-invariant lattices
the work projection is thus trivial in the field-even channel of semiclassics ($W^{(o)}\equiv0$) and
incorrect for coherent dynamics; the main text uses it only as a negative control.

\paragraph{E-linear projection.} For symmetric currents of the form
$j_i=E_j\partial_i\partial_j\Psi(A)$ one has $\jpar=E\partial_A^2\Psi$ and
$\jperp=E\partial_A(A^{-1}\partial_\theta\Psi)$. With $V=\omega^{-1}\int\jpar\,du$ and
$\Psi=\int V\,dA$, this gives $A^2\tilde H=A^2\jperp-E(V'A-\Psi')$, with $'\equiv\partial_\theta$. Its
residual, RMS$(A^2\tilde H-A^2j_B)/\max|A^2j_B|$ over $|A|>0.1A_{\max}$, is $0.23\%$ at $F_0=0.02$ and
$\varepsilon=1.3\%$, as for the single-$\frac13$ angular projection, but $0.19$ at $\varepsilon=0$, against
$0.013$ for the single-$\frac13$ angular projection. At $F_0=0.3$ it is $0.030$ ($\varepsilon=1.3\%$) and
$0.014$ ($\varepsilon=0$), against $0.038$ and $0.016$. It is not shown in Fig.~\ref{fig:proj}.

\paragraph{Field-odd null test.} In strained lattices the linear interband polarization current
$\propto\dot{\mathbf E}$ is anisotropic and enters $\jperp^{(o)}$ at generic angles. It is not a
function of $A$, so $H^{(o)}=0$ is not a valid null test there. Numerically $H^{(o)}/\max|Aj_B|=310$
($F_0=0.02$) and $8.6$ ($F_0=0.3$) at $\theta_0=\pi/4$, $\varepsilon=1.3\%$, against $3\times10^{-5}$
at $\varepsilon=0$.

\section{Continuum model: hysteresis, dephasing and crossover}
\begin{figure}[!htbp]
\includegraphics[width=0.75\textwidth]{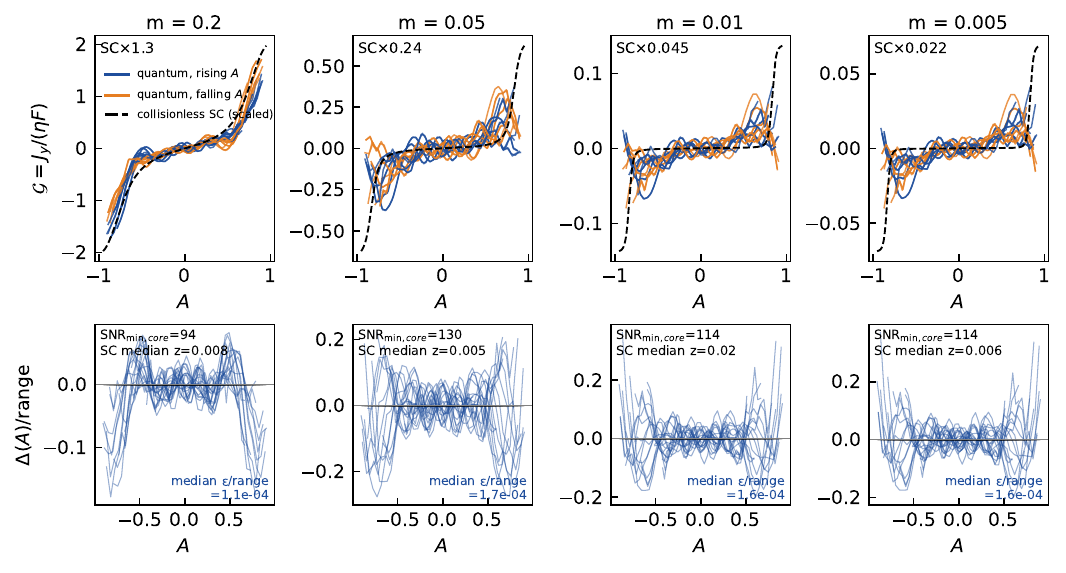}
\caption{Continuum tilted-Dirac model at $F_0=1$: coherent transverse current versus matched vector
potential on rising and falling branches, with pointwise absolute error estimates; the collisionless
semiclassical control is single-valued.}
\end{figure}
\begin{figure}[!htbp]
\includegraphics[width=0.85\textwidth]{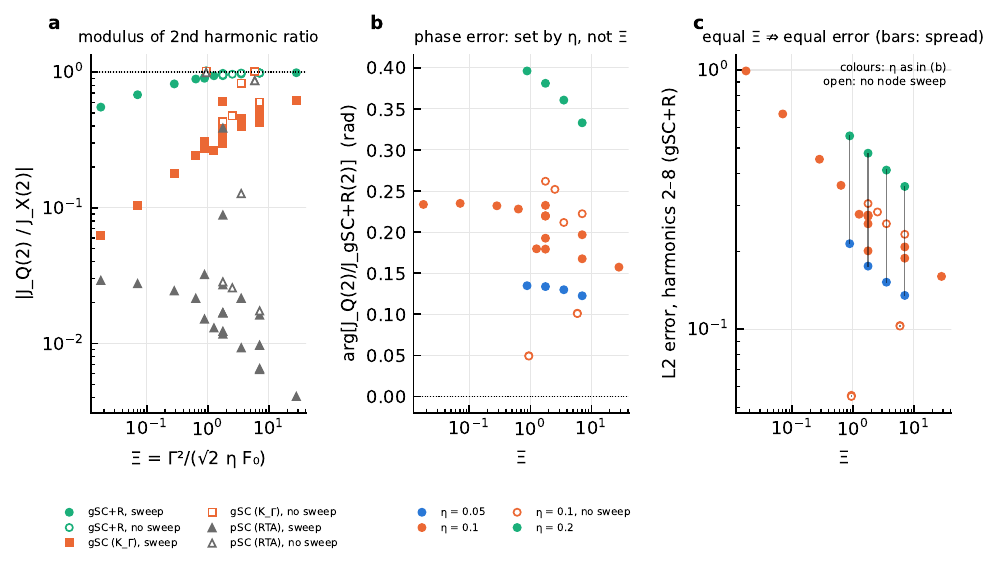}
\caption{Continuum crossover: (a) $|J_2^{\rm Q}/J_2^{\rm gSC+R}|$ versus $\Xi$; (b) its phase,
colored by $\eta$; (c) error spread at equal $\Xi$.}
\label{fig:cross}
\end{figure}

In the continuum model with dephasing $\Gamma$, the generalized semiclassics with
dephasing-assisted transitions recover the modulus $|J_2|$ of the coherent calculation to
$\le1.5\%$ at equal $\Xi=\Gamma^2/(\sqrt2\eta F_0)$, and to $0.96$--$1.01$ for $\Xi\gtrsim2$. The
complex error is not organized by $\Xi$: a phase lag of $0.12$, $0.16$--$0.24$, and $0.33$--$0.40$~rad
remains at $\eta=0.05$, $0.1$, and $0.2$ (Fig.~\ref{fig:cross}). The two-valley continuum model obeys
the exact identity $W(\theta+\pi)=W(\theta)$, which makes its field-even work projection trivial. A
single valley breaks the identity: there the coherent field-even projected current deviates from the
Berry current by $5$--$57\%$ of the latter, beyond numerical error.

\section{Kane--Mele model details}\label{sec:kmSM}
Each spin block is $H_s=d_0+\mathrm{Re}f\sigma_x+\mathrm{Im}f\sigma_y+[\Delta-s\Delta_{\rm SO}g(\bk)]\sigma_z$,
with $g=\tfrac{2}{3\sqrt3}[\sin\bk\cdot\mathbf R_1+\sin\bk\cdot\mathbf R_2-\sin\bk\cdot\mathbf R_3]$. Here
$\Delta_{\rm SO}$ is the Dirac mass, and the Kane--Mele hopping is $\lambda_{\rm SO}=\Delta_{\rm SO}/(3\sqrt3)$.
Under strain $\lambda_{\rm SO}$ is held fixed, which is a model choice. The checks are: spinful time
reversal $H_\downarrow(-\bk)^*=H_\uparrow(\bk)$; the mirror with spin flip
$H_{-s}(-k_x,k_y)=H_s(\bk)$; inversion at $\Delta=0$; and $C_3$ at $\varepsilon=0$. Fukui link
variables~\cite{S-Fukui2005} in the periodic gauge agree with curvature integration ($<10^{-4}$) and
are converged for $N=48$--$192$. The second-order kernel is the zero-field Taylor kernel with
generalized jets including $\partial d_z$; it reduces exactly to the spinless kernel for constant
mass. The charge tensor obeys the spinless selection rules, while the spin-resolved tensors carry
mirror-odd components that cancel in the sum. The mirror $M_y$ maps $(\Delta,\Delta_{\rm SO})\to(-\Delta,-\Delta_{\rm SO})$,
so the charge response is odd in $\Delta$.

Spin Chern numbers from gauge-invariant link variables agree with direct curvature integration. The
critical point is located from the gap closing at $\Delta_c=0.0399806$ for $\varepsilon=1.3\%$. At
matched minimum gap topological and trivial partners differ in every nonlinear coefficient, and in the
insulating state some coefficients reverse sign (main-text Fig.~\ref{fig:km}a). The two
non-topological controls lie in the same parameter space. The first is a trivial Kane--Mele layer
with the roles of $\Delta$ and $\Delta_{\rm SO}$ interchanged and the same form factor $g(\bk)$:
$(\Delta',\Delta_{\rm SO}')=(0.04,0.02)$ and $(0.04,0.03)$ for the topological points
$(\Delta,\Delta_{\rm SO})=(0.02,0.04)$ and $(0.03,0.04)$, with valley masses $(0.02,0.06)$ and
$(0.01,0.07)$, the same magnitudes as the topological partners but without mass inversion. The second
is the reversal of the staggered potential at $\Delta_{\rm SO}=0$, $\Delta=-0.03,\dots,0.03$, which
inverts all four Dirac cones without changing any spin Chern number. Valley additivity
holds in the coherent dynamics at $F_0=0.3$ to $0.4\%$ (doped) and $2\%$ (insulating). In the doped layer at $\varepsilon=1.3\%$
the in-phase correction $(K_\Q-K_\SC)/K_\SC$ lies between $0.7476\%$ and $0.7602\%$ over all computed $\Delta$, on both sides of
the transition and for the controls ($0.748$--$0.760\%$). For germanene at
zero perpendicular field, inversion symmetry makes the second-order charge response vanish
identically.

The germanene-motivated parameters are $t=1.31$~eV and $\Delta_{\rm SO}=12$~meV, from the
first-principles gap of $23.9$--$24.2$~meV of free-standing germanene~\cite{S-Liu2011PRL,S-Liu2011PRB,S-Villarreal2026}.
For supported germanene the field-driven transition has been observed~\cite{S-Bampoulis2023}. The
nominal tight-binding intrinsic coupling of Ref.~\cite{S-Liu2011PRB} overestimates the first-principles
gap and is not used as the Dirac mass.

\begin{figure}[!htbp]
\includegraphics[width=0.85\textwidth]{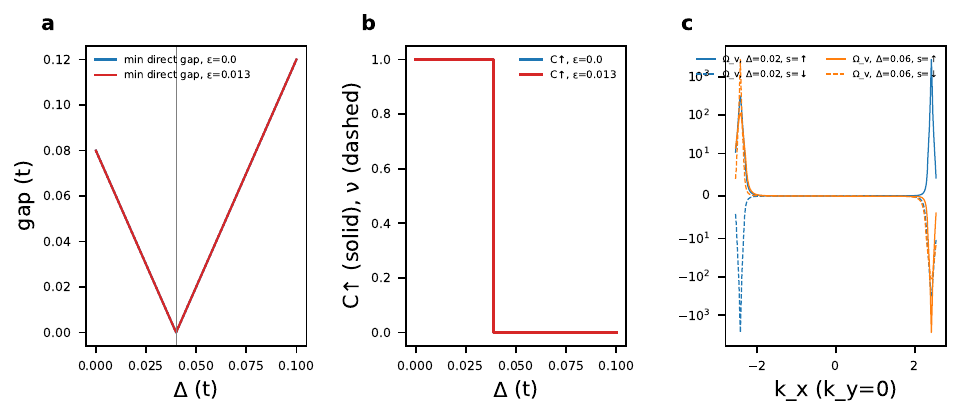}
\caption{Kane--Mele phase diagram: (a) minimum direct gap versus $\Delta$; (b) spin Chern number;
(c) spin-resolved valence Berry curvature along $k_y=0$ for the topological ($\Delta=0.02$) and
trivial ($\Delta=0.06$) partners.}
\label{fig:kmtopo}
\end{figure}

\begin{figure}[!htbp]
\includegraphics[width=0.85\textwidth]{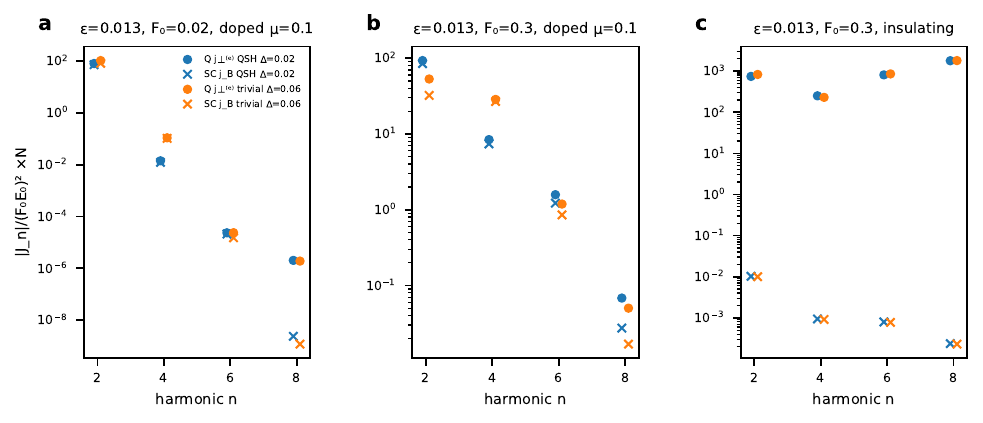}
\caption{Kane--Mele coherent dynamics: harmonic moduli of the field-even transverse current for the
matched topological/trivial pair at $\varepsilon=1.3\%$: (a) $F_0=0.02$, doped; (b) $F_0=0.3$,
doped; (c) $F_0=0.3$, insulating.}
\label{fig:kmdyn}
\end{figure}

\emph{Parameter sets.} All Kane--Mele figures of the main text and Figs.~\ref{fig:kmtopo}--\ref{fig:kmdyn} use the illustrative
set $\Delta_{\rm SO}=0.04\,t$. The germanene-motivated set $\Delta_{\rm SO}=0.0092\,t$ (12~meV at
$t=1.31$~eV) was evaluated only at the weak-field level, for the matched pair $\Delta=0$ (topological)
and $\Delta=2\Delta_{\rm SO}$ (trivial).
\begin{itemize}
\item At $\Delta=0$, inversion symmetry makes all second-order responses vanish.
\item At $\Delta=2\Delta_{\rm SO}$ and $\varepsilon=1.3\%$ (doped, fixed $\mu$): $K_{\rm Q}=-132.73i$
versus $K_{\rm SC}=-131.73i$, an in-phase correction of $0.76\%$.
\item In the insulating state the gap, $0.0183\,t$, lies below $2\hbar\omega=0.02\,t$, so the response
is resonant and acquires large real parts.
\end{itemize}
These values are not quantitative predictions for germanene, since the model omits Rashba coupling,
substrate effects, and calibrated hoppings.


\section{Summary of numerical checks}
Table~\ref{tab:checks} lists the numerical checks, with the number of cases that met the stated
criterion. Checks that did not meet it fall into three classes: known numerical limits
(finite-difference roundoff and truncation, summation roundoff at the mirror angle, small
fourth-harmonic signals), coarse grids kept for comparison, and the inaccuracy of the low-order
truncated amplitude series at $F_0=0.3$ (Sec.~\ref{sec:orders}).
\begin{table}[!htbp]
\caption{Summary of numerical checks (scripts in Ref.~\cite{S-dataset}).}
\label{tab:checks}
\begin{ruledtabular}
\begin{tabular}{p{0.52\textwidth}p{0.12\textwidth}p{0.3\textwidth}}
check & cases & note\\
\hline
Hamiltonian symmetries; BZ area and twist & all & $\le10^{-15}$\\
mirror-forbidden tensor components; $C_3$ relations & 9/9, 2/2 & $\le2\times10^{-9}$\\
angular identity $H=K\cos\theta$ & 9/9 & removal factor $\ge8\times10^8$\\
energy identity, $W_{\rm abs}\ge0$, cyclicity & 13/13 (each) & \\
$W_{\rm abs}$ time/grid convergence ($F_0=0.3$) & met & $0.07\%$ / $0.4\%$\\
tangent vs.\ complex step $<10^{-3}$ / $<10^{-8}$ & 6/6 / 4/6 & summation roundoff at $\theta_0=0$\\
finite-difference Richardson $<10^{-4}$; $h^2$ scaling & 5/6; 4/6 & FD roundoff at $\theta_0=0$\\
order-fit stability ($<3\%$) & 6/8 & two cases at $3.4$--$3.5\%$\\
projection vs.\ rank-5 symmetrization; no $\cos5\theta$ & 4/4; 4/4 & \\
direct octupole vs.\ fitted $j_B^{(4)}$ ($<0.1\%$) & 4/4 & $\le0.05\%$\\
closed form vs.\ exact arithmetic / brute-force resolvent & 60/60; 4/4 & \\
closed form vs.\ general two-band formula of Ref.~\cite{S-HeLaw2026} & 4/4 & all printed digits\\
two-term series within $1\%$ & 31/40 & fails at $F_0=0.3$ and for tiny $n=4$ signals\\
$F_0=1.5$ harmonics: fine, angular-only, extra-fine grids & met & base and radial-only grids unconverged in angle\\
semiclassical identity with exact $\Phi$ / with time-quadrature $W$ & 14/14 / 7/14 & quadrature error in the field-odd channel at $\theta_0=\pi/4$\\
quantum Hall-type sector $=j_B$ within numerical error & 0/4 (refined) & quantum corrections (main text)\\
nonperturbative projection: $H^\SC_{\rm NP}=j_B$ ($<10^{-6}$) & 8/8 & decisive: field-odd, $\theta_0=\pi/4$\\
$H^\Q_{\rm NP}$ vs.\ $H^{(2)}+(F_0/0.01)^2H^{(4)}$ (fit unc.), $F_0=0.01$--$0.1$ & 3/3 & $10^{-6}$ at $F_0=0.01$\\
$H_{\rm NP}$ node change $<$ time-step error & 1/2 & $F_0=1$: comparable (pointwise max)\\
Kane--Mele topology, symmetry, selection rules, additivity & 489/489 & \\
\end{tabular}
\end{ruledtabular}
\end{table}

\clearpage

\makeatletter
\begingroup
\let\SM@label\label
\def\label#1{\SM@label{#1S}}
\def\NAT@bibsetnum#1{\setlength{\topsep}{\z@}\NATx@bibsetnum{\ref{LastBibItemS}}}
\endgroup
\makeatother
\end{document}